\documentclass[twocolumn,aps,prd,superscriptaddress]{revtex4-2}
\pdfoutput=1
\usepackage{amsmath}
\usepackage{anysize}
\PassOptionsToPackage{hyphens}{url}\usepackage[pdftex]{hyperref}
\usepackage{float}
\usepackage{color}
\usepackage{latexsym}
\usepackage{lineno}
\usepackage{graphicx}
\usepackage{comment}
\usepackage{siunitx}
\usepackage{textcomp,gensymb}
\usepackage[normalem]{ulem}  
\usepackage[caption=false]{subfig}
\advance\textheight1cm
\advance\voffset-1cm

\def\subsubsubsection#1{

\vskip .2cm

\noindent{\it #1}

\vskip .1cm

}

\newcommand{\ra}        {\mbox{$\rightarrow$}}

\newcommand{\red} {\color{red} }

\begin{document}

\title{Mu2e-II: Muon to electron conversion with PIP-II \\ Contributed paper for Snowmass}

\author{K.~Byrum}
\author{S.~Corrodi}
\author{Y.~Oksuzian}
\author{P.~Winter}
\author{L.~Xia}
\affiliation{Argonne National Laboratory, Lemont, Illinois 60439, USA}

\author{A.~W.~J.~Edmonds}
\author{J.~P.~Miller}
\affiliation{Boston University, Boston, Massachusetts 02215, USA}

\author{J.~Mott}
\affiliation{Boston University, Boston, Massachusetts 02215, USA; Fermi National Accelerator Laboratory, Batavia, Illinois 60510, USA}

\author{W.~J.~Marciano}
\author{R.~Szafron}
\affiliation{Brookhaven National Laboratory, Upton, New York 11973 USA}

% ...

\today

\begin{abstract}
    An observation of Charged Lepton Flavor Violation (CLFV) would be unambiguous evidence for physics beyond the Standard Model. The Mu2e and COMET experiments, under construction, are designed to push the sensitivity to CLFV in the $\mu\to e$ conversion process to unprecedented levels. Whether conversion is observed or not, there is a strong case to be made for further improving sensitivity, or for examining the process on additional target materials.  Mu2e-II is a proposed upgrade to Mu2e, with at least an additional order of magnitude in sensitivity to the conversion rate over Mu2e. The approach and challenges for this proposal are summarized. Mu2e-II may be regarded as the next logical step in a continued high-intensity muon program at FNAL.
\end{abstract}

\maketitle

\tableofcontents

\section{Overview} 

Charged Lepton Flavor Violation (CLFV) provides an extremely sensitive window into new-physics scenarios, capable of indirectly probing mass scales far beyond the direct reach of both existing and planned colliders. The observation of CLFV would provide clear evidence for phenomena beyond the Standard Model (BSM). Many well-motivated models predict testable CLFV rates involving muons, which lend themselves to very precise measurements. 
We propose to search for the bound-muon to electron conversion process with a significantly improved discovery potential over currently planned projects. This idea has already been outlined in a previous expression of interest~\cite{Abusalma2018}.

The plan is to continue the exploration of the CLFV process, $\mu\to e$ conversion (e.g., \cite{Dinh2013,deGouvea2013}), in the field of a nucleus as an
evolution of the Mu2e experiment~\cite{Ambrose2020a} at FNAL, which we call Mu2e-II. The anticipated single event sensitivity (SES) of Mu2e is
\begin{align}
\begin{split}
 R_{\mu e} &\equiv \frac{\Gamma(\mu^-N(A,Z)\to e^-N(A,Z))}{\Gamma(\mu^-N(A,Z)\to \nu_\mu N(A,Z-1)^*)} \\
 &= 3\times 10^{-17}
 \end{split}
\end{align}
for capture in an aluminum nucleus.
Previous experiments have set limits using various other targets: copper ($R_{\mu e} < 2\times 10^{-8}$)~\cite{Bryman:1972rf,Zyla:2020zbs}, sulfur ($R_{\mu e} < 7\times 10^{-11}$)~\cite{Badertscher:1980bt}, lead ($R_{\mu e} < 5\times 10^{-11}$)~\cite{SINDRUMII:1996fti}, titanium ($R_{\mu e} < 4\times 10^{-12}$)~\cite{SINDRUMII:1993gxf}, and gold ($R_{\mu e} < 7\times 10^{-13}$)~\cite{Bertl2006}. 
Note that the nucleus-dependence of the conversion rate complicates the comparison of these results; nuclear dependence is discussed in section~\ref{sec:theory}.
In addition to Mu2e, the COMET collaboration
 is preparing their apparatus at J-PARC, with planned SES of $3\times10^{-15}$ for Phase-I and $\mathcal{O}(10^{-17})$ for Phase-II~\cite{Lee2018}, also using aluminum. The DeeMe collaboration, also at J-PARC, is preparing for a SES of up to $10^{-14}$ on carbon~\cite{Teshima2019}. The aim of Mu2e-II is a further order of magnitude improvement in sensitivity to $\mu\to e$ conversion over Mu2e.

The observation of neutrino oscillations has shown that lepton flavor is not conserved in nature, which qualitatively predicts CLFV such as muon-to-electron conversion. However, in the simplest Standard-Model extension that allows for neutrino masses, all CLFV rates are suppressed by the neutrino masses, resulting in predicted rates for $R_{\mu e}$ below $10^{-50}$~\cite{Marciano2008}, that is, unobservably small.
The observation of CLFV would therefore be evidence for new physics that goes even beyond neutrino masses, with many well-motivated models leading to testable rates~\cite{Heeck:2016xwg,Lindner:2016bgg,Calibbi:2017uvl}.

Along with $\mu^-\to e^-$ conversion, there are other new-physics signatures that can be investigated in Mu2e-II~\cite{Bernstein2020a}. One example is the $\Delta L = 2$ CLFV process $\mu^-\to e^+$~\cite{Geib2016,Berryman2017}, which generates an approximately monoenergetic positron. A precise measurement of the decay-in-orbit background tail can furthermore reveal other signatures, such as a $\mu\to e X$ decay involving an ultralight new boson~$X$~\cite{GarciaTormo:2011jit,Uesaka:2020okd}.

Construction of the present Mu2e experiment is nearing completion, with data-taking beginning in FY2024 and extending for several years beyond. If a signal for $\mu\to e$ conversion is found, it will be essential to improve the search to help understand the nature of the BSM physics. This can be done by changing the target material, comparing an aluminum target to other targets~\cite{Cirigliano2009,Davidson2018}. We note also that the best existing limit for $\mu^-\to e^+$ is on titanium~\cite{Kaulard1998}, and we include this option. On the other hand, if neither COMET nor Mu2e finds a signal
(or they disagree!) it will be important to improve sensitivity and push the search for new physics to a higher mass scale.
We thus regard Mu2e-II as the next logical step following Mu2e in a future CLFV program with intense muon beams at FNAL. It is currently planned that Mu2e ``Run II'' data-taking will extend into FY2030.
Some construction activities for Mu2e-II could be carried out in parallel with Mu2e running, with access to the Mu2e hall beginning after the Mu2e run is completed.

 Mu2e-II relies on the existence of a more powerful source of protons, the PIP-II linac~\cite{Ball2017},
under construction at Fermilab. This will provide $\sim1.4\times10^{9}$ 800 MeV protons/pulse for Mu2e-II,
compared with $3.9\times10^{7}$ 8 GeV protons/pulse at Mu2e, with 1.7 $\mu$s pulse spacing. An $H^-$ beam is extracted directly from the linac (with additional RF to handle the beam load), stripped of electrons and transported in a new beamline to match the Mu2e  
beamline. The delivery ring and resonant extraction are no longer used,
removing one source of radiation hazard and providing for a much more
stable beam intensity as well as flexibility in time structure. Another improvement is that narrower pulses, $\sim$100 ns, can be delivered, compared with $\sim$250 ns for Mu2e.

We note that the efficiency to produce muons is comparable, for a given beam power, at 800 MeV and 8 GeV. The factor of ten gain in sensitivity and discovery reach over Mu2e is achieved by a combination of higher intensity and higher duty factor.
There are a number of challenges to handling the more powerful beam and
reducing backgrounds sufficiently to achieve the greater sensitivity. The R\&D is, in several cases, already in progress.

The accelerator~\cite{Prebys2020} delivers 100 kW (or more), of which over 20 kW
is deposited as heat in the target. This is too much to remove radiatively, as is done for Mu2e. Thus various alternatives are under investigation~\cite{Neuffer2020}.
Further, the lower momentum beam has increased
curvature in the solenoidal field around the target. A curved target is needed in order to optimize the muon rate. 

Mu2e-II, like Mu2e,  uses a pulsed proton beam to eliminate a limiting background of SINDRUM~\cite{Bertl2006}. After the beam hits the target, produced pions decay to muons, which propagate to the stopping target~\cite{Miller2020} and are eventually captured in atoms. There is a quiet time between beam pulses when the spray of particles from proton interactions with the production target has stopped. It is during this time that the search for conversion electrons is done. However, protons outside of the main beam pulse can produce backgrounds. Thus, the ``extinction'' of protons during this period is very important. For Mu2e, the extinction must be kept to a level of $10^{-10}$. The requirement becomes $10^{-11}$ for Mu2e-II. A combination of techniques is used to obtain this suppression in Mu2e, and similar techniques are available to Mu2e-II. The PIP-II linac pulse is narrower than the Mu2e resonantly extracted beam, aiding in the extinction.

The more intense beam means that radiation and shielding must be re-evaluated. While the delivery ring is no longer an issue, the M4 beam line, production solenoid and associated shielding, as well as downstream components and Mu2e building shielding may all need significant alteration. An advantage of Mu2e-II is that the 800 MeV proton energy is below anti-proton production threshold, eliminating a potentially problematic background.

The Mu2e tracker is a straw tube chamber with 15 $\mu$m aluminized mylar straws. It is crucial to discriminating the monochromatic conversion signal from backgrounds such as muon decays in atomic orbit in the stopping target. Key requirements are on the momentum resolution and pattern recognition. This requires a new chamber, along with re-optimization of other sources of multiple scattering. Ideas are being explored~\cite{Ambrose2020}, and R\&D is
already underway on the possibility of reducing the thickness of straws for a new straw chamber solution.

The Mu2e calorimeter, used for particle identification, triggering, and as a cross check on the tracker momentum, consists of 1348 CsI crystals read out by SiPMs~\cite{Atanov2018}. CsI has a moderate
scintillation light decay time, of the order of 30-40 ns. This is marginal for the higher rates at Mu2e-II so alternatives are needed~\cite{Davydov2020}. Especially promising is the use of BaF$_2$, which has a very fast (sub-nanosecond) component at
around 220 nm. Unfortunately, it also has a very slow component at longer wavelength, and R\&D has begun on suppressing the slow component by
doping with dopants such as yttrium and developing ``solar-blind'' readout~\cite{Zhu2017, Chen2018, Hu2019, Hitlin2020, Hitlin2020a, Hu2020}.

 \begin{table}[t!]
\caption{Selected nominal Mu2e-II quantities}
\label{tab:Mu2eII}
\begin{center}
\begin{tabular}{ l  l  }
\hline\hline
\multicolumn{1}{ c }{Parameter} & \multicolumn{1}{ c }{Value}  \\
\hline\hline
Proton kinetic energy & 0.8   GeV\\
Beam Power &100 kW  \\
Protons/s & $ 7.8 \times 10^{14}$   \\
Pulse Cycle Length &  1.693 $\mu$s (variable) \\
Extinction & $< 10^{-11}$\\
Stopped $\mu$ per proton  & $9.1 \times 10^{-5}$  \\
Stopped $\mu$ per cycle  & $1.2 \times 10^{5} $ \\
Event size & 1 MB\\
Storage & 14 PB/yr\\
Run period & 5  yr\\
Single event sensitivity & $3.25\times 10^{-18}$\\
Total background & 0.47 events\\
$R_{\mu e}$ (discovery) & $2.3\times 10^{-17}$\\
$R_{\mu e}$ (90\% CL) & $6.4\times 10^{-18}$\\
\hline\hline
\end{tabular}
\end{center}
\end{table}

Cosmic rays are a major background consideration and their rejection is crucial to obtaining the desired sensitivity. The higher running duty factor means about a factor of three greater livetime for Mu2e-II compared with Mu2e. The scintillator-based Mu2e cosmic ray veto (CRV) system is not sufficient for Mu2e-II. Additional shielding and different materials can help, but the unavoidable gaps in the Mu2e CRV counters around the solenoids are a limitation. A new geometry to close these gaps is required.
R\&D is needed to develop and optimize the system, and we are considering design choices~\cite{Byrum2020}.

Besides the background from cosmic ray events, which is estimated to dominate in the final sample, there are several other sources of background that need to be controlled. As mentioned earlier, much background is suppressed by timing the trigger away from the primary beam pulse, and by insuring that there is very little out-of-time beam contamination. An important background arises from decays in orbit (DIO) in which a muon orbiting the nucleus undergoes a standard model decay. This is irreducible other than with excellent momentum resolution on the electron. Other background processes include radiative decays of both pions and muons as they capture on the nucleus.    
The estimated background contributions in Mu2e-II are shown in Table~\ref{tab:background_summary} in section~\ref{sensitivity_estimate}.

Mu2e-II will have an order of magnitude higher data rate than Mu2e, as well as higher dosing of the front-end electronics. Thus, there are also challenges for the trigger and data acquisition, and an R\&D program is planned to investigate
possible approaches~\cite{Tran2020a, Tran2020b, Bonventre2020, Gioiosa2020}.

As a natural evolution of Mu2e, Mu2e-II provides the nearest-term next step in a possible major future muon program at FNAL. Considerable infrastructure, expertise, and experience has been and is being developed at Fermilab such that it is logical to build further upon this investment, especially given the construction of the powerful PIP-II accelerator.

The following sections elaborate on the potential physics, challenges, technological options, and required R\&D towards Mu2e-II. Table~\ref{tab:Mu2eII} summarizes a few selected quantities.


\section{Theory}
\label{sec:theory}
\subsection{General motivation for CLFV searches}
\begin{figure} [t]
    \centering
    \includegraphics[width=3in]{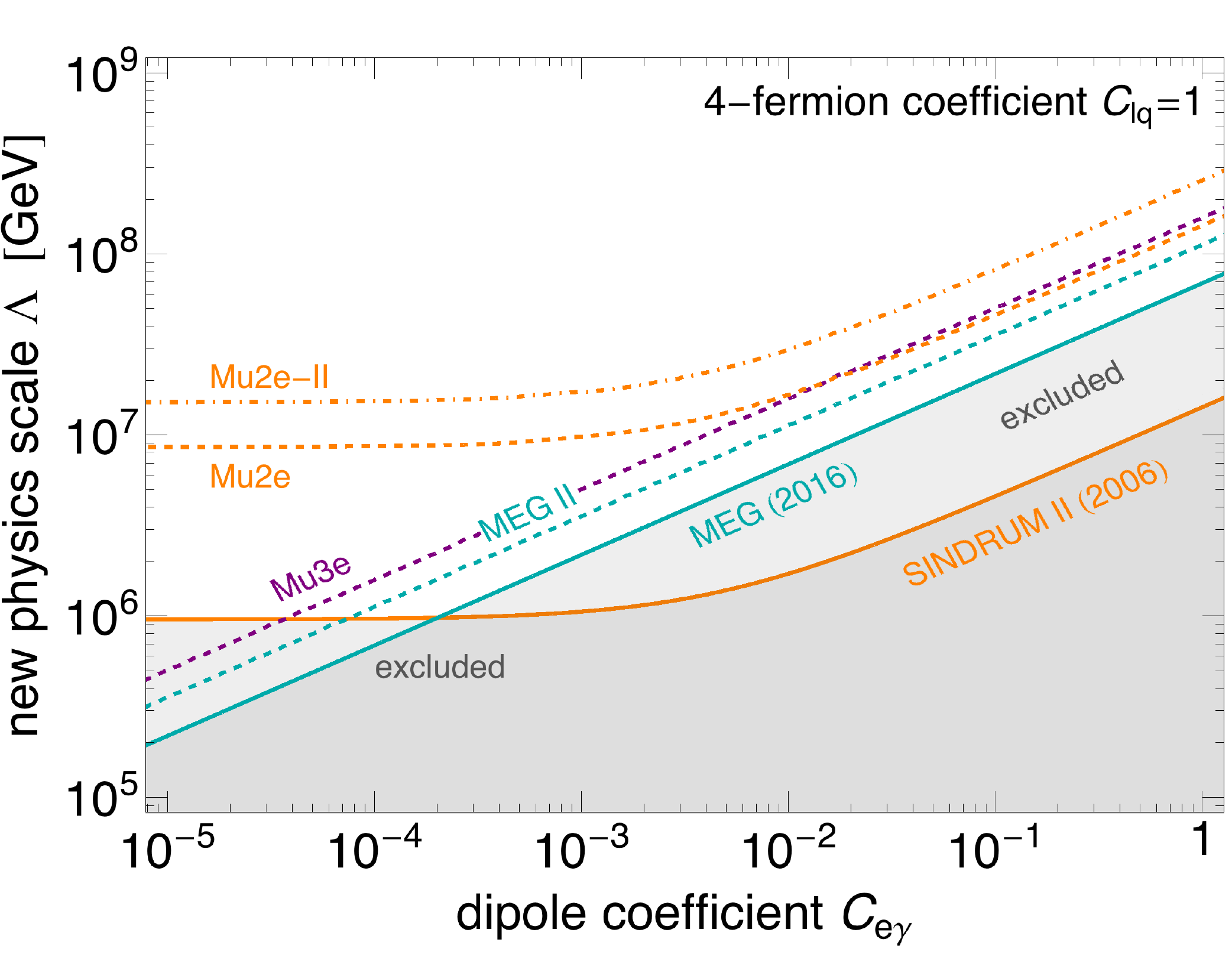}
    \caption{Current and future 90\% CL bounds from $\mu\to e \gamma$~\cite{MEG:2016leq,MEGII:2018kmf} (cyan lines), $\mu \to eee$~\cite{Blondel:2013ia} (purple line) and $\mu\to e$ conversion (orange lines) on the effective CLFV operators $\frac{C_{e\gamma}}{\Lambda^2}\langle H \rangle \,\overline{e}_L\, \sigma^{\mu\nu} \mu_R\, F_{\mu\nu} +\frac{C_{\ell q}}{\Lambda^2}\,(\overline{e}_L\, \gamma^{\mu}  \mu_L)\,(\overline{Q}\, \gamma_{\mu} Q)$, where $Q=(u_L,d_L)^T$ and $\langle H \rangle\simeq 174\,$GeV is the vacuum expectation value of the Higgs field. The limits are displayed as functions of the new-physics scale $\Lambda$ and the Wilson coefficient of the dipole operator $C_{e\gamma}$, while the coefficient of the 4-fermion operator is set to $C_{\ell q}=1$.}
    \label{fig:eff-ops}
\end{figure}

It was realized long ago that searches for CLFV are among the cleanest and most sensitive tests for physics beyond the Standard Model, cf.~the reviews in~\cite{Kuno:1999jp,Lindner:2016bgg,Calibbi:2017uvl}. Current limits constrain the cutoff scale of CLFV operators up to $10^3-10^4$~TeV, see e.g.~\cite{Calibbi:2017uvl} and Fig.~\ref{fig:eff-ops}. This means that searches for CLFV processes are potentially sensitive to virtual effects due to the presence of new particles whose masses are several orders of magnitude larger than the energies accessible at our present and future colliders.

The observation of neutrino oscillations has provided evidence that lepton family numbers are not conserved and the Standard Model needs to be extended to account for neutrino masses. In general, one can expect non-standard contributions to CLFV processes in the context of any extension of the Standard Model that involves new fields coupling to leptons, in particular those addressing the origin of neutrino masses. 
Thus new sources of CLFV beyond those stemming from neutrino oscillation are generically to be expected unless the new physics sector is protected by the same global flavor symmetries of the Standard Model~\cite{Heeck:2016xwg}.
However, in most new-physics scenarios, one can not predict a ``minimum-guaranteed'' amount of CLFV, because, besides depending on the flavor structure of the new-physics interactions, CLFV rates are also suppressed by the unknown new-physics scale. Only very specific models, where both ingredients (mass scale and flavor structure of the couplings) are set by additional phenomenological requirements, can give definite predictions. An example can be found in Ref.~\cite{Vicente:2014wga}, where such requirements include reproducing the measured neutrino mixing and providing a dark matter candidate with the observed properties. Ref.~\cite{Heeck:2018ntp} gives another example of a predictive model where the new-physics scale is fixed by requiring an explanation for the $B$-physics anomalies (see below) and the flavor structure is set by addressing the observed neutrino masses and mixing.

The physics case for CLFV searches has been recently further reinforced by the first results of the FNAL Muon~{g-2} experiment~\cite{Abi:2021gix} and by the so-called $B$-physics anomalies~\cite{Bifani:2018zmi,London:2021lfn,Crivellin:2021sff}. The Muon g-2 experiment  confirmed the long-standing discrepancy between the measurement of the anomalous magnetic moment of the muon, $(g-2)_\mu$, performed at BNL and the theoretical prediction~\cite{Aoyama:2020ynm,Bennett:2006fi}. The two measurements are statistically compatible and their combination currently deviates from the Standard Model prediction by $4.2\sigma$. This makes it very unlikely that the discrepancy is due to a statistical fluctuation or some overlooked systematical effects in the old BNL experiment. Arguably, the only explanations left are (i)~an underestimation of the Standard Model prediction~\cite{Aoyama:2020ynm}, in particular of the leading hadronic contribution, as the recent lattice result by the BMW collaboration may suggest~\cite{Borsanyi:2020mff}; (ii) the presence of additional contributions due to new particles coupling to muons, see Ref.~\cite{Lindner:2016bgg} for a review. The persistent and coherent pattern of anomalies reported in semileptonic $B$ meson decays, especially those of the kind $b\to s \mu\mu$~\cite{Bifani:2018zmi,London:2021lfn}, also seems to point to a new-physics sector coupling preferably to muons. In particular, LHCb has recently released an updated measurement of the theoretically clean lepton flavor universality (LFU) ratio $R_K$ reporting a $3.1\sigma$ deviation from the SM prediction~\cite{Aaij:2021vac}.
In order to be accounted for by new physics, both the $(g-2)_\mu$ discrepancy and the $B$ anomalies would require the existence of new fields coupling to muons at scales $\lesssim 100$~TeV~\cite{Allwicher:2021jkr,DiLuzio:2017chi}, that is, in the range accessible to upcoming CLFV experiments. 
Moreover, any new physics interacting with muons is not in general expected to exhibit a flavor structure aligned to the Standard Model one. In other words, the presence of the flavor-violating counterparts of the new-physics operators responsible for $(g-2)_\mu$ and $b\to s \mu\mu$ is difficult to avoid unless very peculiar flavor symmetries are imposed~\cite{Calibbi:2021qto,Isidori:2021gqe,Glashow:2014iga}. 
Notice, for instance, that any photon-penguin diagram contributing to the muon magnetic moment would  unavoidably contribute to $\mu\to e\gamma$ as well (and also to $\mu\to eee$ and $\mu \to e$ conversion via a virtual photon exchange) if the new-physics fields in the loop couple not only to muons but also to electrons.
Therefore, CLFV rates at observable level are very likely if these experimental anomalies will be confirmed to be signal of new physics. 
In particular, addressing LFU violation in $B$ decays requires a new-physics sector coupled to both quarks and leptons (the paradigmatic example being scalar or vector leptoquarks~\cite{Dorsner:2016wpm}), hence $\mu\to e$ conversion is typically the dominant CLFV channel of such scenarios~\cite{Heeck:2018ntp,Hati:2019ufv}.

\subsection{Specific motivation for a Mu2e upgrade}
Whether a conclusive signal of $\mu\to e$ conversion is found at Mu2e and COMET or not, Mu2e-II is arguably a logical continuation of the present CLFV search campaign.
In case of no evidence of $\mu\to e$ conversion, a further order of magnitude improvement in sensitivity would approximately double the reach in terms of the new physics scale, see Fig.~\ref{fig:eff-ops}. The figure also shows that Mu2e-II would be the most sensitive CLFV probe even if the dominant contribution is given by the dipole operator, while the Mu2e reach is comparable but slightly weaker than the expected final sensitivity of the search for $\mu\to eee$ that will be performed by the Mu3e collaboration~\cite{Blondel:2013ia}. In general, Mu2e-II seems to be capable of testing any source of $\mu\to e$ transitions better than any other experiment, with the exception of those 4-lepton operators that directly induce $\mu\to eee$ and thus are better probed by Mu3e.
The enhanced sensitivity of Mu2e-II to new physics described above would be an important achievement, leading to a likely discovery, in particular if the existence of new physics coupling to muons will be confirmed at Muon g-2 or in semileptonic $B$ decays, and even more so if a signal is found in other CLFV channels by MEG-II~\cite{MEGII:2018kmf} or Mu3e~\cite{Blondel:2013ia}.

\begin{figure}[t]
    \centering
    \includegraphics[width=3in]{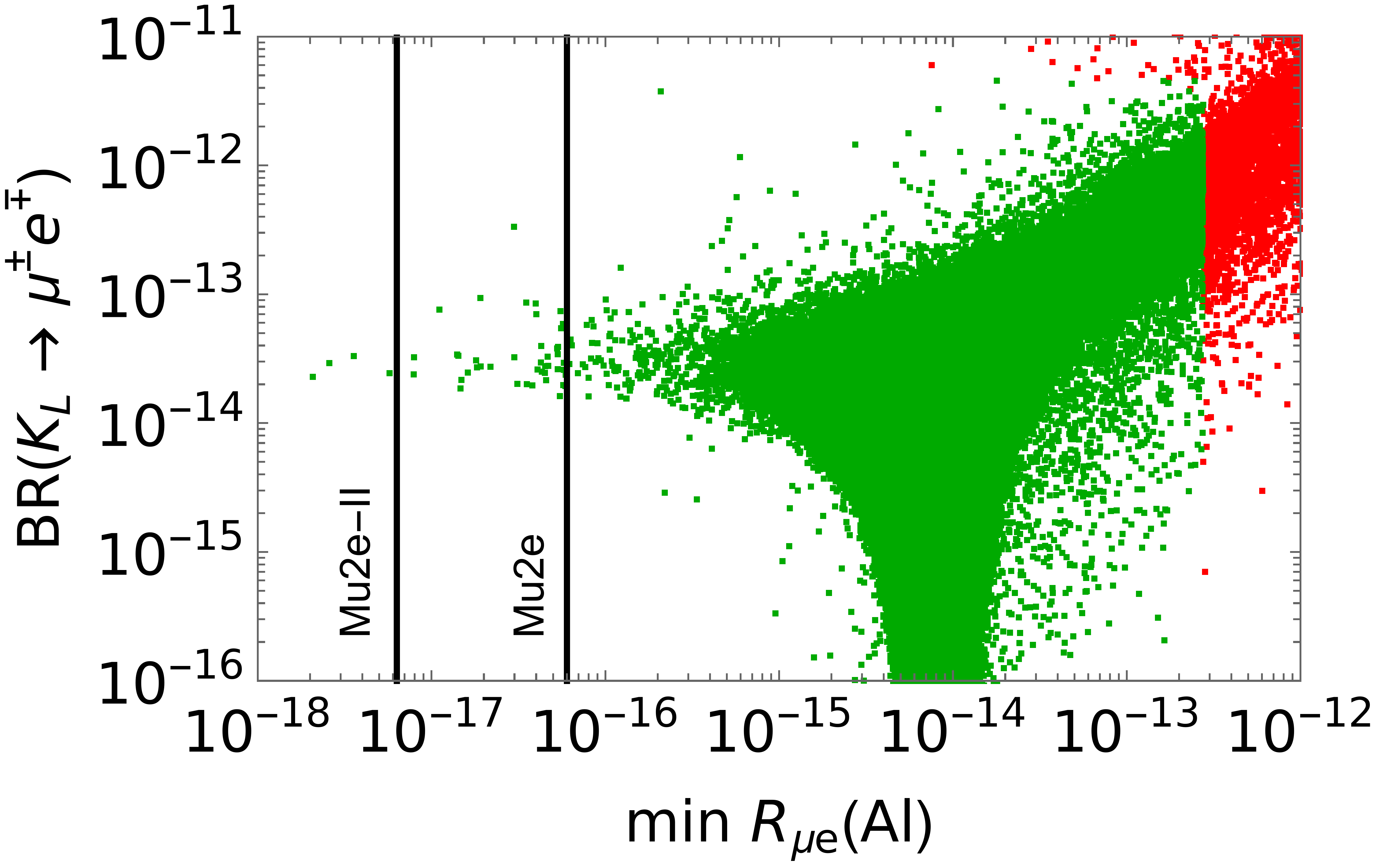}
    \caption{CLFV predictions in the Pati--Salam model of Ref.~\cite{Heeck:2018ntp} that explains neutrino masses together with the neutral-current $B$-meson anomalies. Red points are excluded, green points are currently allowed. The vertical lines denote the $90\%\,$C.L.~reach of Mu2e and Mu2e-II.}
    \label{fig:Pati-Salam-CLFV}
\end{figure}
In some cases, the additional order-of-magnitude sensitivity Mu2e-II strives for can be the determining factor to exclude new physics models. As an example, we show CLFV predictions of Ref.~\cite{Heeck:2018ntp} in Fig.~\ref{fig:Pati-Salam-CLFV}. In this Pati--Salam gauge extension of the Standard Model, leptoquarks were used to explain the neutral-current $B$-meson anomalies -- which fixes the new-physics scale -- and whose flavor structure was determined by that of the neutrino mass matrix due to the Pati--Salam symmetry, providing a lower bound for muon-to-electron conversion that is almost entirely within the reach of Mu2e-II.

\begin{figure}[t]
    \centering
    \includegraphics[width=3in]{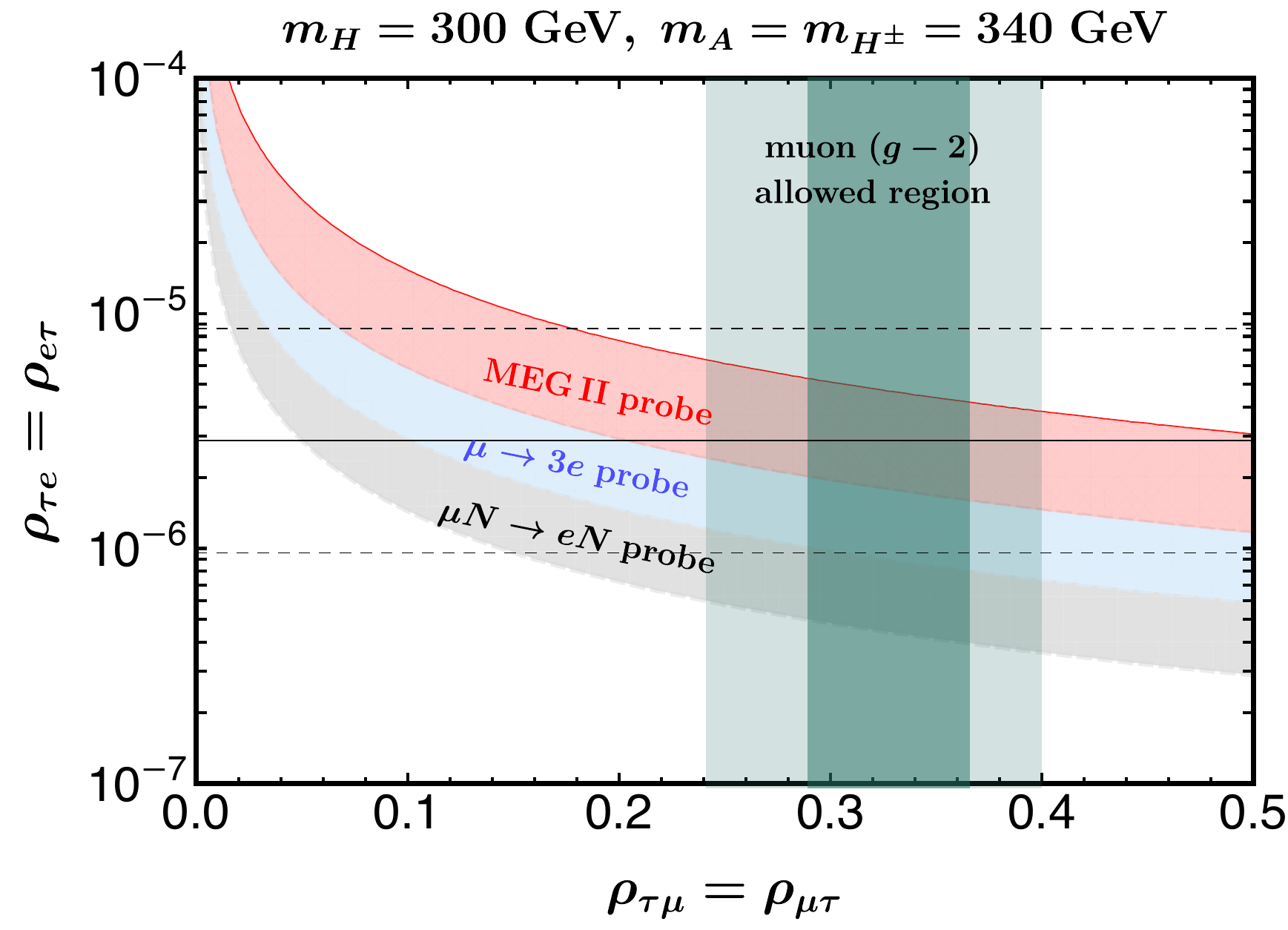}
    \caption{Bounds on the CLFV couplings of the extra Higgs fields in a two-Higgs-doublet model able to explain the   $(g-2)_\mu$ anomaly. Taken from Ref.~\cite{Hou:2021qmf}.}
    \label{fig:2HDM-CLFV}
\end{figure}

Another example is shown in Fig.~\ref{fig:2HDM-CLFV} where the prediction of the two-Higgs-doublet model studied in Ref.~\cite{Hou:2021qmf} is displayed. Here, a non-standard contribution to $(g-2)_\mu$ arises from loops involving the extra Higgs particles and the $\tau$ lepton, thus addressing the Muon g-2 result sets the mass of these fields and their couplings to muons and taus. The coupling of the extra Higgses to electrons and taus are, on the other hand, tightly constrained by CLFV transitions. Fig.~\ref{fig:2HDM-CLFV} shows that a search for $\mu \to e$ conversion with a sensitivity at the $10^{-17}$ level will be able to probe such coupling even below the naturally expected value of the order of the electron Yukawa coupling (horizontal lines).

\subsection{Isotope dependence of muon-to-electron conversion and identification of next targets}
\label{sec:target_isotope_theory}
If $\mu\to e$ conversion (and thus new physics!) is discovered at a previous experiment or in Mu2e-II, it will be of paramount importance to collect more data in the attempt of identifying the kind of new physics at the origin of such a signal.
In particular, it was realized long ago that one can discriminate among different CLFV effective operators using complementary target nuclei due to the dependence of the $\mu \to e$ conversion rate on the target nucleus~\cite{Kitano:2002mt,Cirigliano:2009bz,Bartolotta:2017mff,Davidson:2017nrp,Davidson:2018kud,Davidson:2020ord,Davidson:2020hkf}.
The model-discriminating power of searches for $\mu\to e$ conversion in nuclei can be illustrated by considering the following dimension-6 effective Lagrangian~\cite{Davidson:2018kud}:
\begin{align}
&\mathcal{L}_{\mu e} = -\frac{4 G_F}{\sqrt{2}} \sum_{X = L,R}\left[ m_\mu C_{D,X}\, \overline{e} \sigma^{\alpha\beta} P_X \mu \,F_{\alpha\beta} \right.\nonumber\\
&+\sum_{N=p,n}\left( C_{S,X}^{N}\, \overline{e} P_X \mu \, \overline{N} N
+ C_{P,X}^{N}\, \overline{e} P_X \mu \,  \overline{N} \gamma_5 N\right.\nonumber\\
&+ C_{V,X}^{N}\, \overline{e} \gamma^\alpha P_X \mu \,  \overline{N} \gamma_\alpha N
+ C_{A,X}^{N}\, \overline{e} \gamma^\alpha P_X \mu \,  \overline{N} \gamma_\alpha \gamma_5 N\nonumber\\
&+ C_{Der, X}^{N} \,\overline{e} \gamma^\alpha P_X \mu \,  (\overline{N} \overleftrightarrow{\partial}_\alpha\text{i} \gamma_5 N)\nonumber\\
&+ \left. \left. C_{T, X}^{N}\, \overline{e} \sigma^{\alpha\beta} P_X \mu \,  \overline{N} \sigma_{\alpha\beta} N \right)\right] + \text{h.c.} ,
\label{eq:lagr}
\end{align}
where $P_{L,R}$ are chiral projection operators and the $C_a$ are dimensionless Wilson coefficients.
As long as the scale of new physics $\gg $\,GeV, the above Lagrangian provides a model-independent description of any possible CLFV interactions involving muons, electrons, and nucleons.

Even though each operator in $\mathcal{L}_{\mu e}$ generates $\mu\to e$ conversion, those that are independent of the nuclear spin are expected to be enhanced due to a coherent conversion on all nucleons.
Ultimately, $\mu\to e$ conversion needs to be measured on nuclei with and without spin to fully determine the underlying operator composition~\cite{Davidson:2017nrp}, but for now we assume that the spin-\emph{independent} operators dominate. The corresponding $\mu\to e$ conversion rate can then be written as~\cite{Davidson:2017nrp,Davidson:2018kud}
\begin{align}
R_{\mu e} = \frac{32 G_F^2 }{\Gamma_\text{capture}} \left[  
|\vec{v}\cdot \vec{C}_L|^2+|\vec{v}\cdot \vec{C}_R|^2\right] ,
\label{eq:BR2}
\end{align}
where
\begin{align}
\vec{v} \equiv \left( \frac{D}{4}, V^{(p)}, S^{(p)}, V^{(n)}, S^{(n)}\right) 
\label{eq:v}
\end{align}
is a vector consisting of overlap integrals specific to the $\mu\to e $ conversion target -- calculated in Refs.~\cite{Kitano:2002mt,Heeck2022} --  and 
\begin{align}
\vec{C}_{L} \equiv \left({C}_{D,R}, {C}^{p}_{V,L}, {C}^{p}_{S,R}, {C}^{n}_{V,L}, {C}^{n}_{S,R}\right)  
\label{eq:C}
\end{align}
(and similar with $L\leftrightarrow R$ for $\vec{C}_{R} $) contains linear combinations of the Wilson coefficients appearing in Eq.~\eqref{eq:lagr}, i.e.~all new-physics information.

For a given new-physics model one can calculate the vectors $\vec{C}_{L,R}$ and obtain the conversion rate on a given target. Staying model independent, we can say that by measuring $\mu\to e$ conversion on different nuclei we effectively determine $\vec{C}$ along different \emph{directions}. To get the maximum amount of information about $\vec{C}$, i.e.~the new physics model, it is then necessary to measure muon conversion in targets that probe $\vec{C}$ along different directions, i.e.~have overlap vectors $\vec{v}$ that have large angles with respect to each other.
For complementarity with respect to aluminium, the relevant angle is quantified as~\cite{Davidson:2018kud}
\begin{align}
\theta_\text{Al} = \arccos \left( \frac{\vec{v}\cdot \vec{v}_\text{Al}}{|\vec{v}|| \vec{v}_\text{Al}|}\right) .
\label{eq:theta}
\end{align}
As pointed out long ago and confirmed in Refs.~\cite{Davidson:2018kud,Heeck2022}, light and heavy targets provide good complementarity, so an ideal second target after aluminium would be heavy, say gold or lead.
Within the Mu2e-II experimental setup, this is problematic due to the short muon lifetime in heavier elements. As such, we restrict ourselves to targets with $Z<25$ here, which gives muon lifetimes in excess of 250\,ns~\cite{Suzuki:1987jf} that should be suitable for an experiment like Mu2e-II.

\begin{figure} [t]
    \centering
    \includegraphics[width=3in]{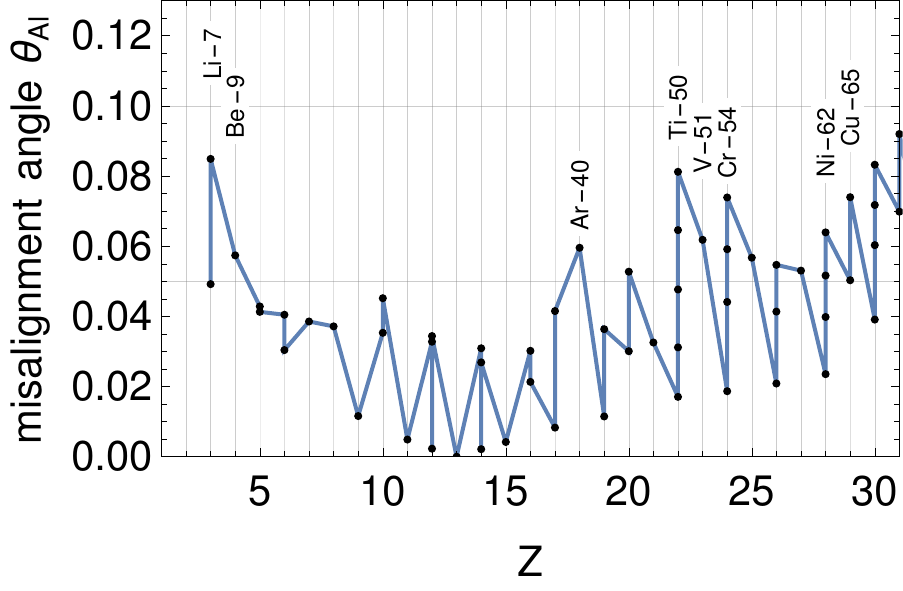}
    \caption{Misalignment angle with Al, as calculated with Eq.~\eqref{eq:theta}, taken from Ref.~\cite{Heeck2022}. The misalignment angle increases with the number of neutrons in isotopes.}
    \label{fig:Al_misalignment_isotopes_zoom}
\end{figure}

For these low-$Z$ targets, the misalignment angle $\theta_\text{Al}$ is shown in Fig.~\ref{fig:Al_misalignment_isotopes_zoom}, where we see that 
among low-$Z$ targets, lithium-7 and titanium-50 show the largest complementarity with respect to aluminium~\cite{Heeck2022}, followed by chromium-54 and vanadium. They have larger $N/Z $ ratios, $ 2.33$ and $2.27$ for lithium-7 and titanium-50, respectively, compared to Al's $N/Z\simeq 2.08$, which might ultimately help to distinguish CLFV operators involving protons from those involving neutrons~\cite{Davidson:2018kud}. 
Lithium has already been identified as a promising target in Ref.~\cite{Davidson:2018kud}.
Titanium has long been proposed as a suitable second target for aluminum-based experiments and the analysis of Ref.~\cite{Heeck2022} shows that the isotope Ti-50 would be particularly useful; aside from the conversion rate and the background from muon decay in orbit, different isotopes of an element are expected to behave essentially identically experimentally, notably because the conversion energy depends only weakly on the number of neutrons~\cite{Heeck2021}.
Some details about the isotopes of interest are collected in Tab.~\ref{tab:B_and_binding}.
The theoretically interesting isotopes Ti-50, Ti-49, and Cr-54 have a low natural abundance and are difficult to  enrich in the large quantities necessary for Mu2e-II; Li-7 and V-51, on the other hand, are the dominant isotopes and hence practically preferable.

\begin{table*}[ht!]
\renewcommand{\baselinestretch}{1.0}\normalsize 
\caption{\label{tab:B_and_binding}
Parameters for the DIO electron spectrum near the endpoint of Eq.~\eqref{eq:spectrum} for select isotopes together with the isotope spin and natural abundance (NA), taken from Ref.~\cite{Heeck2021}.
$E_\text{end}$ also corresponds to the electron energy in $\mu\to e$ conversion.
Muon lifetime $\tau_\mu$ and capture rate $\Gamma_\text{cap}$ are taken from Ref.~\cite{Suzuki:1987jf}; the numbers for titanium refer to a natural-abundance composition dominated by Ti-48.
}
\renewcommand{\baselinestretch}{1.3}\normalsize 
\centering
\begin{tabular}[t]{ lccclcc}
\hline\hline
 & spin & NA/\%& $E_\text{end}/$MeV & $B/\text{MeV}^{-6}$ & $\tau_\mu$/ns & $\Gamma_\text{cap}/s^{-1}$ \\
\hline\hline                    
${}^{6}_{3}\text{Li}$ & $1$ & 7 & $104.64$ & $1.3\times 10^{-19}$ & 2175.3 & 4680\\
${}^{7}_{3}\text{Li}$ & $\tfrac{3}{2}$ & 93 & $104.78$ & $1.3\times 10^{-19}$ & 2186.8 & 2260 \\
${}^{27}_{13}\text{Al}$ & $\tfrac{5}{2}$ & 100 & $104.97$ & $8.9\times 10^{-17}$ & 864 & $662\times 10^3$\\
${}^{46}_{22}\text{Ti}$ & $0$ & 8 & $104.25$ & $5.2\times 10^{-16}$ & & \\
${}^{47}_{22}\text{Ti}$ & $\tfrac{5}{2}$ & 7 & $104.26$ & $5.3\times 10^{-16}$ & & \\
${}^{48}_{22}\text{Ti}$ & $0$ & 74 & $104.26$ & $5.3\times 10^{-16}$ & 329.3 & $2.59\times 10^6$\\
${}^{49}_{22}\text{Ti}$ & $\tfrac{7}{2}$ & 5 & $104.26$ & $5.4\times 10^{-16}$ & & \\
${}^{50}_{22}\text{Ti}$ & $0$ & 5 & $104.26$ & $5.4\times 10^{-16}$ & & \\
${}^{51}_{23}\text{V}$ & $\tfrac{7}{2}$ & 100 & $104.15$ & $6.3\times 10^{-16}$ & 284.5& $3.07\times 10^6$\\
${}^{50}_{24}\text{Cr}$ & $0$ & 4 & $104.04$ & $7.1\times 10^{-16}$ & 233.7 & $3.82\times 10^6$\\
${}^{52}_{24}\text{Cr}$ & $0$ & 84 & $104.04$ & $7.2\times 10^{-16}$ & 256.0 & $3.45\times 10^6$\\
${}^{53}_{24}\text{Cr}$ & $\tfrac{3}{2}$ & 10 & $104.05$ & $7.1\times 10^{-16}$ & 266.6 & $3.30\times 10^6$ \\
${}^{54}_{24}\text{Cr}$ & $0$ & 2 & $104.05$ & $6.9\times 10^{-16}$ & 284.8 & $3.06\times 10^6$\\
\hline\hline 
\end{tabular}

\end{table*}

\subsection{Isotope dependence of muon decay in orbit background}
\label{sec:isotopes_DIO}

Measuring $\mu\to e$ conversion over a wide range of nuclei to pin down the underlying new-physics operator/model requires experimental adjustments due to the nucleus dependence of muon lifetime, capture rate, conversion energy, and certain backgrounds. 
This includes, in particular, the irreducible background of muon decay in orbit (DIO), which features an electron-energy tail up to the signal region $E_e\sim m_\mu$ due to nuclear recoil.
This DIO has been calculated to sufficient precision for aluminium~\cite{Czarnecki:1998iz,Czarnecki:2011mx,Czarnecki:2011ei,Czarnecki:2014cxa,Szafron:2015kja,Szafron:2015mxa,Szafron:2015wbm,Szafron:2016cbv}, Mu2e's first target, but not for other nuclei.
Ref.~\cite{Heeck2021} has recently provided an approximate expression for the relevant DIO electron spectrum near the kinematic endpoint, i.e.~in the signal window for Mu2e-II.

The DIO electron spectrum~\cite{Heeck2021} -- normalized to the leading-order free muon decay rate $\Gamma_0 = G_F^2 m_\mu^5/(192\pi^3)$ -- is parametrized as
\begin{align}
\frac{1}{\Gamma_0} \frac{\text{d} \Gamma }{\text{d} E_e} \Big|_{E_e \sim E_\text{end}} =  B \, E_\text{end}^5 \left( 1 - \frac{E_e}{{E_\text{end}}}\right)^{5 + \delta} ,
\label{eq:spectrum}
\end{align}
improving the endpoint expansion of Ref.~\cite{Czarnecki:2011mx} by including soft-photon radiation through $\delta = \alpha (2\log [2m_\mu/m_e] -2)/\pi\simeq 0.023$~\cite{Szafron:2015kja}  and a shifted endpoint energy~\cite{Szafron:2016cbv}
\begin{align}
E_\text{end} &\equiv m_\mu - E_b - E_\text{recoil}  \\
&\quad +\frac{\alpha  m_\mu \left( Z \alpha\right)^2 }{\pi} \left( \frac{11}{9} - \frac{2}{3} \log \left[ \frac{2 m_\mu Z \alpha}{m_e}\right]\right) \nonumber
\label{eq:endpoint_shift}
\end{align}
that includes vacuum-polarization effects, the muon's binding energy $E_b\simeq \alpha^2 Z^2 m_\mu/2$ and the nuclear recoil $E_\text{recoil}= ( m_\mu - E_b)^2/(2 m_N)$ in a nucleus with charge $Z$ and mass $m_N$.
For a target consisting of different isotopes or elements, the final electron spectrum is the sum of the Eq.~\eqref{eq:spectrum} spectra weighted by the relative abundance of each isotope.
The prefactor $B$ and binding energy $E_b$ are obtained numerically by solving the Dirac equation for a given nuclear charge distribution~\cite{Heeck2021}.
The resulting parameters for a small selection of Mu2e-II-relevant isotopes are given in Tab.~\ref{tab:B_and_binding}. 
The uncertainty on the endpoint energy $E_\text{end}$ is estimated to be at the permille level, whereas the uncertainty on $B$ is around $5\%$ for the small-$Z$ isotopes of interest here~\cite{Heeck2021}.
For aluminium, the parameters in Tab.~\ref{tab:B_and_binding} agree with Refs.~\cite{Czarnecki:2011mx,Szafron:2015kja}.
The uncertainty is dominated by the nuclear charge distribution, which could be improved prior to Mu2e-II measurements with dedicated electron--nucleus scattering experiments in the relevant momentum-transfer region $q\sim m_\mu$, e.g.~at Jefferson Lab.

\subsection{Motivation for other searches}
 Although the main target of Mu2e-II will undoubtedly be the measurement of $\mu^-\to e^-$ conversion in nuclei, the experiment will be also able to address different, more `exotic' CLFV processes, given the unprecedented number of stopped muons ($\sim 10^{19}$) that it will collect. 

\noindent\emph{$\mu^-\to e^+$ conversion:}
In addition to $\mu^-\to e^-$ conversion, Mu2e and Mu2e-II can also be sensitive~\cite{Diociaiuti:2020fjx} to the lepton-flavor-violating \emph{and} lepton-number-violating ($\Delta L=2$) process 
\begin{align}
    \mu^- + N(A,Z)\to e^+ + N^\prime(A,Z-2).
\end{align}
For a recent review, see Ref.~\cite{Lee:2021hnx}.
The current best limit on the conversion rate, $1.2\times 10^{-12}$, was set by the SINDRUM II collaboration employing a titanium target~\cite{Kaulard1998}.

Lepton number violation (LNV) is a necessary ingredient of models of Majorana neutrino masses and may be at the origin of the matter-antimatter asymmetry observed in the universe~\cite{Fukugita:1986hr}. This makes searches for $\mu^-\to e^+$ conversion extremely interesting, although the LNV scale is typically constrained to be very high by searches for other processes such as nuclear neutrinoless double-beta decays~($0\nu\beta\beta$)~\cite{Agostini:2022zub}. Furthermore, $\mu^-\to e^+$ conversion is mediated by higher dimensional operators (e.g.~dimension 7 and dimension 9) and therefore this process is much more suppressed than the typical CLFV observables~\cite{Domin:2004tk,Geib2016,Berryman2017,DeGouvea:2019wnq,Plestid:2020irv,Sato:2022vny}. Therefore, future searches for $\mu^-\to e^+$ conversion will be only sensitive to new physics somehow disentangled from $0\nu\beta\beta$ and at scales below 100~GeV~\cite{Berryman2017,DeGouvea:2019wnq}.
The discovery of $\mu^-\to e^+$ conversion would therefore not only be the first evidence of LNV and a strong hint of the Majorana nature of neutrinos, but it would also point to a very specific new-physics sector characterized by a non-trivial set of symmetries and couplings, such that not only the $0\nu\beta\beta$ bounds but also collider searches at LEP and the LHC are evaded~\cite{DeGouvea:2019wnq}.

\noindent\emph{$\mu\to e X$:}
Every decay channel of a muon in orbit comes with a distribution tail of electron energies, potentially up to $E_\text{end}$. This allows Mu2e-II to probe non-standard muon decay channels, as long as they are not too suppressed. A very well-motivated example is the decay $\mu\to e X$ with a new, light, long-lived boson $X$ that escapes the detector unseen. 
This new particle could be a scalar or pseudoscalar~\cite{Calibbi:2020jvd,Escribano:2020wua}, or even an ultralight gauge boson~\cite{Heeck:2016xkh,Ibarra:2021xyk}. 
In particular, a wide range of models predict the existence of a light pseudoscalar (often called axion-like particle) that is the pseudo-Nambu--Goldstone boson (thus naturally expected to be very light) of a spontaneously broken global $U(1)$ symmetry, typically connected to possible solutions of certain puzzles of the Standard Model: the strong CP problem, the origin of neutrino masses, the hierarchical structure of fermion masses and mixing etc. Depending on the the symmetry in question, the particle takes different names (axion, majoron, familon) but its phenomenology is to large extent similar~\cite{Calibbi:2020jvd}. Interestingly, flavor-violating couplings to leptons are an unavoidable consequence of this family of models if the leptons of different generations carry different charges under the $U(1)$ symmetry (which is unavoidable in certain cases, such  as in the context of models of fermion masses and mixing). Even in cases where the charges are the same, such as the lepton number in Majoron models, loop effects generally induce lepton-flavor-violating couplings~\cite{Heeck:2019guh}. Therefore, the process $\mu\to e X$ is often among the most sensitive tests of this class of models.
\begin{figure}[t]
\includegraphics[width=3in]{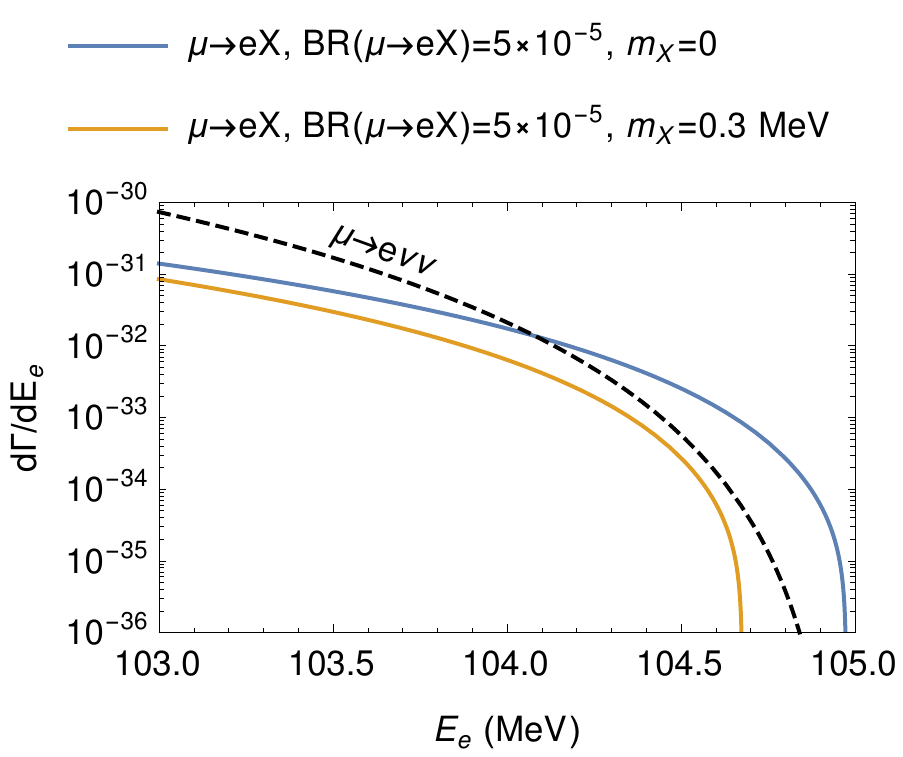}
\caption{The tail of the $\mathrm{d} \Gamma(\mu \to e \nu\nu)/\mathrm{d} E_e$ distribution (black, dashed) near the endpoint~\cite{Szafron:2015kja}.
Following Ref.~\cite{GarciaTormo:2011jit} we also show,  for two values of $m_X$, the tail of $\mathrm{d} \Gamma(\mu \to e X)/\mathrm{d} E_e$ (with $X$ coupling only to left-handed leptons) corresponding to $\text{BR}(\mu\to e X)=5\times 10^{-5}$, that is, just below the current limit~\cite{Bayes:2014lxz}.}
\label{fig:mueX}
\end{figure}

The current bounds on this decay are rather weak and crucially depend on the chirality of the coupling of the invisible boson $X$ to leptons~\cite{Calibbi:2020jvd} (which determines the angular distribution of the signal) as well as on the mass of the new particle, $m_X$. For instance, for an ultralight $X$ coupling only to left-handed leptons (such as in typical Majoron models) the current limit is $\text{BR}(\mu\to e X) < 5.8\times 10^{-5}$~\cite{Bayes:2014lxz}. These bounds can be substantially improved by Mu3e~\cite{Perrevoort:2018ttp} or MEG-II~\cite{Calibbi:2020jvd}, but $\mu \to e$ conversion experiments such as Mu2e-II could have some sensitivity due to the large number of collected muons.
$\mu\to e X$ plus nuclear recoil leads to an electron spectrum with tail up to $E_\text{end}$ and a different \emph{shape} ($(E_e-E_\text{end})^3$ compared to the standard $(E_e-E_\text{end})^5$)~\cite{GarciaTormo:2011jit,Uesaka:2020rrm,Uesaka:2020okd}. For non-vanishing $m_X$, the \emph{endpoint} is different, too (see Fig.~\ref{fig:mueX}). 
If Mu2e-II can measure the DIO spectrum precisely enough it may be sensitive to the unusual shape coming from the $\mu\to e X$ decay.


\section{Accelerator and beam line}

\subsection{Overview of PIP-II capabilities}

Construction has begun on the PIP-II accelerator and is expected to be completed by 2028. The PIP-II Linac will provide a 800 MeV proton beam with continuous wave (CW) capability, with beam power up to $\sim1.6$ MW (2 ma, 800 MeV beam) available for user experiments \cite{Ball2017}. The Mu2e-II experiment would use ~100 kW of the PIP-II beam in our initial baseline scenario. This could be increased if the Mu2e-II components can handle more intensity. The  Mu2e-II beam will require post-construction upgrades of PIP-II that enable CW operation, which will include installation of CW rf power sources. It will also require construction of the proton beam line to the Mu2e target hall, including beam switching magnets and possibly an rf separator to enable beamlines for other experiments.

 \begin{table*}[t]
\caption{Mu2e and Mu2e-II Proton beam parameters}
\label{tab:MuParam}
\begin{center}
\begin{tabular}{ l  l  l  l }
\hline\hline
\multicolumn{1}{ c }{Parameter} & \multicolumn{1}{ c }{Mu2e} & \multicolumn{1}{ c }{Mu2e-II} & \multicolumn{1}{ c }{Comment} \\
\hline\hline
Proton source & Slow extraction from DR~~~~& PIP-II Linac&   \\
Proton kinetic energy & 8 GeV & 0.8 GeV &  \\
Beam Power for expt. &8 kW & 100 kW & Mu2e-II can be increased \\
Protons/s & $ 6.25 \times 10^{12}$ &  $ 7.8 \times 10^{14}$ &   \\
Pulse Cycle Length &1.693 $\mu$s &  1.693 $\mu$s  & variable for Mu2e-II \\
Proton rms emittance & 2.7 &0.25 & mm-mrad, normalized\\
Proton geometric emittance & 0.29 & 0.16& mm-mrad, unnormalized \\
Proton Energy Spread ($\sigma_{E}$ )~~~~& 20 MeV & 0.275 MeV&  \\
$\delta p/p$ & $2.25 \times 10^{-3}$ &  $2.2 \times 10^{-4}$ &  \\
Stopped $\mu$ per proton & $1.59 \times 10^{-3}$  & $9.1 \times 10^{-5}$  &\\
Stopped $\mu$ per cycle &  & $1.2 \times 10^{5} $ &\\
\hline\hline
\end{tabular}
\end{center}
\end{table*}

Table~\ref{tab:MuParam} presents proton beam parameters for Mu2e-II, which is based upon use of the PIP-II linac, with comparison numbers for the Mu2e proton beam, which is based on protons slow extracted from the 8 GeV Delivery Ring (DR). PIP-II can provide very high-quality beam with small emittances and energy spreads. The transverse emittances and energy spreads are smaller than that of the 8 GeV beam, even after considering adiabatic damping. The geometric emittance is a factor of 2 smaller and the relative momentum spread ($\delta p/p$) is a factor of 10 smaller.

\subsection{Proton economics and Mu2e-II bunch formation}

\begin{figure} [!h]
    \centering
    \includegraphics[width=3in]{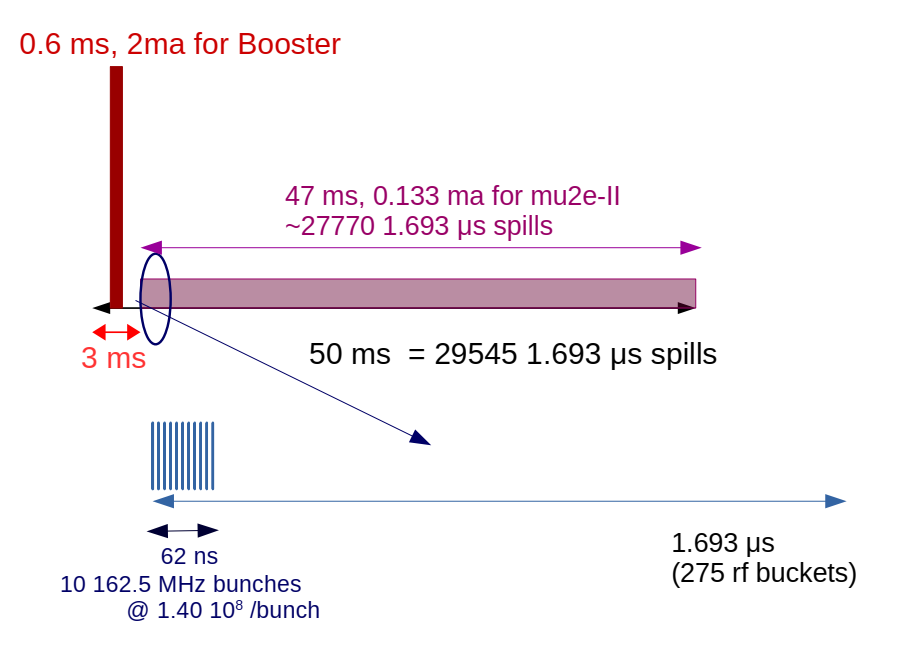}
    \caption{Schematic view of an example of bunch train formation for Mu2e-II. The beam spill occurs over 47 ms of each 20 Hz cycle of PIP-II, with the first 3 ms reserved for Booster injection. The beam spill is split into 1.693 $\mu$s periods, with beam occupying only the first 10 162.5 MHz buckets in each period.}
    \label{fig:10bunch}
\end{figure}

The PIP-II linac begins with a 162.5 MHz radio frequency quadrupole (rfq) and includes a chopper system that can produce an arbitrary pattern of filled or empty 162.5 MHz buckets. The maximum current per bucket is $\sim 5$ mA ($1.93\times 10^8$ particles). For Booster injection the intensity per bucket is limited to $\sim 1.4\times 10^8$, and concurrent Mu2e-II injection would be constrained by that limit. The PIP-II timeline must reserve time for the pulsed injection into the 20 Hz Booster; this requires $\sim$3ms out of every 50 ms. The desired Mu2e spill is a relatively short beam spill followed by a gap matched to the muon lifetime in the stopping target. Fig. ~\ref{fig:10bunch} shows a bunch spill pattern for Mu2e-II modelled on the initial Mu2e plan, and designed to provide $\sim$ 100 kW of beam on target.  The time between bunch spills is $\sim 1.693$ $\mu$s (similar to the Mu2e period of 1.695 $\mu$s). Only 10  buckets are required in each spill; the resulting beam pulse is $\sim 62$ ns.  This is much shorter than the $\sim250$ ns of the beam spill / turn for mu2e. This should provide a cleaner separation between primary beam arrival and the later captured $\mu$ decay. 

The initial example shown in figure~\ref{fig:10bunch} is simply one of many possible configurations, since the PIP-II Linac has broad flexibility in beam formation. The 1.7 $\mu$s beam spill is matched to the mu decay time in Al; a higher-Z stopping target such as Ti could use a shorter beam spill. Beam sharing with other experiments could also modify the pattern.

\subsection{Beam switching options}

\begin{figure*} 
    \centering
    \includegraphics[width=5in]{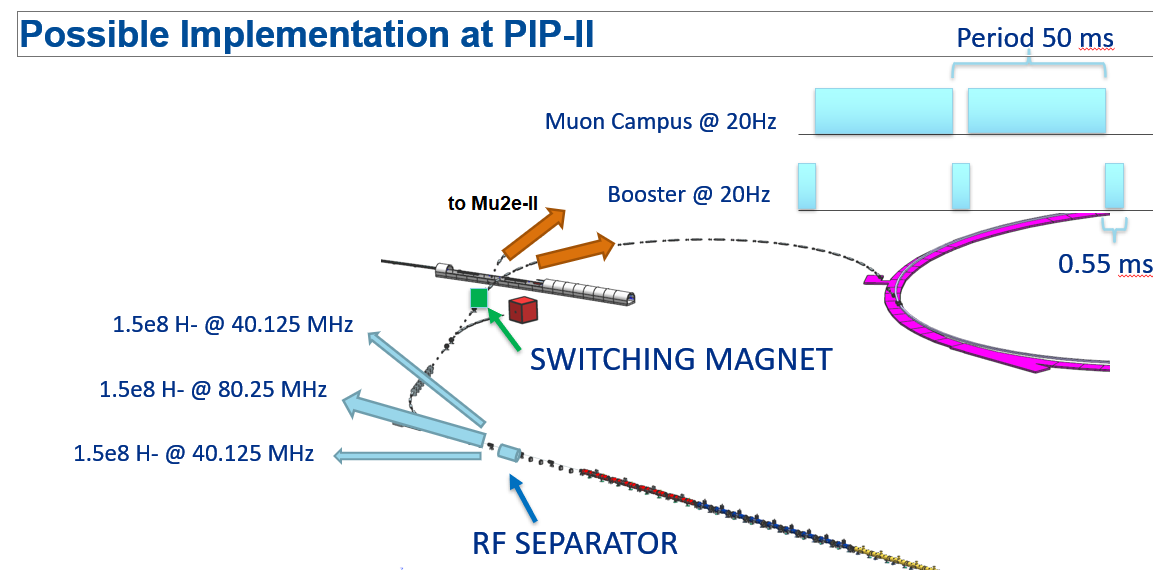}
    \caption{Layout of a possible implementation of multiple beam delivery at PIP-II. An rf separator is placed at the end of the Linac and a switching magnet separating beam for Mu2e from beam to the booster are shown.}
    \label{fig:PIPIIsplit}
\end{figure*}

\begin{figure} [!h]
    \centering
    \includegraphics[width=3in]{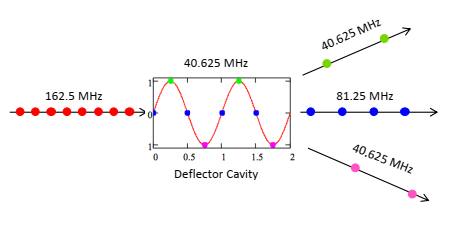}
    \caption{RF split of a 162.5 MHz beam into 3 beams by a 40.625 MHz deflector cavity. One of these beams would be for Mu2e-II, with the others directed toward other experiments.}
    \label{fig:3split}
\end{figure}

Mu2e-II will use only a fraction of the potential output of PIP-II and may be required to share beam with other experiments that have not yet been developed. It is likely that beam will be shared with other beam lines using an RF deflector for beam separation. Figure~\ref{fig:PIPIIsplit} shows a possible layout for multiple beam delivery from PIP-II, including an RF separator for producing  3 separate CW beams. Figure~\ref{fig:3split}  shows a three-way split using a 40.625 MHz deflector for RF beam separation. Use of the 81.25 MHz beam for Mu2e-II would stretch out the 10 bunch injection pulse to 125 ns in Fig. ~\ref{fig:10bunch}. 

The other beam lines would not be required to match the pulsed timeline of Mu2e-II. For example, a 40.625 MHz beam with $1.4\times 10^8$ protons in each bunch would provide ~0.7 MW of beam power to an experimental area, and could operate simultaneously with Booster injection and Mu2e-II.

\subsection{Beam line design}

\begin{figure} [!h]
    \centering
    \includegraphics[width=3in]{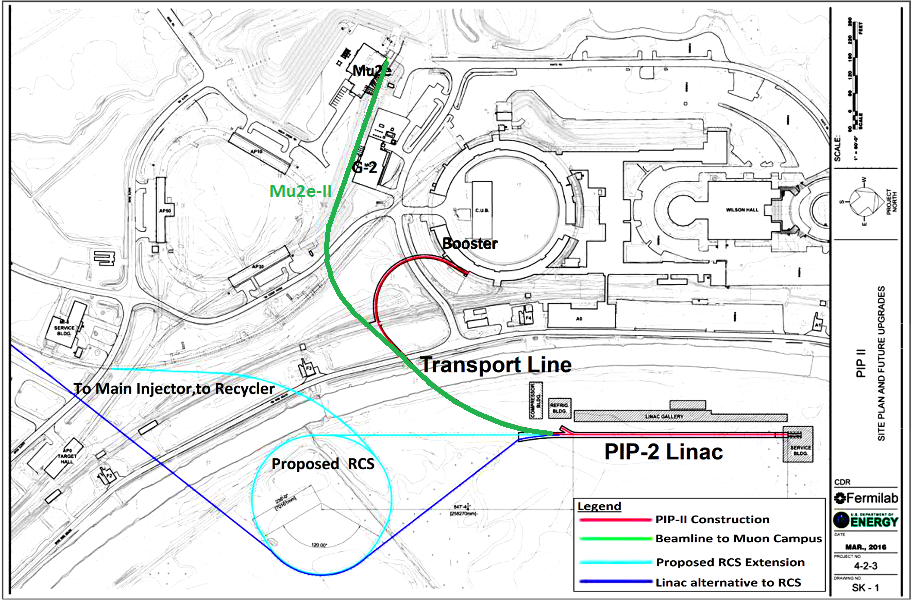}
    \caption{PIP-II linac with transport lines to Booster (brown) and Mu2e-II (green) indicated.}
    \label{fig:mu2eline}
\end{figure}

The Mu2e-II experiment will require a new beam transport from the end of the PIP-II linac into the M4 beamline that continues into the Mu2e experimental hall. An initial design of that beam line was included in the PIP-II design report, and its trajectory is shown in figure~\ref{fig:mu2eline}. The initial PIP-II construction project only includes a stub for this beam line, and the remainder must be completed later. The beam line may be modified to include H$^-$ stripping and a beam switchyard for beam sharing with other experiments.

\subsection{H$^-$ to proton stripping}

The beam from the PIP-II linac is an 800 MeV H$^-$ beam. The 800 MeV H$^-$ beam would be magnetically stripped to hydrogen atoms in the production solenoid, complicating the beam delivery on target. Instead, foil stripping to obtain 800 MeV protons from H$^-$ should be incorporated into the transport line toward Mu2e-II to avoid this complication. A preliminary assessment indicates that foil heating from the Mu2e-II-directed  H$^-$ beam will be manageable. \cite{ref:Neuffer2019}

\subsection{Extinction}

\begin{figure} [!h]
    \centering
    \includegraphics[width=3in]{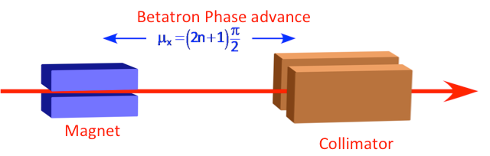}
    \caption{ Conceptual illustration of the active extinction system. A resonant set of
    magnets deflect the beam such that only the in-time beam is transmitted through a
    downstream collimator.
     } 
    \label{fig:extinction-concept}
\end{figure}

The extinction requirement for the Mu2e Experiment is $<10^{-10}$. For the Mu2e-II experiment, that requirement will be increased to $<10^{-11}$.  Like the Mu2e experiment, this will be achieved in two stages.  In the Mu2e experiment, the initial stage will be the bunch formation in the Recycler and Delivery Ring, which is estimated to have an extinction of about $10^{-5}$.  The second stage will be achieved via a system of resonant dipoles and collimators, phased such that only the in-time bunches pass through, as illustrated in Figure~\ref{fig:extinction-concept}.  That stage is designed for $10^{-7}$, for a two order of magnitude safety margin. 

\begin{figure} [!h]
    \centering
    \includegraphics[width=3in]{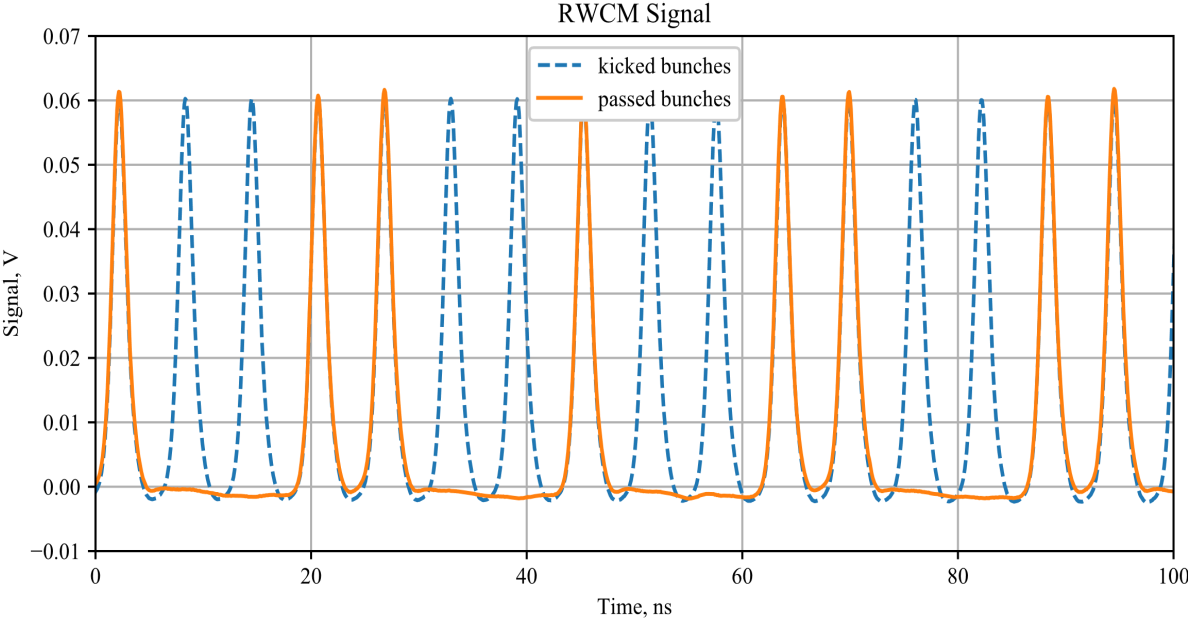}
    \caption{ Example of measured bunch-by-bunch extinction from the PIP-II linac.  The
    extinction specification for the linac is $10^{-4}$, but it is likely much better.
     } 
    \label{fig:PIP-II-extinction}
\end{figure}

In Mu2e-II, the first stage of extinction will come from the PIP-II linac itself.
Figure~\ref{fig:PIP-II-extinction} shows the bunch pattern coming out of the linac with and without representative extinguished bunches.  The nominal specification for this stage is $10^{-4}$, but it is likely to be better.

The plan is to continue to use the Mu2e extinction system for Mu2e-II.  The beamline from PIP-II to Mu2e-II passes through the M4 beam extinction line for Mu2e, which is where the secondary extinction elements are located. The extinction elements can be readily modified to provide extinction for Mu2e-II. Following the same guidance as the original design, we would need $10^{-9}$ extinction from this system at this point -- a two order of magnitude improvement over the original specification.  Detailed simulations are under way, but there are three reasons to have confidence this challenge can be met:

\begin{enumerate}
    \item The beam rigidity at 800 MeV is only 1/6 the beam rigidity at 8 GeV, so the magnets can deflect the beam further at the boundaries of the transmission window.
    \item The beam halo caused by the slow extraction septum will be absent in this beam.
    \item The lower energy beam will have dramatically lower punch-through at the collimator.
\end{enumerate}

\subsection{Beam trajectory to target}
\begin{figure} [!h]
    \centering
    \includegraphics[width=3in]{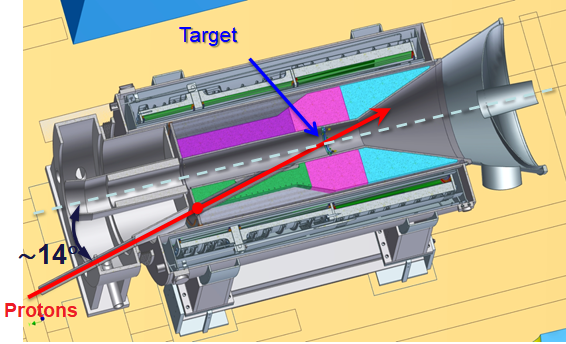}
    \caption{Trajectory of protons entering the Mu2e production solenoid (PS) through the HRS. The protons enter along a 14$^\circ$ horizontal angle toward the target. 8 GeV protons would be relatively undeflected while 800 MeV protons would be deflected vertically into the HRS. (see also Fig.~\ref{fig:SolTraj} )
     } 
    \label{fig:targettoPS}
\end{figure}

The proton beam is injected into the production solenoid (PS), which has high solenoidal fields. 8 GeV protons are relatively undeflected by the fields, but 800 MeV protons are significantly deflected \cite{ref:Neuffer2018}.
Calculations have shown that 800 MeV protons entering along the same trajectory as the 8 GeV Mu2e beam would be deflected into the HRS (heat and radiation shield) and not reach the production target. (See Fig. ~\ref{fig:targettoPS}) The Mu2e-II beam must be redirected, and the HRS must be modified, possibly by increasing the diameter of the beam port through the HRS. 

In addition to modifying the solenoids and HRS, modifications will need to be made to the proton delivery line in the neighborhood of the PS (see also Section~\ref{sec:beamTrajectory}).  In Mu2e, there is sufficient flexibility to move the beam completely off-target: the target has a core diameter of 6.3 mm, while the beam can be moved transversely by 10 mm in any direction.  Due to the lever arm between the steering magnets and the target, the beamline aperture is only large enough to enable a 0.15 degree rotation of the beam path about the target center.  For Mu2e-II, we will need to increase the transverse and angular degrees of freedom.  This will not require completely novel ideas: during early design efforts on Mu2e, a system of movable magnet stands and flexible bellows was designed to allow much larger beam manipulations, but these options were abandoned due to cost and schedule implications. 

\subsection{Target design}
The Mu2e experiment will use a radiatively cooled tungsten target which is limited to ~8 kW of 8 GeV beam. The larger proton beam current associated with 100 kW of 800 MeV beam will require an actively cooled target.

\begin{figure} [!h]
    \centering
    \includegraphics[width=3in]{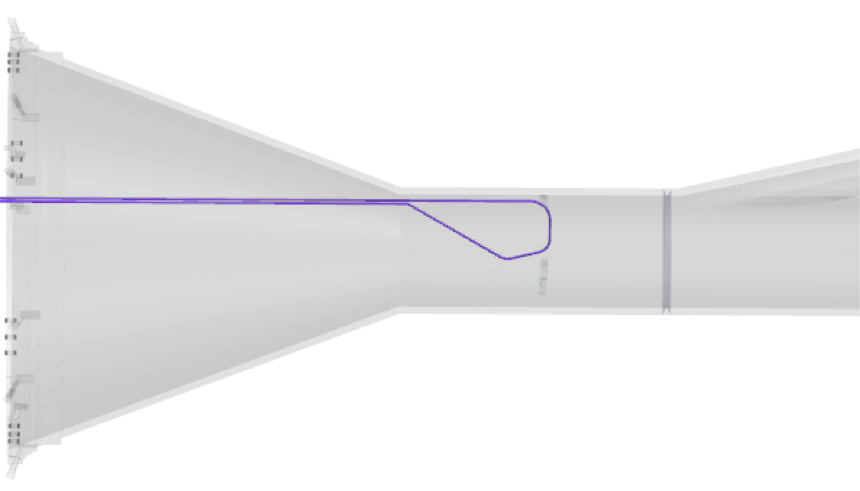}
    \caption{A sketch of the conveyor target.
     } 
    \label{fig:conveyor_target_geom}
\end{figure}

A Fermilab-based LDRD project \cite{ref:Fang2020} is exploring possible configurations for a Mu2e-II production target that can be inserted into the production solenoid. The currently preferred design is based on a conveyor movable target (target elements are either carbon,
tungsten, or tungsten carbide spheres moving within the channel situated inside
the HRS inner bore), Fig.~\ref{fig:conveyor_target_geom}.

Based on MARS15 (see Appendix~\ref{sec:MARS15}) optimization studies, we found that the optimal
number of target spherical elements in the beam would be 11 for tungsten or WC and 28 for carbon. 

To cross check and compare results obtained with different Monte Carlo codes, two target designs for Mu2e-II have been modeled with FLUKA (Appendix~\ref{sec:FLUKA}) and MARS15:
\begin{itemize}
\item a target made out of 28 carbon spheres (0.75 cm radius each)
\item a target consisting of 11 tungsten spheres (0.5 cm radius each) 
\end{itemize}

For both target designs, the surrounding HRS and PS structures have been included based on the Mu2e geometry (the HRS inner bore was enlarged to 25~cm radius to accommodate the tungsten target design). 
The total number of spherical elements in the HRS inner bore in the simulations was about 285.

\begin{figure} [!h]
    \centering
    \includegraphics[width=3in]{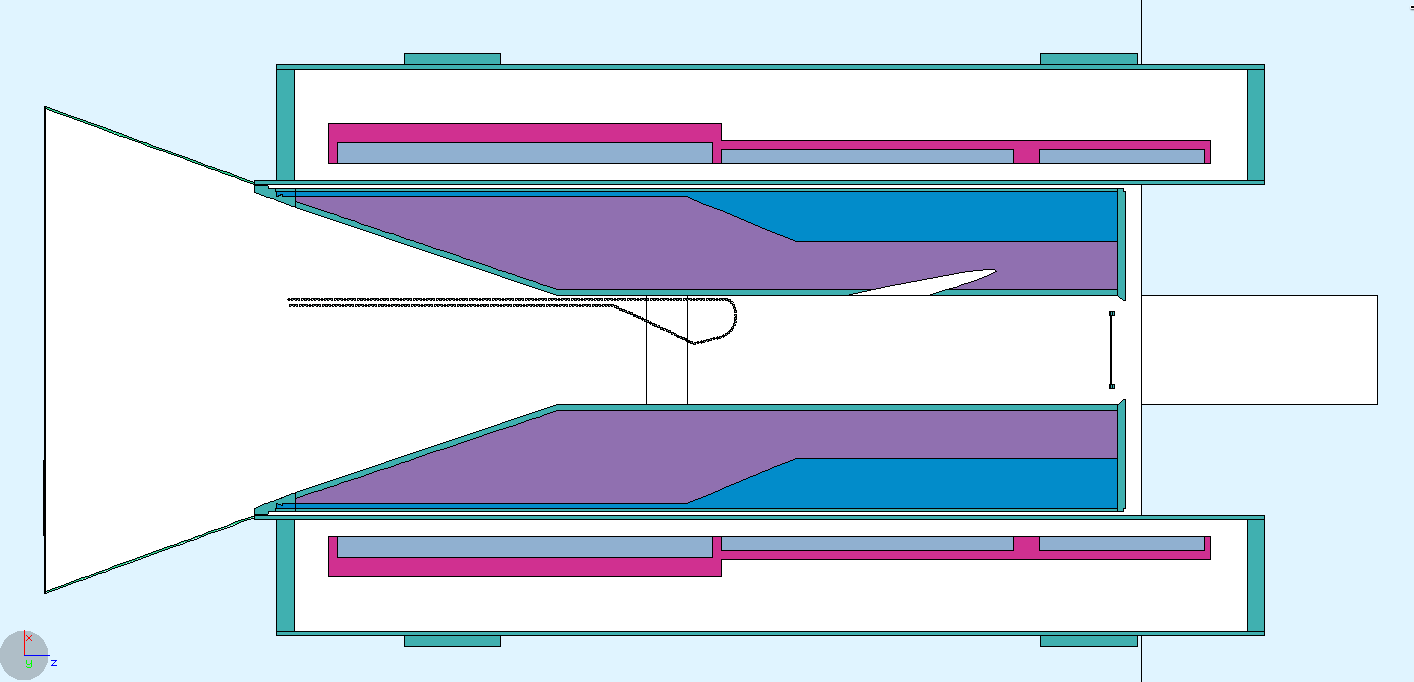}
    \caption{FLUKA geometrical model of the tungsten target design inside the HRS and PS sturctures.
     } 
    \label{fig:FLUKAgeom_Tungsten}
\end{figure}

\begin{figure} [!h]
    \centering
    \includegraphics[width=3in]{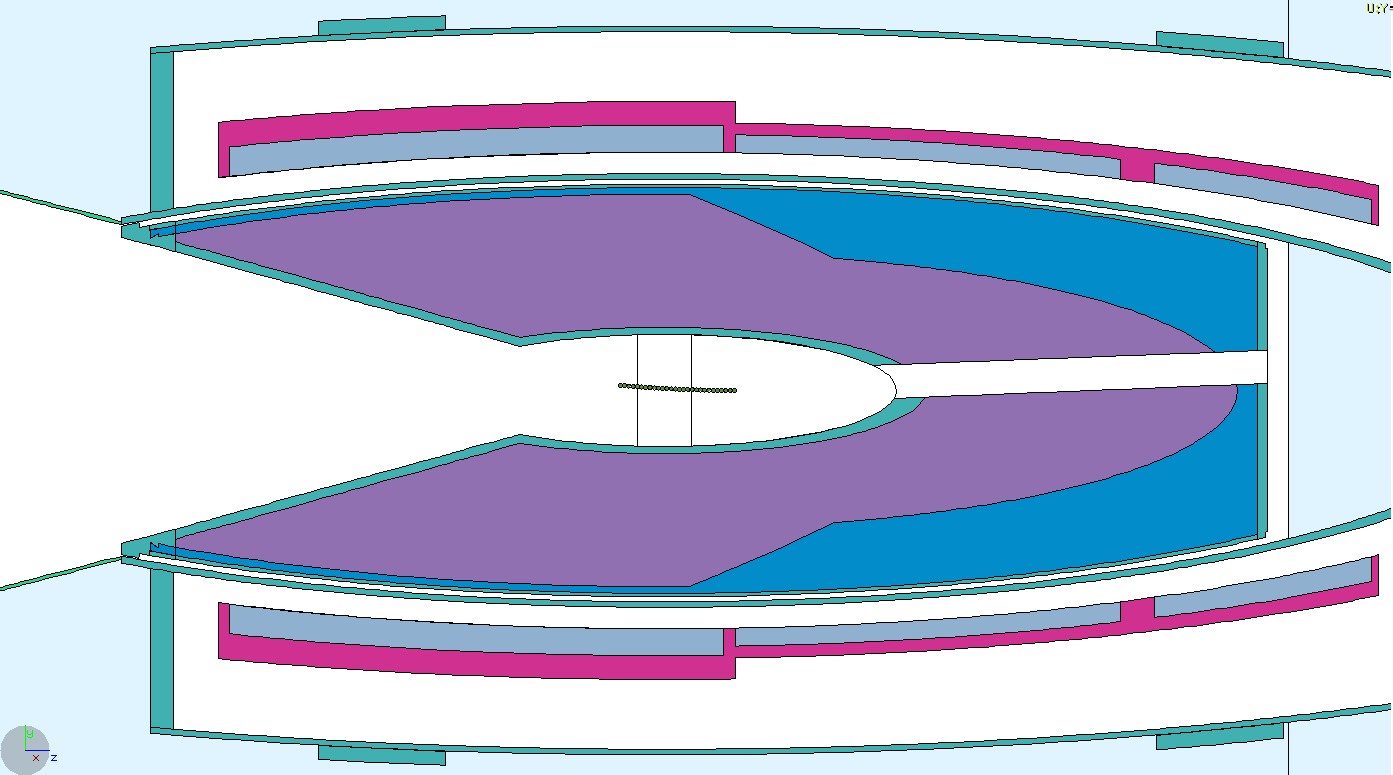}
    \caption{FLUKA geometrical model of the carbon target design as seen in the plane of deflection of protons in the magnetic field.
     } 
    \label{fig:FLUKAgeom_Carbon}
\end{figure}

\begin{figure} [!h]
    \centering
    \includegraphics[trim=47 87 44 54,clip,width=3in]{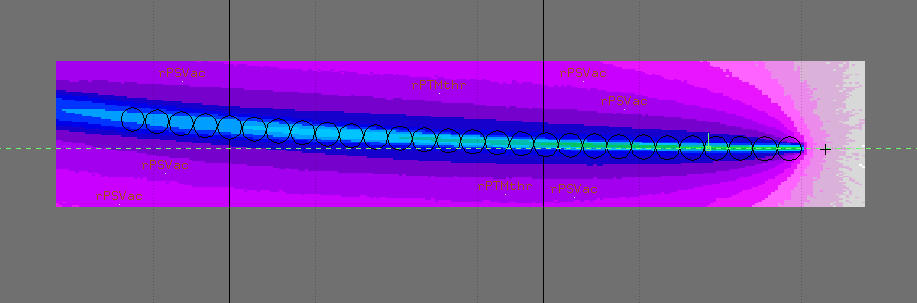}
    \caption{Proton fluence obtained with FLUKA for 800 MeV proton beam hitting the carbon target.
     } 
    \label{fig:FLUKAgeom_ProFlu}
\end{figure}

Figures~\ref{fig:FLUKAgeom_Tungsten} and~\ref{fig:FLUKAgeom_Carbon} show the FLUKA geometrical models for both target designs. Figure~\ref{fig:FLUKAgeom_ProFlu} shows the proton fluence obtained with FLUKA for a 800 MeV proton beam hitting the carbon target. It can be noted how the location of the carbon spheres follows the proton trajectories in the magnetic field in order to maximize the resulting particle yield.  

While the current FLUKA implementation of the target designs and the surrounding HRS and PS structures allows an estimate of the radiation in the vicinity of the PS region, work is underway to include additional geometry structures with the goal to perform a more accurate global shielding and radiation analysis with FLUKA and compare with the results obtained with MARS15.  

Regardless of the final design choices, a plan for radiation-hard beam alignment and target health instrumentation will be needed early in Mu2e-II Project planning.  In Mu2e the target is uninstrumented, and the nearest beam instrumentation is many meters from the target, providing only indirect measures of target health and performance.  Given the much higher beam power and corresponding power deposition in the target, the higher availability and rapid feedback of in situ instrumentation will be critical to ensure trouble free operation.

\subsection{Extinction monitor}

\begin{figure*} 
    \centering
    \includegraphics[width=5in]{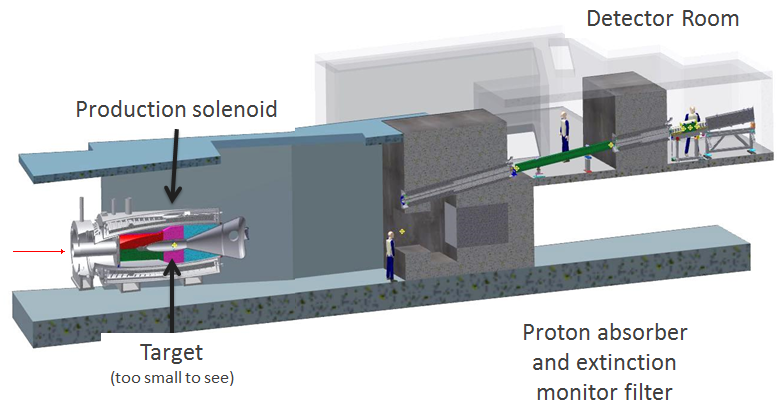}
    \caption{Layout of the extinction monitor.  A small acceptance spectrometer detects approximately one scatter for every million particles on target.
     } 
    \label{fig:Extinction-monitor-overview-drawing}
\end{figure*}

The Mu2e Extinction Monitor relies on a small acceptance spectrometer integrated into the beam dump target shielding~\cite{ref:extinction-monitor}, as illustrated in Fig.~\ref{fig:Extinction-monitor-overview-drawing}.  It is designed to see approximately one $\sim 4$ GeV/c momentum scattered particle for every million protons on target, and to use this to build up a statistical picture of the out of time beam to the required precision over the course of a few hours.  

While the spectrometer itself should be easily adaptable to lower energy scatters, the acceptance of the monitor channel will present a significant challenge.  The current channel is designed for 4 GeV/c particles from the target, taking into account the effect of the magnetic field on their trajectory.  Obviously, there will be no such particles from an 800 MeV primary beam, and the lower energy scatters will follow a very different path. Reworking the monitor acceptance for this scenario will require significant R\&D effort.

\section{Solenoids}
\begin{figure*} 
    \centering
    \includegraphics[width=5in]{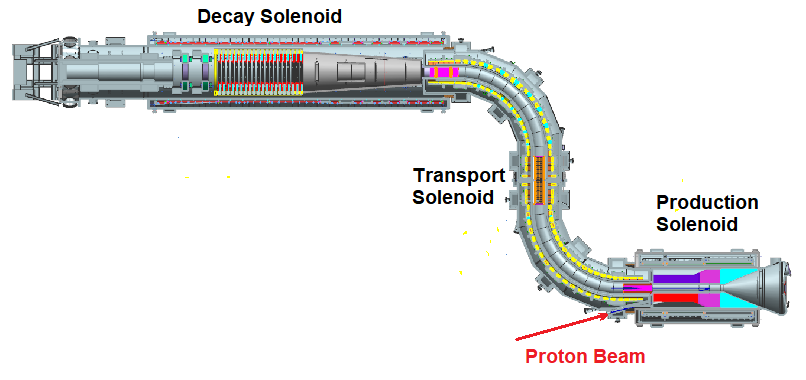}
    \caption{Cutaway view of the production solenoid (PS), transport solenoid (TS), and detector solenoid (DS) for Mu2e. }
    \label{fig:PSTSDS}
\end{figure*}

The Mu2e-II muon beamline is very similar to the Mu2e muon beamline, with a production solenoid (PS), a transport solenoid (TS), and a detector solenoid (DS) as the principal components (see Fig.~\ref{fig:PSTSDS}). Mu2e-II will reuse as much as possible of this beamline from Mu2e in order to save on cost and time. 

\subsection{Requirements for Mu2e-II}
\subsubsection{Beam power}
The proposed Mu2e-II experiment requires an increase from an 8 kW to a 100 kW proton beam. The increase in beam power creates several challenges for the production solenoid, the peak power density, the total radiation-induced displacements per atom (DPA), and the total absorbed dose on the insulating materials. Reducing the DPA on the conductor can be achieved by making the HRS out of tungsten instead of bronze. Other options for the redesign of the HRS include making the shield out of both tungsten and bronze to reduce weight, making the HRS asymmetric to cover areas of high heat and rates and smaller where the rates are lower. With an all-tungsten HRS, studies have shown that the additional heat from the increased peak power density in Mu2e-II could in principle be managed by lowering the helium coolant temperature to ~3.7 K to maintain the same temperature margin as in Mu2e. ~\cite{Kashikin2018}

\subsubsection{Beam Trajectory}
\label{sec:beamTrajectory}
\begin{figure}[t]
\includegraphics[width=3in]{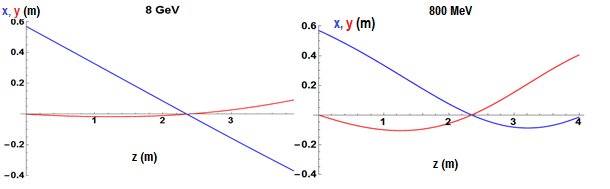}
\caption{Horizontal (x, blue) and vertical trajectories (y, red) of protons passing through the PS at 8 GeV (left) and 800 MeV (right) The target is at $z =2.35$ m. Eight GeV protons are relatively undeflected while an 800 MeV beam is deflected by $\sim 10$ cm vertically on its trajectory toward the target, and hence would intersect the Mu2e HRS. An 800 MeV beam would also be deflected away from the Mu2e beam dump when exiting the PS ($z > 4$ m)~\cite{ref:Neuffer2020}.}
\label{fig:SolTraj}
\end{figure}

The 8 GeV Mu2e beam passes through the front of the TS cryostat and enters the PS ~0.6 m off axis and about 14 degrees relative to the axis of the PS.  A decrease in beam energy from 8 GeV to 800 MeV would result in a large deflection of the beam, shown in Fig.~\ref{fig:SolTraj}. 
For the beam to hit the target, both vertical and horizontal incoming angles would need to be changed, or the location of the target would need to moved, or the PS field would need to be modified, or some combination of the three ~\cite{ref:Neuffer2020}. The outgoing beam would also be deflected and would miss the existing position of the beam dump~\cite{ref:Pronskikh2020}. An additional dipole may need to be added to direct the outgoing beam towards the beam dump, which may need to be moved from the Mu2e beam dump location. The TSu (Transport Solenoid, upstream half) magnet will also need to be modified to accommodate the incoming beam’s new angle and position.

\subsection{Production solenoid options} 
The two paths forward for the PS would be: use the existing PS “as is” at the Mu2e-II radiation load with or without the HRS upgrade and with upgrades to the cryosystem or rebuild the PS entirely or at least substantially.
\subsubsection{Existing PS}
At the end of the Mu2e experiment, the existing Production Solenoid will have absorbed a substantial fraction of its absorbed dose budget and will have become activated. Although only one coil will see the peak radiation level, it will be difficult or impossible to remove or replace the HRS, transport the magnet to a vendor for updates, or disassemble the vacuum vessel and replace the coils. Depending on the activation level, it may be possible to recycle the vacuum vessel, thermal shield, and cold mass supports, but the cold mass will likely need to be replaced. 

\subsubsection{Replacement PS}
Replacing and fabricating new coils would allow a new cable to be designed that could handle the maximum expected radiation load.
Currently the  DPA causes degradation in the residual resistivity ratio (RRR, Al and Cu), which requires a yearly thermal cycle for annealing repair. Increasing the radiation would mean more thermal cycles will be needed and also would speed up any irreversible RRR degradation of the copper. Superconducting and resistive coil designs are under consideration.

\subsubsubsection{Superconducting options}

Several conductor options are being considered for a new superconducting PS coil ~\cite{Kashikin2018}. The cable-in-conduit conductor (CICC) option provides direct cooling of the superconductor by liquid helium, which allows for larger power dissipation. CICC cable and magnet technology is relatively well developed and used by the fusion community. CICC uses high-density materials, with Cu for the stabilizer and Stainless Steel for the conduit, which would triple the heat dissipation compared to Al-stabilized conductor. The higher thermal load may force the use of Nb$_3$Sn, instead of NbTi, which is more expensive and difficult to work with. The electrical conductivity of Cu also permanently degrades under irradiation and there are few vendors capable of making the cable and winding the coils. Another option is an internally-cooled aluminum stabilized cable. The cable would provide nearly direct cooling of the superconductor by liquid helium similar to the CICC, but with lower density materials, reducing the heat load to the cryogenic system and it would not permanently degrade under irradiation. This cable concept is new and would require R\&D and a willing vendor. High-temperature superconducting (HTS) coils made from REBCO or Bi-2212 could also be considered. The higher critical temperatures may tolerate a higher heat dissipation and higher operating temperature, but these materials are still extremely expensive and difficult to work with. More extensive R\&D would be required for HTS coils.

\subsubsubsection{Resistive options}

A water-cooled resistive copper coil could extend into the HRS space, and potentially replace the HRS. It would eliminate the need for cryogenics and simplify the coil design and fabrication. There are several vendors which could fabricate this coil and it would be at relatively low cost with little required R\&D, mostly for inorganic insulating materials. The coil would require a lot of electrical power, around 5 MW, to create the same field as the Mu2e PS magnet. The copper or aluminum coil could also be cooled by liquid nitrogen, which would reduce the resistivity by a factor of 6--10 at 77 K. The required power would then be about 1 MW. There would be some R\&D required for irradiation of aluminum and copper at LN$_2$ temperatures. A LN$_2$ cooled coil would require around 20,000 liters/hour of LN$_2$.

\subsection{Transport and detector solenoids}
The Mu2e beam passes through the cryostat at the upstream end of the TSu, and the Mu2e-II proton beam will pass through a similar trajectory. Depending on the choice of PS and beam trajectory, the cryostat and and first coil or two of the TSu may need to be modified. Larger modifications may be required if the incoming trajectory is greatly modified.  

There are no anticipated upgrades required in the Detector Solenoid, and no modifications are expected.

\subsection{R\&D}
The creation of a new superconducting cable will require design, prototyping, and finding a vendor willing to produce the cable. A complete design of the PS cold mass, based on the magnetic, thermal, structural and quench protection analyses with the new conductor will also need to be produced. A prototype magnet using the new cable should also be fabricated and a heat load test performed. The projected timeline from cable design to magnet prototype production is estimated to take between four and five years.

A new HRS will need to be designed, and manufacturing R\&D with a vendor will be needed to produce tungsten with appropriate geometries. Simulations will need to be performed to determine the optimal proton beam trajectory for both the target and the dump. The simulations will drive the engineering modifications required in the TSu.

\section{Radiation}

\subsection{Radiation Environment at Fermilab Site}
An important aspect of the shielding assessment is the determination of the prompt dose above the PS and DS hatches. Explicit MARS15 (Appendix~\ref{sec:MARS15}) simulations have been performed for the 8 GeV beam of the baseline Mu2e experiment. In order to estimate the order of magnitude of the projected prompt dose above the berm (above the PS Hall) for the 800 MeV 100 kW proton beam of Mu2e-II, the following assumptions have been made: the proton beam intensity at Mu2e-II was taken to be $7.8\times 10^{14}$ p/s; and based on previous experience at Fermilab, we supposed that the prompt dose is proportional to $E_p^{0.8}$, where $E_p$ is the primary proton beam energy. In the case of the residual dose, we made a similar assumption on the energy dependence to scale the Mu2e baseline dose to the Mu2e-II one. 
In the baseline Mu2e, the following typical radiation quantities were calculated: 1) the residual dose in air 1 foot away from the west wall in the PS hall will be $\sim$ 300 mrem/hr~\cite{Grebe:2016mcn} (expected to be $\sim$ 6.2 rem/hr in Mu2e-II based on the assumptions discussed above); 2) the peak prompt dose above the berm (PS Hall area) will be $\sim$ 3-10 mrem/hr in baseline Mu2e~\cite{Pronskikh:2016} (206 mrem/hr in Mu2e-II); the prompt dose above the berm away from the peak ($\sim4$ meters South) can become $\sim$ 0.01 mrem/hr into Mu2e~\cite{Pronskikh:2016} ($\sim0.2$ mrem/hr in Mu2e-II). Among mitigation strategies one can consider increasing
the berm and/or fencing in the controlled zone area.

\subsection{Radiation Environment Around the Production Target}
A simulation study has been performed to assess the energy deposition and DPA damage to the coils of the PS under Mu2e-II conditions (see~\cite{Pronskikh:2017tR} for details). It is assumed that the PS coils determine the power density and DPA constraints and are structurally and chemically similar to those of the Mu2e magnets. The figure of merit, which is defined here as the ratio of the number of muon stops in the stopping target to the DPA rate in the hottest spot of the PS coil for the 800 MeV beam was found to be close to that of a 1 GeV proton beam and is close or slightly better than that of the Mu2e baseline's 8 GeV beam.

The DPA constraint was determined to be at the level of $4\times 10^{-5}$ DPA/yr; that level would allow the experiment to run without shutdown for annealing
for about a year. In the case of a bronze HRS with an inner bore radius of 20 cm, the baseline coil design would tolerate 10 kW beam power of a 800 MeV proton
beam; for the tungsten HRS, this increases to $\sim40$ kW. However, if one increases the inner bore radius to $\sim25$ cm, the tolerable beam power for the
tungsten HRS would be higher than 100 kW.

\subsection{Radiation Environment at the Detector Locations}
The prompt effective dose levels are of particular importance for electronics in regions of the DS Hall such as the equipment alcove and the electronics alcove.
In the Mu2e baseline experiment those levels have been simulated as 10 mrem/hr and 6 mrem/hr, respectively ~\cite{Pronskikh:2016-2}. After applying the aforementioned corrections for the beam energy and beam intensity, the estimated doses for the Mu2e-II conditions are found to be $\sim$ 206 mrem/hr (equipment alcove) and $\sim$ 125 mrem/hr (electronics alcove).

In the baseline Mu2e, the radiation quantities in rack electronics are expected to be 0.24 Rad/yr (absorbed dose), $2.9\times10^7$ n/cm$^2$/yr
(1-MeV equivalent neutron flux), $29.2\times10^3$ h/cm$^2$/yr (hadron flux $E > 30$ MeV).
The peak radiation quantities in the tracker and calorimeter electronics are 9 kRad/yr (absorbed dose), $1\times10^4$ n/cm$^2$/yr (1-MeV equivalent neutron
flux), and 60  h/cm$^2$/yr (hadron flux $E > 30$ MeV)~\cite{Pronskikh:2017-2}. For the case of Mu2e-II, if one scales with the beampower, the hadron flux
$E > 30$ MeV, for example, can be expected to become $3.7\times10^5$ h/cm$^2$/yr (rack electronics), and 750 h/cm$^2$/yr
(tracker and calorimeter electronics).

Detailed studies will be required to improve the accuracy of these numbers. However, from the preliminary assessments it is clear that radiation-tolerant electronics will be necessary.

\subsection{Residual Mu2e Radiation and Access for Mu2e-II Construction}

\begin{figure} [t]
    \centering
    \includegraphics[width=3in]{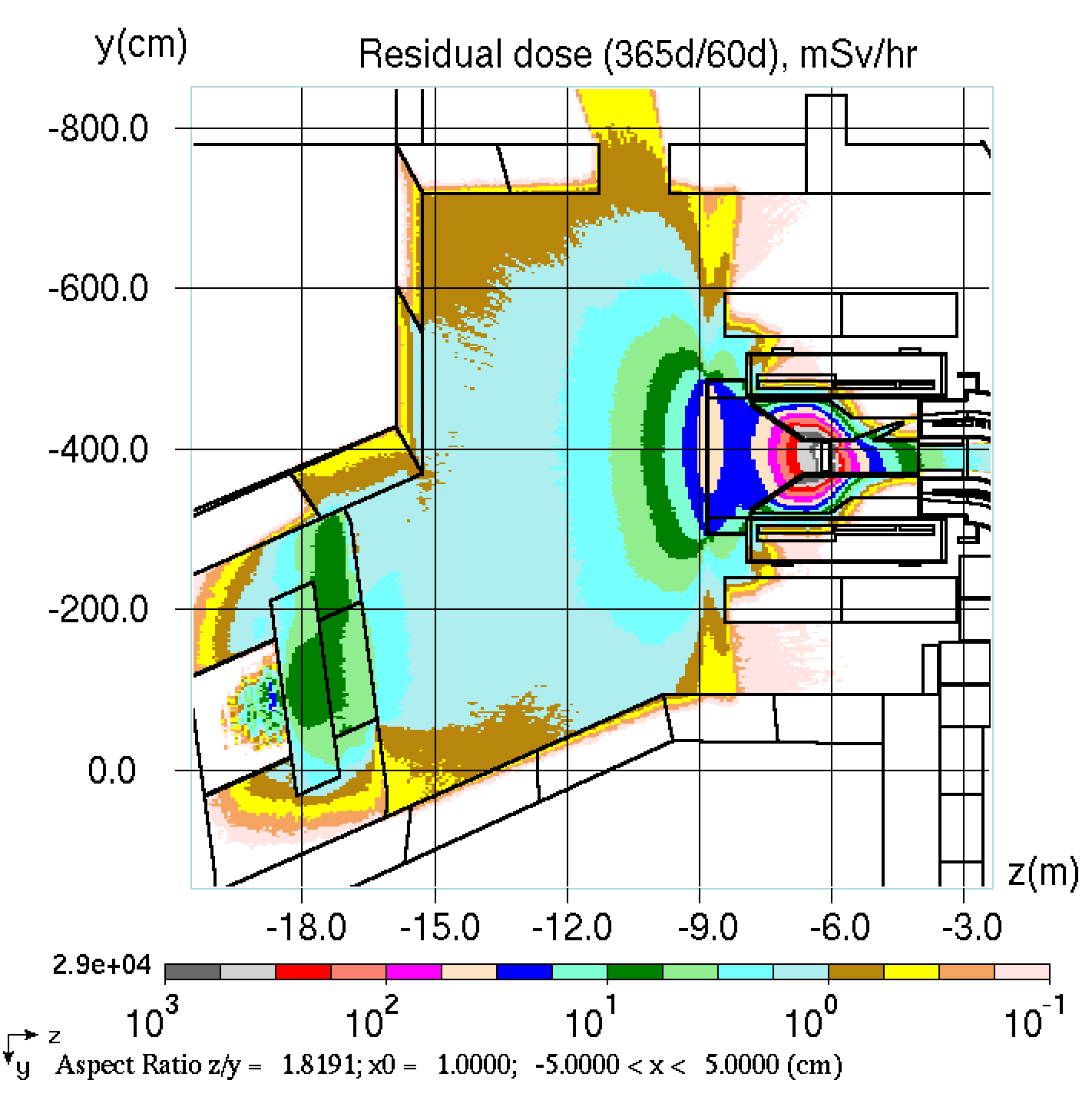}
    \caption{Residual dose in the PS Hall after 1 year of irradiation
    and 2 months of cooling, mSv/hr}
    \label{fig:resdose}
\end{figure}

A preliminary assessment of the residual Mu2e radiation in the
 PS Hall (the hottest area) can be based on the MARS15 and FermiCORD~\cite{Grebe:2016mcn} simulation
results shown in Figure~\ref{fig:resdose}. The residual dose
from several sources (target, endcap, beam dump, walls, the floor,
the ceiling) on the target elevation after 1 year of irradiation
and 60 days of cooling will be $\sim10$ mSv/hr (1 Rem/hr)
at a distance of 1 m West from the endcap and $\sim1$ mSv/hr (100 mrem/hr) near the West wall. Among the  ways to mitigate the dose
after the Mu2e run that can be considered are the removal of
the endcap (technical feasibility would need to be determined) and the
target.

\section{Tracker}
\subsection{Introduction}
The Mu2e-II tracker will provide the primary momentum measurement for charged particles originating in the stopping target.
The tracker must accurately and efficiently identify and measure 105 MeV/c conversion electrons while rejecting backgrounds.
The Mu2e-II tracker will face all of the issues encountered by the Mu2e experiment, with additional challenges arising from Mu2e-II’s increased beam intensity and the need to improve its precision to reject the background from the DIO tail.
The increased muon intensity in the Mu2e-II experiment introduces new challenges of increased DIO backgrounds, radiation, and detector occupancy.
In this tracker section, we address issues unique to the Mu2e-II tracker, for shared challenges we refer you to Mu2e literature (\cite{Mu2e_TDR,Pezzullo:2019Uk}).  

This tracker section is split into 3 parts: descriptions, critical issues, and R\&D.
The first part starts with a general description of the detector designs and the simulation being used to identify critical issues.
We will then present the critical issues facing the detector.  
Finally, we end with a review of R\&D work being done both in material studies and in simulations, and with an overview of required research for the anticipated detector.

\subsection{Descriptions}
\subsubsection{Detector}
The Mu2e tracker is a low mass array of straw drift tubes aligned transverse to the axis of the detector solenoid.
The basic detector element is a straw made of a \SI{25}{\micro\metre} sense wire inside a 5 mm diameter tube made of \SI{15}{\micro\metre}  thick metalized Mylar.
The detector will have about 20k straws evenly distributed among 18 measurement stations along a length of about 3 meters.
The most minimal design change in going to the Mu2e-II tracker is reducing the straw wall thickness from $15$ to \SI{8}{\micro\metre}.
We use this as a baseline tracker design for simulations studies.  

A possible alternative construction approach of the tracking system is also under evaluation, a brief description is reported in \ref{sec:altracker}.

The geometry and material of the stopping target, inner proton absorber (IPA), and tracker, all interact with the conversion electrons and must be treated as a whole system in determining the momentum resolution and requirements on the tracker.
Studies on the stopping target and descriptions of its geometry can be found in section \ref{stopping_target_studies}. 
The IPA is currently assumed to be the same as the design in Mu2e (\cite{Mu2e_TDR}): a \SI{0.5}{\milli\metre} thin, conical frustum made of Polyethylene. 
The IPA is needed to stop or degrade protons released from muon capture in the stopping target.
Without the IPA, these protons would overload the tracker leading to aging effects.
Protons also induce very large signals which could cause cross-talk and dead time in channels.
The idea of changing the geometry of the IPA to a non-radially uniform geometry, where the material is positioned to intercept positively charged particles but be absent from the vast majority of negatively charged paths, has been suggested.
This idea is suggested for further study if resolution needs additional improvement.

\subsubsection{Simulation}
To understand the process of muon production from protons hitting the production target, we analyzed data from a detailed simulation.
This simulation was based on the existing Mu2e Geant4 simulation, with modifications of the proton beam energy and timing, the production target, and the detector systems to reflect the changes in Mu2e-II. See Appendix~\ref{sec:GEANT4} for details.
The standard Mu2e production procedures were used to produce a large dataset for further studies.  

To better understand what aspects of the Mu2e tracker design need further optimization for the higher rates and sensitivity goals of Mu2e-II, we used a hybrid Monte Carlo \cite{TrackToy}.
This takes as input the muon beam entering the detector region, as calculated by a Geant4 beam simulation.
A parameterized model of the passive and active materials is then used to predict the trajectories of these beam muons, their stopping in the stopping target, the production of daughter electrons, the propagation of those daughters through the tracker, and the production of tracker and calorimeter signals.
The tracker hits are then fit using the KinKal \cite{KinKal} kinematic Kalman filter track fit package, and the reconstructed momentum from that is used to estimate the sensitivity to the conversion electron signal.
While the absolute predictive power of this ``TrackToy'' Monte Carlo has limited accuracy, the change in predicted sensitivity due to parameter changes provides a robust estimate of which parameters are most important in the design optimization.
Details of the TrackToy simulation are presented in Appendix \ref{sec:TrackToy}.

\subsection{Critical Issues}
The increased muon intensity and improved sensitivity of the Mu2e-II experiment leads to three critical issues facing the detector: improving momentum resolution, handling detector occupancy, and surviving higher radiation rates.
While these issues are not solved here, by analyzing these issues we will outline the anticipated requirements of the tracker and suggested paths of future R\&D.

\subsubsection{Resolution}
While the CE signal is mono-energetic, all detectors impart resolution degradation through interaction with the electron.
Reducing the mass and atomic number of the detector will naturally reduce the observed momentum spread.
Reduced mass also reduces the cross-sections for generating background hits from photons.
Mass reduction is limited by the need to have a structurally sound detector, and by the need to shield the tracker active volume from highly ionizing radiation.

Resolution is also degraded by uncertainties and errors in the track reconstruction algorithms.  
Improvements to the track reconstruction through implementations of machine learning processes could help attain the needed resolution for Mu2e-II.

\subsubsection{Occupancy}
Increased muon rate, reduced IPA shielding, and increased timing windows could all lead to more hits, cross-talk, and dead-time in channels.
The $4\times$ increase in muon rate is needed to increase sensitivity, so the tracker must accommodate this rate of muon absorption protons and DIO electrons.
Muon capture protons can be reduced by increasing the IPA mass but it will smear the electron momentum.
We are investigating if the impact of IPA material on signal electron resolution could be mitigated by a more sophisticated design of the IPA geometry.

The crux of the issue of occupancy in the detector is the pattern reconstruction algorithm for reconstructing the tracks.
It remains to be determined if improvements to pattern recognition can be used to overcome the inefficiencies created from additional hits due to increased occupancy.

\subsubsection{Radiation and rates}
Increased beam intensity will lead to a larger radiation exposure and charge deposition onto the tracker material and electronics during the beam flash and over the lifetime of the experiment.
In addressing the critical issue of radiation there are two categories, damage to the tracker components and to the electronics recording the measurements.

Radiation aging studies performed on the Mu2e straws showed that there was no observable degradation in gain or cathode resistivity after a total charge deposition of \SI{0.9}{\coulomb\per\centi\meter} \cite{Mu2e_straws}.
The Mu2e-II simulation studies estimating the expected charge deposition on the straws are ongoing.
We expect the value to be larger than the total seen by Mu2e tracker, greater due to the larger beam intensity but less because of the lower mass of the straw materials.
It remains to be tested if the thinner straws will degrade after a large charge deposition.

The Mu2e tracker design was able to add additional bronze shielding to reduce the radiation exposure for the electronics.
Mu2e-II would expect to keep the same shielding.
In addition, Rad-hard ASICs and other components would be developed to address this issue.

\subsection{R\&D}
\subsubsection{Material Studies}
A path towards improving momentum resolution is to reduce the mass of the tracker.
The largest component of the mass in the detector's active region is the Mylar of the straw tubes.
Research has been conducted to push the limits of how thin the walls of a straw tube can be made.

We were able to construct Mylar straw tubes \SI{8}{\micro\metre} thick.
These straws were constructed as a double helical wrap with two layers of \SI{3}{\micro\metre} mylar and \SI{2}{\micro\metre} of a mylar like adhesive.
Prototype straws did not have a metalization layer on either the inside or the outside.
The next batch of prototype straws will have an aluminum metalization.
Without the metalization, we focused on measurements of mechanical components of physical size, density, pressure limits, and the elastic limit.
A comparison of these prototype straws and Mu2e straws can be found in table \ref{tab:Straw_Comparison}.
Preliminary mechanical studies of these straw tubes show that these would be feasible to use in the Mu2e-II tracker.

\begin{table}[t]
\renewcommand{\baselinestretch}{1.0}\normalsize 
\caption{\label{tab:Straw_Comparison}
Comparison for properties of the Mu2e and Mu2e-II straw tubes.}
\renewcommand{\baselinestretch}{1.3}\normalsize 
\centering
\begin{tabular}[t]{ lcc }
\hline\hline
 & Mu2e & Mu2e-II \\
\hline\hline                               
Wall thickness (\SI{}{\micro\meter}) & $18.1$ & $8.2$ \\
Al thickness (\SI{}{\micro\meter}) & $0.1$ & $0.2$  \\
Au thickness (\SI{}{\micro\meter}) & $0.02$ & $0.0$ \\
Linear Density (g/m) & $0.35$ & $0.15$ \\
Pressure limits (atm) & $0$--$5$ & $0$--$3$ \\
Elastic Limit (gf) & $1600$ & $500$ \\
\hline\hline 
\end{tabular}
\end{table}

Additional measurements are needed to determine expectations of a tracker constructed with these thin straws.
Long term sag studies are needed to determine the material creep rate and whether these straws can be installed at a sufficient tension such that the eventual sag does not affect HV stability on the wires.
Other measurements require prototype straws with metalization layers.
The metalization layers are important for reducing the gas leak rate, so while the prototype straws have proven the ability to hold pressure for a day, we have not measured the leak rates of CO$_{2}$ gas.
Finally, radiation aging studies need to be performed on metalized straws to determine if there is degradation in gain or conductivity.

\begin{figure}[!hb]
\centering

\includegraphics[width=1\columnwidth]{"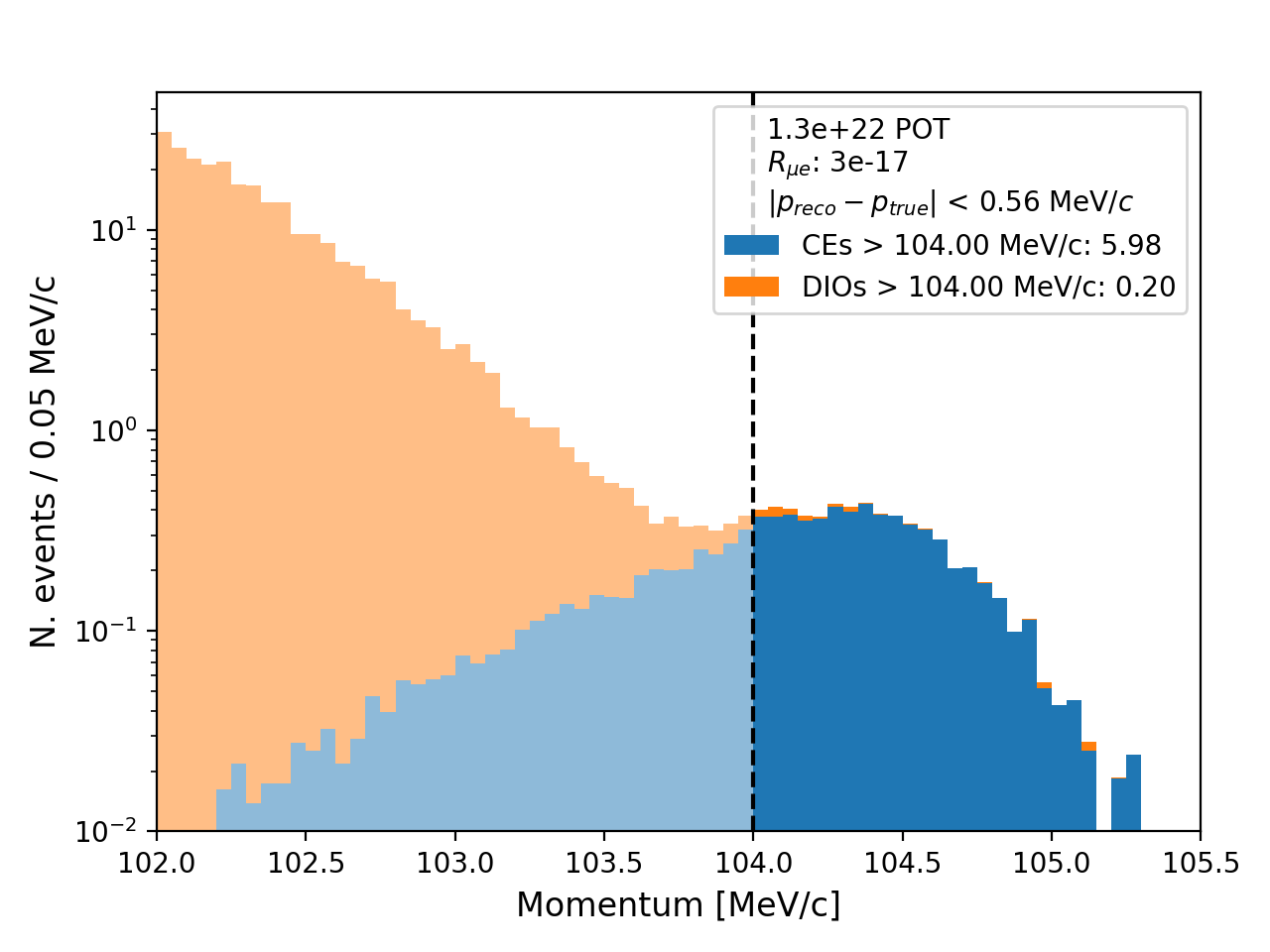"}
\includegraphics[width=1\columnwidth]{"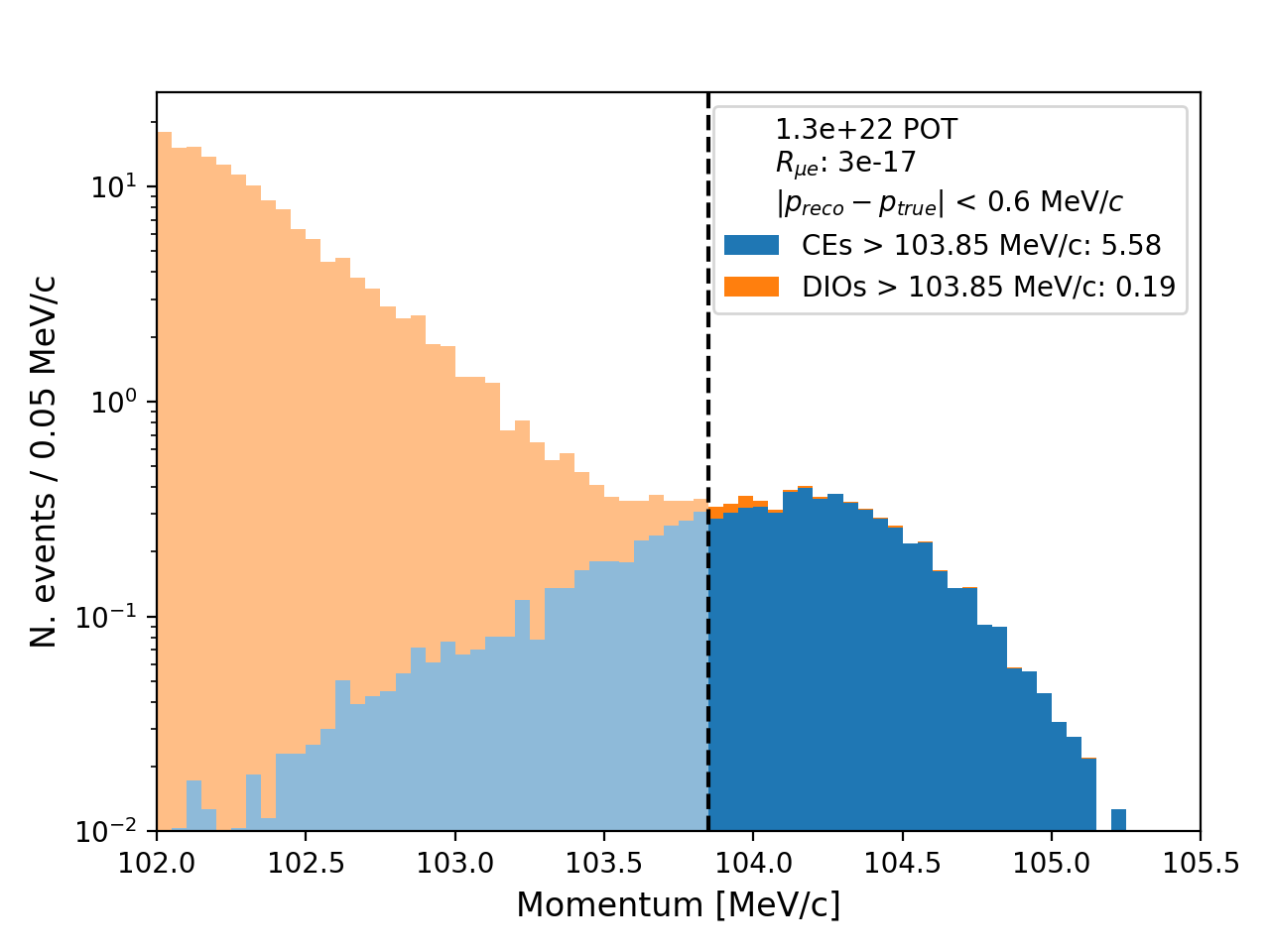"}

\caption{Stacked histograms of the simulated reconstructed momentum spectra for CEs (blue) and DIOs (orange), with 8~$\mu$m (top) and 15~$\mu$m (bottom) thick
straw tubes.} \label{fig:straws}
\end{figure}

The TrackToy Monte Carlo simulation software has been used to compare the reconstructed momentum spectra obtained with \SI{8}{\micro\meter}- and \SI{15}{\micro\meter}-thick straw tubes. Figure \ref{fig:straws} shows that \SI{8}{\micro\meter}-thick straw tubes achieve $\mathcal{O}(10\%)$ signal efficiency increase, while keeping the number of expected DIOs below 0.2, a number similar to the expected Mu2e DIO background.
The absolute distance of the reconstructed momentum from the true momentum $|p_{\mathrm{reco}}-p_{\mathrm{true}}|$ has been estimated by training a multi-layer perceptron neural network.
This parameter helps to selected well-reconstructed tracks and the value of the cut has been chosen in both cases to optimize the number of CEs while keeping the same background acceptance. 

\subsubsection{Simulation Studies}

The TrackToy Monte Carlo simulation software has been used to estimate the effect of detector parameters on the experiment sensitivity. 

\begin{figure} [!h]
    \centering
    \includegraphics[width=1\columnwidth]{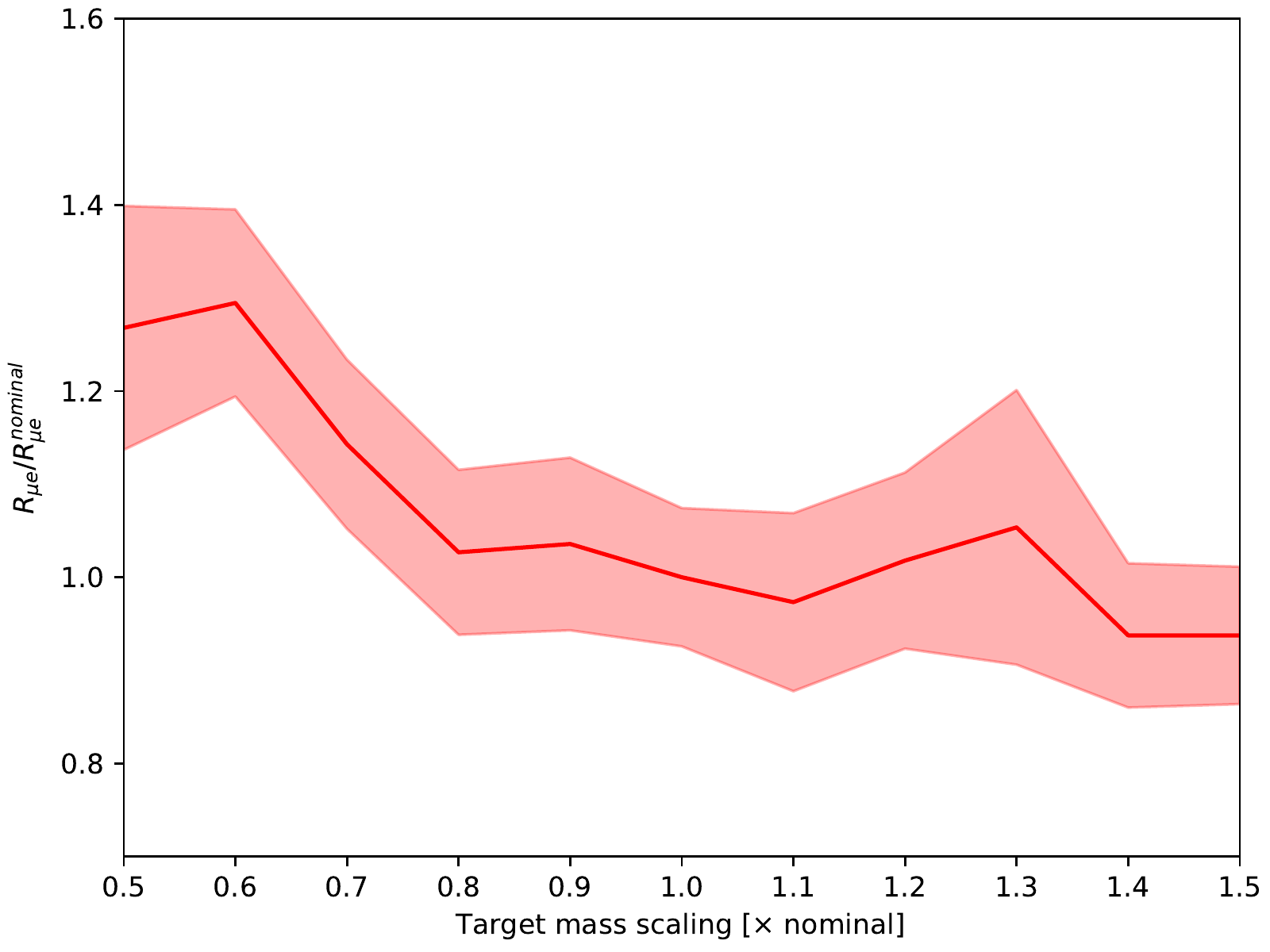}
    \caption{$R_{\mu e}$ discovery potential for a scan of the target mass values using TrackToy. The shaded region corresponds to the statistical uncertainty.}\label{fig:targetmass}
\end{figure}

Figure \ref{fig:targetmass} shows the $R_{\mu e}$ $5\sigma$ discovery potential for a scan of the target mass values.
There are two competing effects at play: a heavier stopping target increases the muon stopping rate, thus improving $R_{\mu e}$ sensitivity, but the additional energy straggling smears the momentum spectrum, producing non-gaussian tails that reduce signal efficiency and increase DIO background acceptance.
 The plot shows that these two effects roughly cancel each other around the nominal value for the target mass and that the sensitivity plateaus with heavier targets. 
 A heavier target also increases the background from cosmic rays which interact with the target.  It will also increase the background hit rate in the tracker due to beam electron bremsstralung.  Those effects, which will favor a lighter target, were not included in this study.

\begin{figure} [!h]
    \centering
    \includegraphics[width=1\columnwidth]{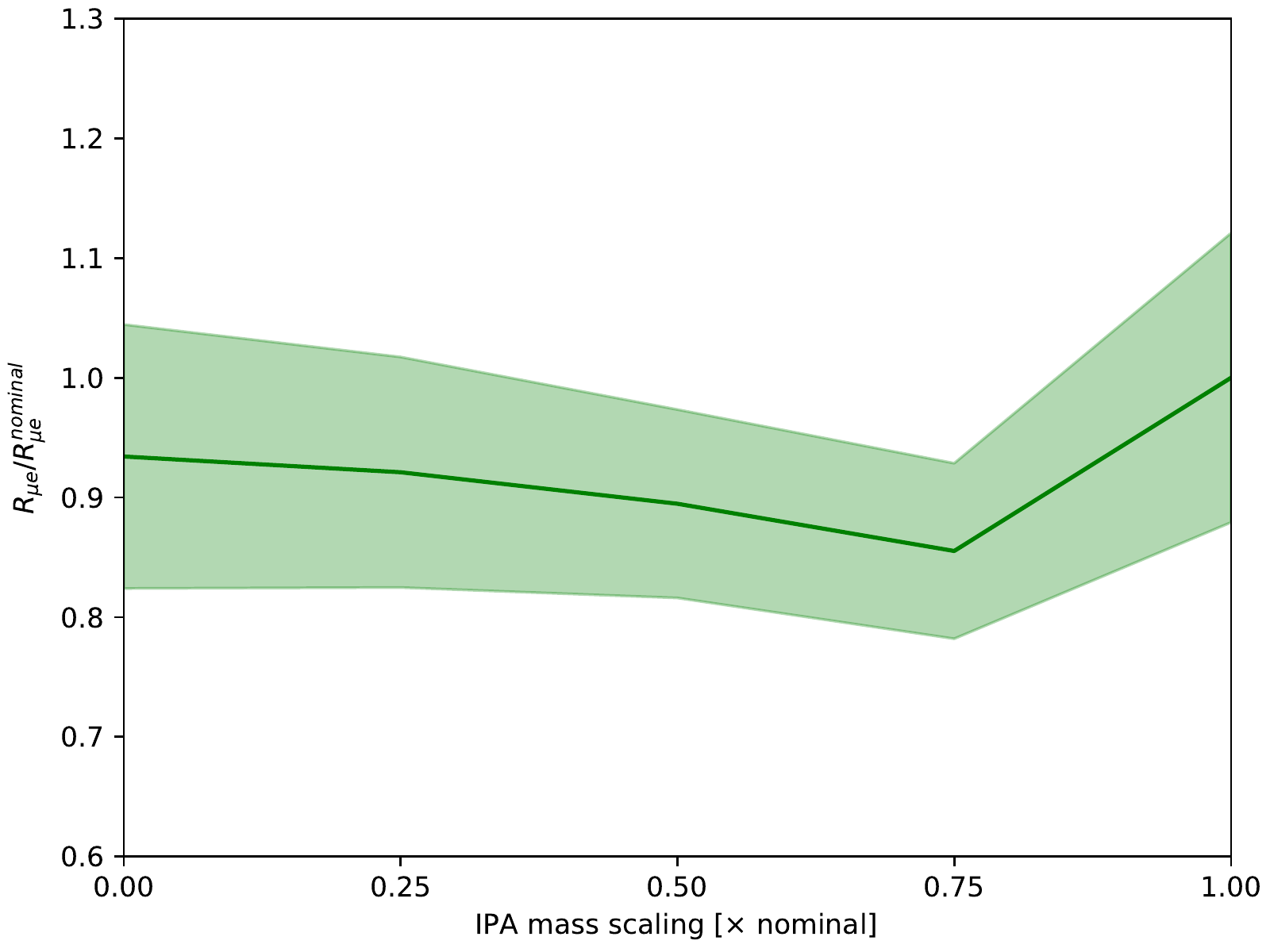}
    \caption{$R_{\mu e}$ discovery potential for a scan of the IPA mass values using TrackToy. The shaded region corresponds to the statistical uncertainty.}\label{fig:ipamass}
\end{figure}

An analogous study has been performed with a halved IPA mass and without a IPA. Figure \ref{fig:ipamass} shows that these variations have a limited effect on the experiment sensitivity. 

A scan of the number of tracker measurement stations was also performed using TrackToy, holding the tracker dimensions constant and simply increasing the density of straws. We found no significant improvement in the discovery sensitivity with more than 18 stations. 

\subsubsection{Alternative design of the tracking system\label{sec:altracker}}
\begin{figure} [H]
	\centering
	\includegraphics[height=4cm,keepaspectratio]{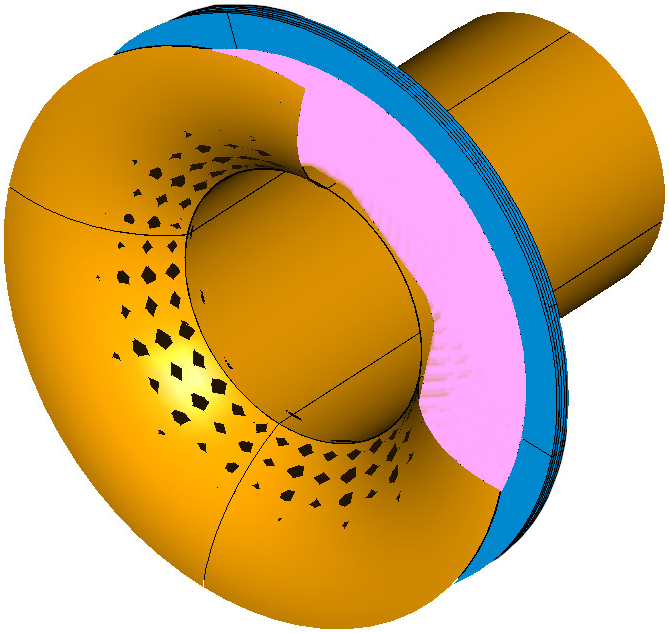}
	\caption{Pictorial views of tracker alternative, a station of a Mu2e tracker like inserted in the C-fiber gas vessel.}
	\label{fig:trackerAlt}
\end{figure}

An alternative tracker proposal for Mu2e-II can be envisioned by applying the concept of separating the gas containing function from the holding structures.
Following this paradigm, it is possible to imagine enclosing a drift tracker, similar to the Mu2e one, in an ultra-light gas vessel, see Fig.~\ref{fig:trackerAlt}.
\begin{figure} [ht]
	\centering
	\includegraphics[height=2.5cm,keepaspectratio]{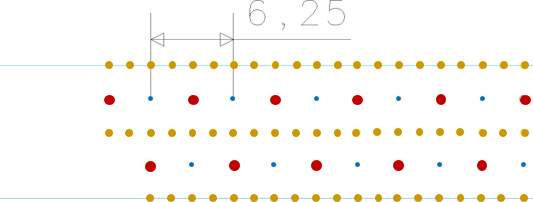}
	\caption{Sketch of square drift cells of a panel of a Mu2e-II tracker option. In the figure the blue points are 20 $\mu$m W wires, red points are 50 $\mu$m Al wires and orange points are 40 $\mu$m Al wires.}
	\label{fig:panelSchemeAlt1}
\end{figure}
The ultra-light gas vessel proposed for the I-Tracker~\cite{project2012mu2e,grancagnolo:2013,ASSIRO2013443}, made of C-fiber, was able to sustain the differential pressure of 1 atm by using an equivalent material thickness of $0.8\times 10^{-3} X/X_{0}$ for the inner cylinder and of $0.3\times 10^{-3} X/X_{0}$ for the end caps.

With an external gas vessel, the gas leakage requirements on the single straw tube are released and possible alternatives can be identified.
With the assumption of preserving the Mu2e tracker layout, electronics, structures etc., the drift cells can be placed on analogous arched structures to form panels (Fig.~\ref{fig:trackerAlt}) with the analogous dimensions and straw disposition.
In this configuration the straw walls are necessary only to define the electric field of the drift cells and can, thus, be reduced as much as possible provided that they are strong enough to resist the electrostatic attraction and the gravitational sag.
An even further reduction of the straw wall material and a simplified construction can be envisaged by replacing the straw tubes with squared drift cells with the same pitch. 
Figure~\ref{fig:panelSchemeAlt1} shows a sketch of a possible configuration of the drift cells arrangement obtained by using only thin metallic wires arranged in layers with only field wires and layers with sense and field wires.
As a starting point, we may assume using the same wires, respectively $20~\mu$m W wires, $50~\mu$m and $40~\mu$m Al wires, and the same wire arrangement as the one adopted for the construction of the MEG-II drift chamber~\cite{MEGII:2018kmf}.
It is important to stress, however, that wire dimensions, materials, and wire arrangement have to be optimised.

Alternatively, it is possible to replace the three field wire layers with three thin foils of aluminized mylar.
Since these are not subject to differential pressure the foils have to sustain only the electrostatic and the gravitational forces. By hand calculations point out that the electrostatic attraction should be managed by using 5 $\mu$m (or 2.5 $\mu$m) thick mylar foils (a detailed evaluation on the stretching force required is to be performed). The metallic coating will be kept as small as possible, at the same level of the one foreseen for the 8 $\mu$m straws, 500 \AA~of Al on one or both sides for the foil in the middle. This mixed configuration (mylar foils and wires) has an advantage with respect to the configuration with only wires that an eventual broken wire will remain confined within a layer of the panel, causing little or no harm to the entire tracker.

The construction procedure for the panel of this alternative tracker design can be based on the construction technique used for MEG-II drift chamber~\cite{MEGII:2018kmf}.
Using a wiring robot, wires can be soldered onto PCBs (designed for this purpose) creating multi-wire frames. Then the panel construction can be achieved by overlapping a multi-wires frame with field wires, or a mylar foil, a spacer, a multi-wire frame with sense and field wires, a spacer and so on. Moreover, since the panel does not need to be sealed, no glue would be required between the single layers but they can be properly positioned by screws and dowel pins. This construction approach guarantees a high precision on the wire position, as was proven by MEG-II, the wires can be located with a precision of 20 $\mu$m in one direction (along the PCB) and of about 50 $\mu$m in the other direction (it depends on the machining precision of the spacers). These precision levels on the wires positioning reduce the problem of the electrostatic cell stability and allow the construction of 5 mm, or even smaller, cells.

For this alternative configuration the use of a helium based gas mixture is mandatory to maintain the gas multiple scattering contribution at the $10^{-3} X/X_{0}$ level. Considering a 90\% He -- 10\% $i$-$C_{4}H_{10}$ gas mixture as reference point, the single drift cells are, respectively for the all wire configuration and for the mixed configuration, equivalent to $1.9\times 10^{-5} X/X_{0}$  and $3.5\times 10^{-5} X/X_{0}$.
Assuming an average value of 35 hits per track plus the contribution of the gas in the non-active areas and of the inner wall of the gas vessel the expected material budget for this tracker alternative is expected to be, respectively for the two configurations considered, about $3.8\times 10^{-3} X/X_{0}$ (equal to the expected material budge for the straw tracker) or $4.3\times 10^{-3} X/X_{0}$.
Moreover, an even lighter tracker, up to $3.8\times 10^{-3} X/X_{0}$, could be designed by supposing a construction scheme that allows use of a gas mixture inside the active area of the panel and pure helium inside the rest of the volume defined by the gas vessel.

We performed a simulation of the alternative tracker configurations with TrackToy. This indicates that: the configuration with all wires has the same tracking resolution as the 8 $\mu$m straw tracker; the one with mylar foils with the same gas inside the entire gas vessel has a little poorer resolution and the one with mylar foils but with pure He outside of the cells has a little better resolution. It is important to notice that the drift velocity in a He based gas mixture is slower than the Ar based ones and so the maximum drift time for 5 mm cells can be in the range of 80--100 ns. This is about 2 time slower than the straw tracker and this will reduce the rate capability and potentially the pattern recognition performance, and this has to be evaluated with more care.

However, the construction technique can allow the possibility to reduce the cell dimension to about 3 mm. Apart from the assembly features, by hand calculations show that the electrostatic stability can be reached with 3 mm cells, but a better analysis will be performed using Garfield++ simulations. Having 3 mm cells allows reducing the maximum drift time back to about 50 ns and to increase by about 30\% the rate capability with respect to the straw case. The drawback of the reduction of cell size to 3 mm is that there will need to be two times more cells to cover the panel sensitive area, and so twice the electronics channels are needed. If the increase of the number of readout channels is an affordable option, as for example by using an ASIC for the front-end readout, the tracker alternative configurations can improve a bit the tracker performance with respect to the straw tracker. Moreover, using 3 mm cells inside a panel (preserving the straw tracker dimensions), 3 layers of cells can fit. In this case, apart from the further increase of the electronics readout channels, the tracking capability can be improved, especially on the pattern recognition, as indicated by TrackToy simulations.

\subsubsection{Future plans}
In this section we highlight a few additional areas of research that could be important in the overall design of the tracker.

Pattern reconstruction is going to be critical toward improving the track reconstruction efficiency in the higher occupancy environment of Mu2e-II.
The development and use of machine learning algorithms are quickly becoming standard for many HEP experiments.
It is believed that further research into this field will be fruitful for this tracker.

The prototype straws were significantly more difficult to handle without bending or crinkling than the thicker Mu2e straws.
Detector construction techniques on how to build the detector that preserve straw integrity are being developed.
Building the detector while the straws have an internal pressure would make them easier to handle and could also identify damage to a straw during the construction process.
Additionally the idea of self-centering wire terminations on the straws could be used to significantly simplify the building procedure.

The use of a drift gas other than Argon CO$_2$ has possible benefits to the tracker.
A faster gas could reduce the dead time on cells and help reduce occupancy issues.
It is also known that Mylar is particularly permeable to CO$_2$.
Many other gases would have a lower leak rate through the straws.


\section{Calorimeter}
\subsection{Introduction}
The calorimeter provides an alternative measurement of the conversion electron candidate's energy, as well
as a fairly precise measurement of the time of energy deposit that is useful in track-finding and cosmic ray rejection.\\
The	Mu2e calorimeter design consists of	pure Cesium Iodide (CsI) crystals comprising two disks.	The	calorimeter	has	robust	performance	at	Mu2e rates	but	may	be	challenged	by Mu2e-II	instantaneous rates	that are	two to	three	times	higher.	The	x10	integrated	radiation	dose	on	the	calorimeter	readout	electronics also	motivates	study	of	appropriate	rad-hard	readout	electronics	at	a	level	informed	by	the	HL-LHC	detector	upgrades. 

\subsubsection{Mu2e-II requirements}
The Mu2e-II calorimeter should have the same energy ($<$ 10\%) and time ($<$ 500 ps) resolutions as in Mu2e, aiming to provide a standalone trigger,  track seeding and PID as before. The Mu2e-II calorimeter must also withstand a higher radiation environment:
\begin{itemize}
    \item 10$^{12}$ - 10$^{13}$ 1 MeV eq./cm$^2$ neutron flux on the photosensors
and $\sim$ 0.1 - 1 Mrad fluence on crystals;
    \item high rate and high pile-up probability;
    \item 1 Tesla magnetic field.
\end{itemize}
With respect to the dose, the second disk can perhaps be left as it is, with CsI crystals and SiPM readout, but the first disk will need a drastic change. It will be necessary to replace the CsI crystals with crystals capable of sustaining higher levels of radiation and performing at higher rates. BaF$_2$ crystals meet these criteria, and are a leading candidate for the Mu2e-II calorimeter.

\subsubsection{Design to meet Mu2e requirements}
An alternative	calorimeter	design	has	been	developed	based	on	barium fluoride (BaF$_2$) crystals	readout	with	 solar-blind	UV-sensitive silicon photomultipliers that	efficiently	collect the	very fast UV component	($\sim$220	nm)	of	the	scintillation	light while	suppressing	the	slow	component	near	300	nm.		This	design	is	considerably	more	robust	against	Mu2e-II	rates but	requires	the	development	and	commercialization	of	the	required	solid	state	photosensors,	which	is	ongoing.
\subsection{Choice of crystal}
\subsubsection{Options}

Improving on the decay time and radiation hardness of pure CsI is likely necessary to meet the more stringent requirements of Mu2e-II. LYSO:Ce is brighter, more dense and more radiation hard than CsI, but has a 40 ns decay time which is slower than CsI. LYSO:Ce is also more expensive because of the Lu$_2$O$_3$ raw material used and the higher melting point. PbWO$_4$ has a similar decay time to CsI, but a light yield of less than 10\%  of CsI. The radiation damage in PbWO$_4$ recovers at room temperature, requiring continuous light monitoring {\it in situ} to maintain calorimeter precision. Other bright and fast inorganic scintillators, such as LaBr$_3$:Ce and CeBr$_3$, are highly hygroscopic which presents a technical challenge for calorimeter construction. Table~\ref{tab:crystals} compares basic properties for three fast scintillating crystals which are candidates for the Mu2e-II calorimeter, where light yield is shown relative to NaI:Tl~\cite{Zyla:2020zbs}. 
\begin{table}[htb]
\caption{Properties of three fast scintillating crystals that are practical candidates for the Mu2e-II calorimeter~\cite{Zyla:2020zbs}}
\label{tab:crystals}
\begin{tabular}{ ccccccc}
\hline \hline
 Crystal& $X_0$ & $R_M$  &  $\lambda_I$ & $\tau_{\rm decay}$  & $\lambda_{\rm max}$ & Light\\
 & cm & cm & cm & ns & nm & Yield\\ \hline\hline
 CsI & 1.86 & 3.57 & 39.3 & 30 & 310 & 3.6\\
  &  &  &  & 6 &  & 1.1 \\ 
  BaF$_2$ & 2.03 & 3.10 & 30.7 & 650 & 300 & 36 \\
  &  &  &  & \textless 0.6 & 220 & 4.1 \\
 LYSO:Ce & 1.14 & 2.07 & 20.9 & 40 & 402 & 85 \\
\hline\hline
\end{tabular}
\end{table}

\begin{figure} [!h]
    \centering
    \includegraphics[width=3in]{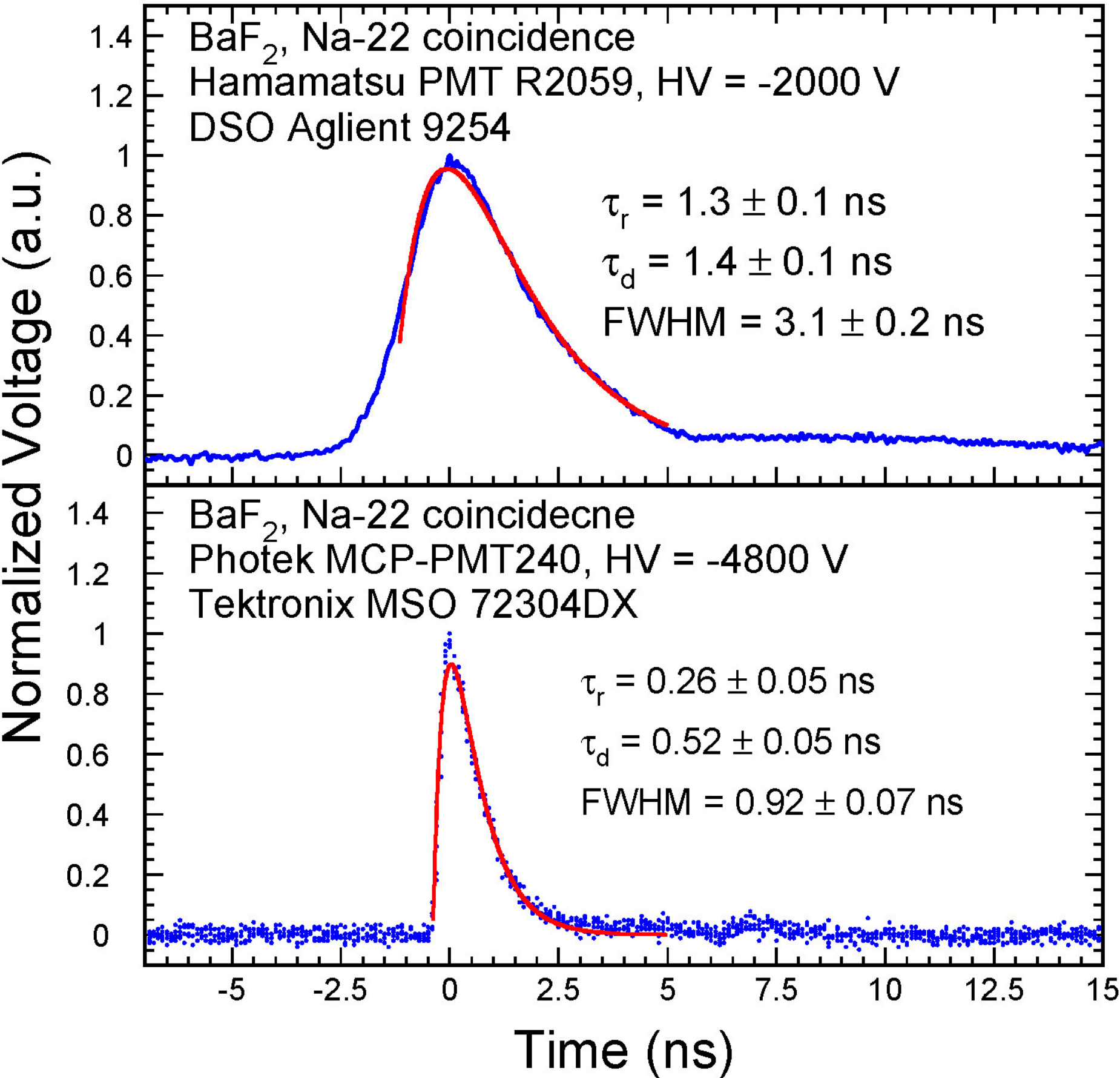}
    \caption{A comparison of BaF$_2$ pulse shape measured with a Hamamatsu R2059 PMT (top) and a Photek MCP-PMT 240 (bottom).}
    \label{fig:BaF2_pulse_shape}
\end{figure}

Barium fluoride (BaF$_2$) stands out as a candidate for its ultrafast scintillation component with $<0.6$ ns decay time and similar light output to  CsI. Figure~\ref{fig:BaF2_pulse_shape} compares the temporal response of the BaF$_2$ scintillation light measured by using a Hamamatsu R2059 PMT (top) and a Photek MCP-PMT 240 (bottom). While the FWHM pulse width and decay time of 3 and 1.5 ns were observed by the PMT, they are about 0.9 and 0.5 ns observed by the MCP-PMT~\cite{Hu2019}.  Such an ultrafast light provides a foundation for an ultrafast BaF$_2$ calorimeter. A TrackToy simulation with the improved time resolution given by BaF$_2$ has been performed, resulting in a $\mathcal{O}(5\%)$ better sensitivity.

Undoped BaF$_2$ also maintains its light output at high ionizing radiation levels after an initial light loss, so is more radiation hard than CsI at a large integrated dose~\cite{Yang2016Gamma}. The main issue to overcome is that its fast scintillation component at 220 nm is accompanied by a slow component at 300 nm with 650 ns decay time and a significantly larger intensity, which results in pileup and readout noise in the high-rate Mu2e-II environment. 

\subsubsection{Development efforts}

\begin{figure} [!h]
    \centering
    \includegraphics[width=3in]{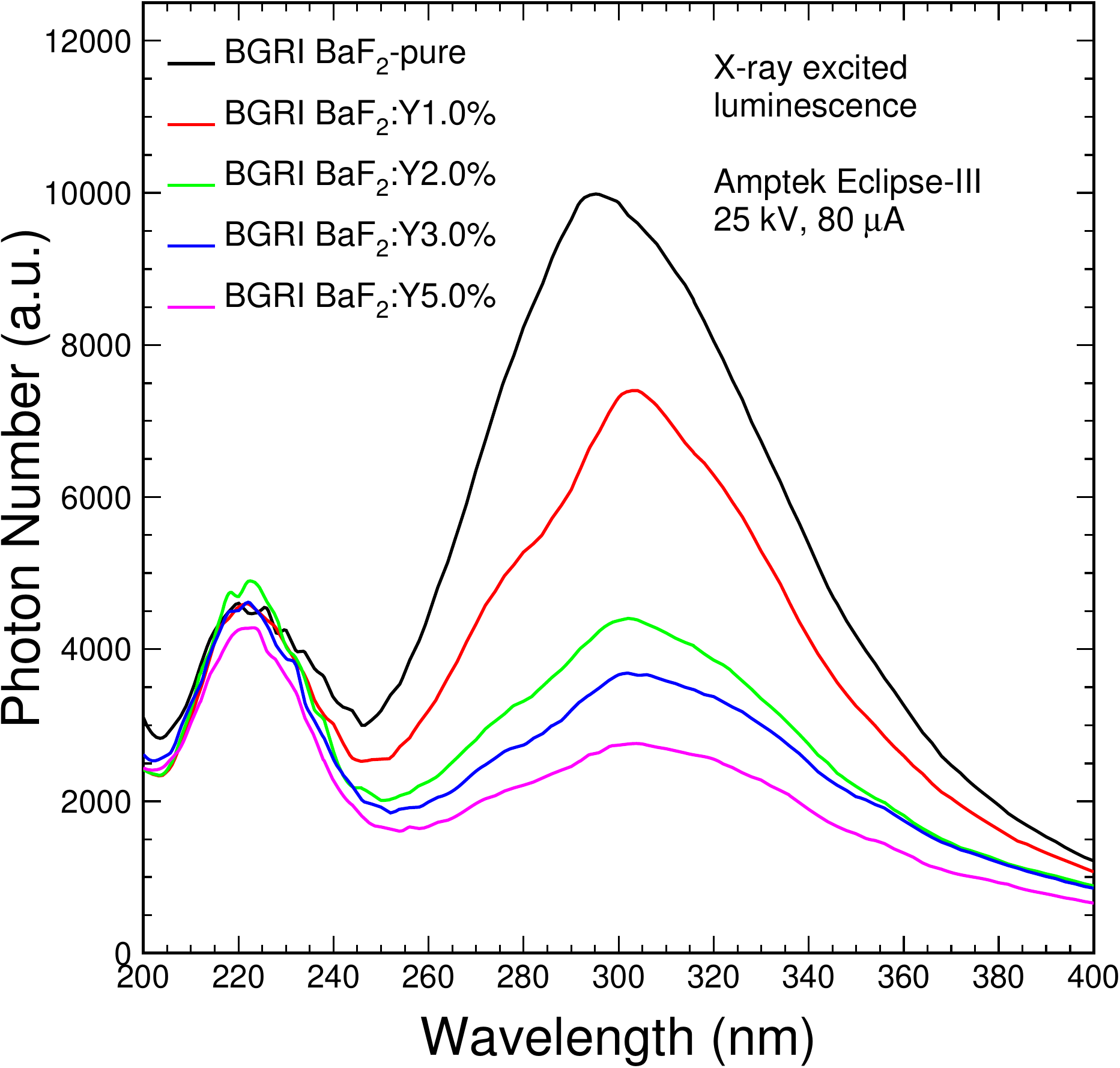}
    \caption{X-ray excited emission spectra measured for BGRI BaF$_2$ crystal samples with different yttrium doping level.}
    \label{fig:BaF2Y_emission}
\end{figure}

Yttrium doping is found effective in suppressing the slow component while maintains the ultrafast  component~\cite{Zhu2017, Chen2018, Hu2019, Hitlin2020}. Figure~\ref{fig:BaF2Y_emission} shows the X-ray excited emission spectra measured for BaF$_2$ samples with different yttrium doping level~\cite{Zhu2017}. 

\begin{figure} [!h]
    \centering
    \includegraphics[width=3in]{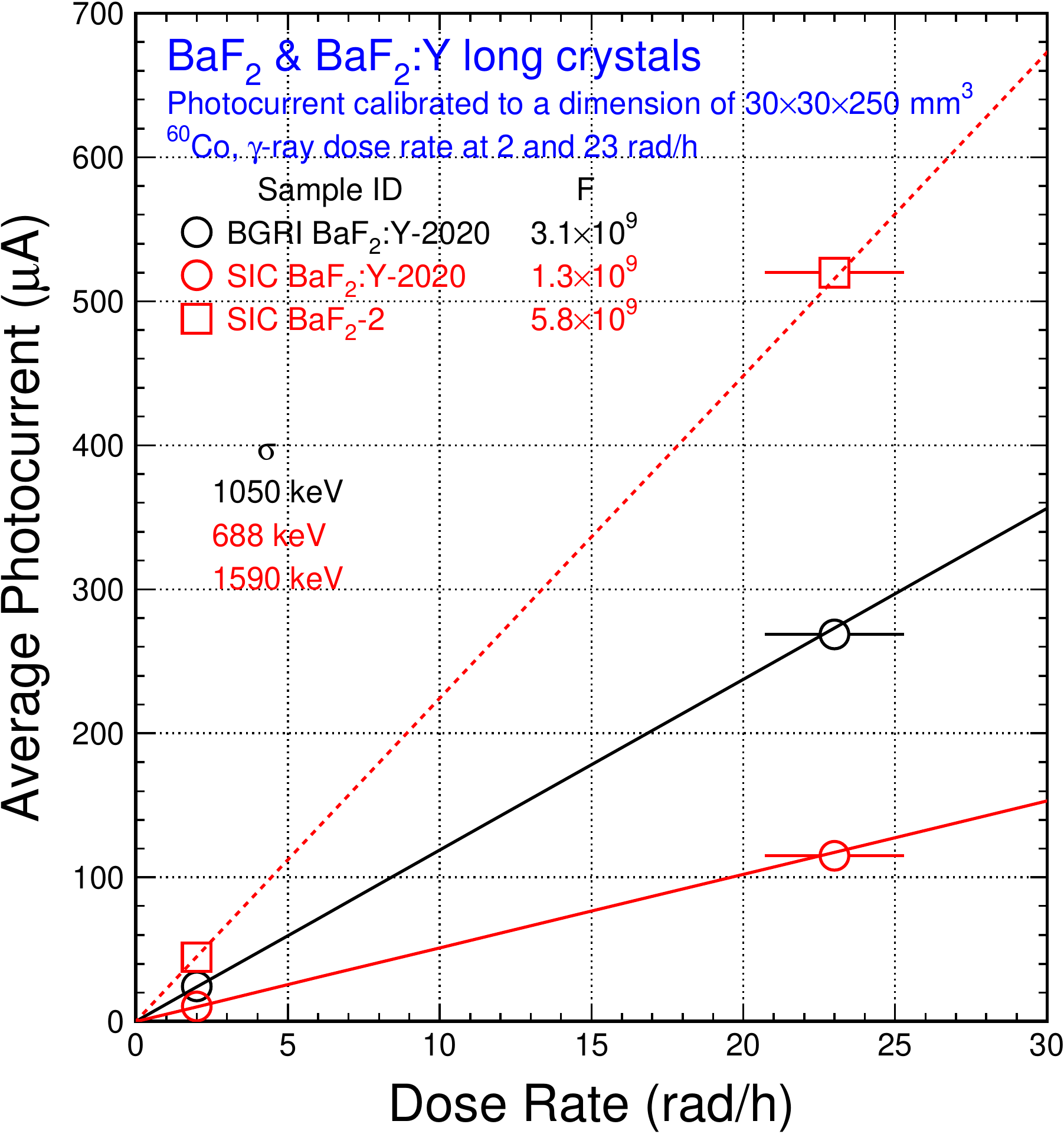}
    \caption{ Photocurrent is shown as a function of the dose rate for an SIC BaF$_2$ and two BGRI and SIC BaF$_2$:Y samples of calorimeter size under ionization dose rate of 2 and 23 rad/h.}
    \label{fig:BaF2_RIN}
\end{figure}

R\&D is on-going in collaboration with crystal producers to develop BaF$_2$:Y crystals of large size~\cite{Hu2020, Hu2021BaF2_RINCR}.  Gamma-ray induced noise under the Mu2e-II environment was measured for large size BaF$_2$ and BaF$_2$:Y crystals. Figure~\ref{fig:BaF2_RIN} shows photocurrent as a function of the dose rate for a BaF$_2$ and two BaF$_2$:Y samples of calorimeter size under 2 and 23 rad/h. Both yttrium doping and solar-blind photodetector are needed to reduce the gamma-ray induced readout noise to less than 0.6 MeV~\cite{Hu2021BaF2_RINCR}.

\subsection{Choice of photosensor}
\subsubsection{Options}
The choice of an appropriate photosensor depends, of course, on the choice of scintillator.
\par
There are already large area SiPMs appropriate for the readout of pure CsI and for LYSO:Ce. The only major concern is radiation hardness in the Mu2e-II environment, particularly for low energy neutrons.

An appropriate large area photosensor for BaF$_2$ for use in the Mu2e-II calorimeter must meet additional criteria: It must have good quantum efficiency for the 220 nm fast component while being insensitive to the 300 nm  slow component, must be radiation hard, must work in an axial magnetic field, and must have as fast a time response as possible.

There are several candidates for such a sensor. All require additional R\&D efforts before a fully appropriate photosensor can be identified. These are discussed  below.

\subsubsection{Development efforts}

There are several potential approaches. A microchannel plate photomultiplier is very fast and works in a magnetic field. It can be equipped with a photocathode such AlGaN, which is UV-sensitive and solar-blind, and thus a good match to the BaF$_2$ fast component, with quantum efficiency as high as 30\%. Such devices have been used in astrophysics for years. The problem is that even with recent advances in MCP longevity due to the application of ALD coatings to the MCP (to lifetimes of tens of coulombs/cm$^2$), they cannot cope, by orders of magnitude, with the integrated radiation dose of Mu2e-II.

LAPPDs with UV-extended photocathodes such as Cs$_2$Te are an attractive, and possibly less expensive, alternative. They could perhaps be developed on the needed time scale, but again the question of longevity of the MCP in the Mu2e-II environment is an issue.

Wavelength-shifting techniques, particularly involving nanoparticles, are being explored. Specific formulations have been applied to MPPCs operating in photovoltaic mode.

\begin{figure} [!h]
    \centering
    \includegraphics[width=3in]{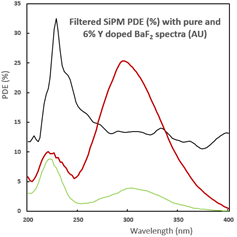}
    \caption{Scintillation spectrum of pure BaF$_2$ and BaF$_2$ doped with 6\% Y, compared with the measured PDE of a $6 \times 6$mm SiPM with a three-layer integrated filter.}
    \label{fig:BaF2w3}
\end{figure}

Atomic-layer-deposition (ALD) bandpass filters  integrated with the silicon structure of the photosensor promise several advantages. These can be either avalanche photodiode \cite{Hitlin2016} or silicon photomultiplier devices \cite{Hennessy2017}; work has been done on both. These have excellent quantum efficiency at 220 nm, strong rejection of 300 nm response, time response superior to existing SiPMs and adequate longevity in the face of exposure to strong UV radiation. The ultimate realization of this concept would be a back-illuminated device with delta doping to improve the time response and resistance to degradation from the incident UV radiation. Figure~\ref{fig:BaF2w3} shows the scintillation spectrum of pure BaF$_2$ and BaF$_2$ doped with 6\% Y, compared with the measured PDE of a $6 \times 6$mm SiPM with a three-layer integrated filter \cite{Hitlin2020}.

\subsection{Data acquisition options}

The Mu2e electromagnetic calorimeter DAQ uses custom waveform digitizers mounted inside the magnet cryostat. The system is qualified for a total integrated dose of up to 20 krad, and samples SiPM signals at 200 MHz. Serial data readout is through a 4.8 Gbit link. The expected Mu2e-II instantaneous event rate will be about three times higher than in Mu2e and will generate a ten times larger data sample; the integrated radiation dose absorbed by the electronics will be
ten times higher as well. Given the expected high rates, 
the shaping time will have to be reduced and therefore the 200 MHz digitizers employed in Mu2e will be inadequate. In the next few years, an intense R\&D campaign will be carried out to evaluate possible alternatives, which will be driven mostly by the choice of crystal and photodetector.
These alternatives include:

\begin{itemize}
     \item A faster waveform digitizer system. Sampling at 1 GHz  will be sufficient to solve pile-up and measure the pulse time and energy. Due to the much higher data flow, raw data will need to be processed inside the digitizer board; only physics related parameters (energy, time, quality, ...) will be transmitted to the central DAQ, a choice that will  reduce the needed bandwidth. Several challenges remain, including availability of rad hard fast ADCs and high performance FPGAs, cost and power dissipation. 

     \item A pure TDC system. CERN has developed a new 64-channel ps resolution TDC, rad hard, named the Pico TDC. In principle, this would solve the bandwidth problem. The required FPGA performance would be limited and would allow the use of commercial low cost rad hard components, such as a Xilinx Kintex or a Microsemi Polarfire. Unfortunately, TDCs do not solve the pile-up problem and energy resolution is quite low. A detailed Monte Carlo simulation will be needed to explore this option. 

     \item A multi-level TDC system. Given the limited number of calorimeter channels, it is possible to transmit the data of one channel to several discriminators with increasing thresholds. A  system with between 4 and 8 thresholds would be possible with less than 200 Pico TDCs. The conversion of the same pulse time at different heights would help resolve the pile-up problem, and also improve the energy resolution. Data could be interpolated on the fly with a limited performance rad hard FPGA.

     \item A mixed system with a TDC plus a relatively slow ADC system. This solution would help to solve the pile-up while retaining optimal time resolution. The ADC speed could be limited to 40 MHz, to be confirmed by Monte Carlo simulations. TDC and ADC data could be combined on the fly by the onboard FPGA which would return only the pulse parameters, thus limiting the employed bandwidth.

\end{itemize}

\section{Stopping target monitor}

In order to measure the denominator of $R_{\mu e}$ the number of captured muons must be determined. In Mu2e this is done by the Stopping Target Monitor (STM) which monitors X- and $\gamma$ - rays emitted at the stopping target during the muon capture process. The detector looks for characteristic emission lines, for an aluminum target:

\begin{enumerate}
    \item A 347 keV emission from the $2p \rightarrow 1s$ transition, which is prompt with the muon stop,
    \item An 1809 keV emission from the nuclear capture, with the characteristic muonic aluminum lifetime of 864 ns, and
    \item An 844 keV emission from the decay of the meta-stable $^{26}$Mg$^\ast$ capture product, with a lifetime of 9.5 min.
\end{enumerate}

In Mu2e the STM consists of a pair of detectors: a high-purity germanium (HPGe) solid-state photon detector, operated at liquid nitrogen temperatures, and a scintillating crystal LaBr$_3$ calorimeter. These detectors complement one-another. The HPGe has an excellent resolution of 1-2 keV and the LaBr$_3$ is capable of handling high rates and has excellent radiation hardness. These detectors are housed in a shielded enclosure, and view the muon stopping target through a collimation system and vacuum window from a distance of about 34 m. The large distance, small collimator openings, and plastic absorber placed between the stopping target and detectors should reduce the photon rate to manageable levels.  These two detectors aim to measure the capture rate to an accuracy of 10$\%$.

The Mu2e-II environment poses significant challenges for the HPGe detector:

\begin{itemize}
    \item HPGe has a slow recovery time. The passage of the beam through the stopping target foils leads to an extremely intense bremsstrahlung flash (``beam flash"), with a high end-point energy of order 60MeV - an order of magnitude larger than our highest signal energy. The system of collimators developed for Mu2e may be not be able to handle the higher rates of Mu2e-II.
    
    \item The resolution of the HPGe detector will suffer from neutron-induced displacement damage. 
\end{itemize}

There are a number of ways to mitigate against the above issues and continue to use the HPGe and LaBr$_3$ in Mu2e-II:

\begin{enumerate}
    \item Reduce the ``beam flash" by increasing the absorber thickness in the STM beamline, this will, however, result in loss of signal rate;
   \item Utilize the high resolution of the HPGe to identify and separate contaminant peaks in the neighborhood of signal lines during special low intensity runs, and use that data to calibrate the LaBr$_3$ detector;
\item The flash is highly directional, while the signal lines are isotropic, moving the detectors off-axis could help, however, there is limited space in the experimental hall, so this may not be feasible;
\item Replace some crystals in the calorimeter with LYSO or LaBr$_3$ , this makes absolute calibration difficult;
\item Create a tertiary photon beam and view that instead. Compton scattering and Bragg diffraction offer two alternatives.
\end{enumerate}

To summarize, the STM provides an in-situ measurement of the muon capture rate at the stopping target, at an accuracy of 10$\%$. In order use the same technology as is used in Mu2e in Mu2e-II significant revisions are required. The more intense environment at Mu2e-II provides higher rates and larger potential neutron damage which can prevent the STM detectors achieving the required resolution. Several potentially extensive modifications are outlined which might help mitigate these issues, however, R\&D is required to understand the capabilities for the future.

\section{Cosmic ray veto}
\subsection{Introduction}
\begin{figure}[htbp]
	\centerline{\includegraphics[width=3.0in]{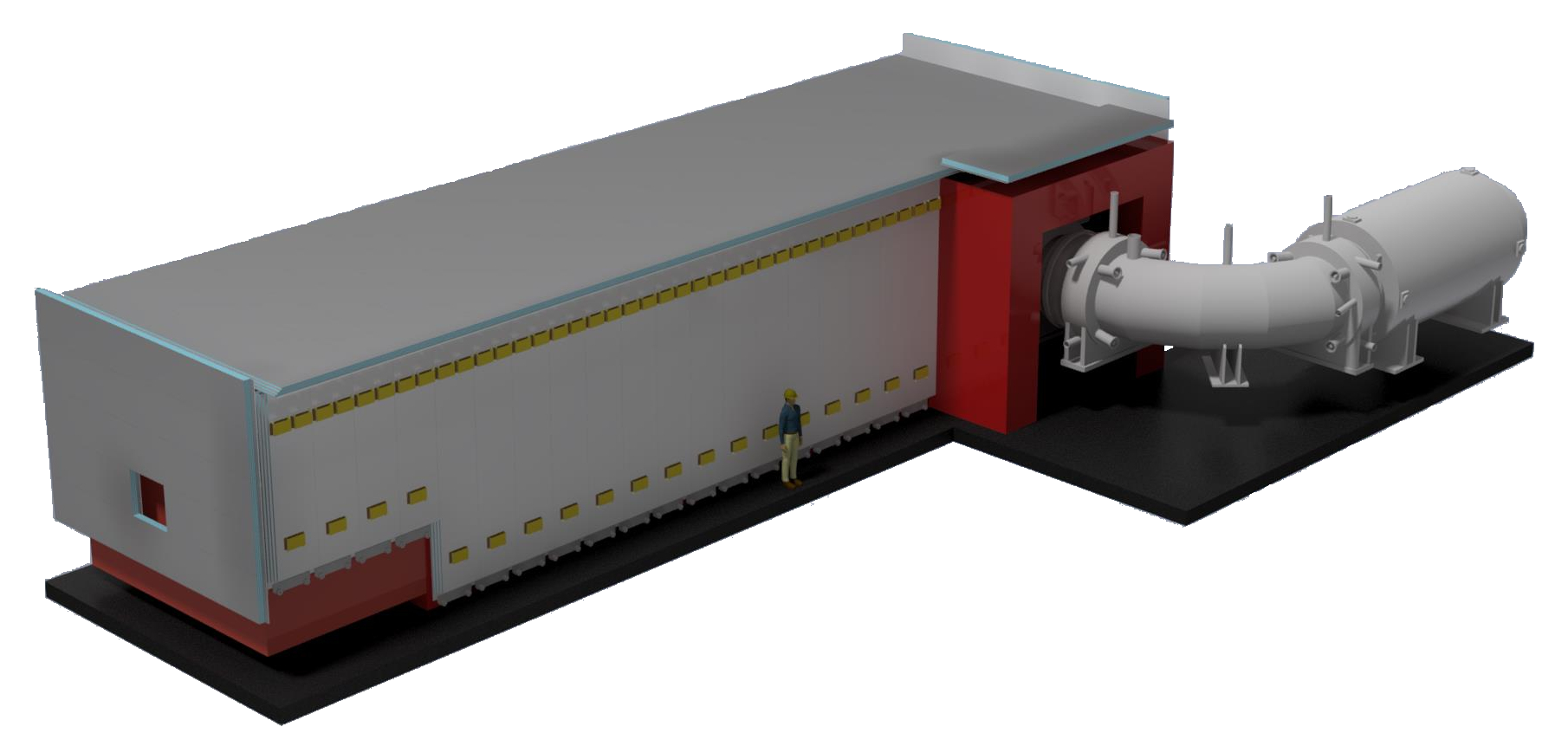}}
	\caption
	{The Mu2e CRV surrounding the detector apparatus.  The solenoids that transport the secondary muon beam to an opening in the CRV can be seen at right.  The detector solenoid which houses the stopping target region and apparatus cannot be seen as it lies under the CRV and concrete shielding.}
	\label{fig:bkgnd-r}
\end{figure}

The Mu2e detector is surrounded by a large-area cosmic-ray veto (CRV), Fig.~\ref{fig:bkgnd-r}, which identifies cosmic-ray muons and vetoes conversion-like events (in the offline analysis) found in coincidence with track-stubs found in the CRV.  The detector consists of four layers of rectangular scintillating counters, each 50 mm wide by 20 mm thick, and with lengths ranging from 1 to 7 meters.  The counters are outfitted with wavelength-shifting fibers placed in channels embedded in the scintillator extrusions, and read out by silicon photomultipliers (SiPMs) situated on each end of the counters \cite{crvdukes}.  The counters envelop the concrete shielding placed around the solenoid that houses the detector elements.

\subsection{Requirements}

The increases in beam intensity and running time for Mu2e-II pose several challenges to the CRV.  First, in order to achieve the proposed Mu2e-II single-event sensitivity goal, the backgrounds need to be kept well below a single event.  The largest anticipated background for Mu2e is induced conversion-like electrons produced from cosmic-ray muons \cite{Kargiantoulakis_2020}.  Such backgrounds scale with live time: the design increase in live time for Mu2e-II of a factor of three means that those backgrounds will increase by the same factor in the absence of any improvements.  Second, non-cosmic-ray induced noise rates in the CRV read-out from particles produced by the primary proton beam as well as from the secondary muon beam are large and are a source of deadtime in the CRV, as well as very high rates in the front-end electronics in certain hot regions of the CRV.  The Mu2e-II increase in beam rate demands measures to keep the deadtime and rates at a reasonable level.  Third, the increase in the total delivered beam means an increased radiation dose to the SiPMs and the front-end electronics.  Finally, aging of the light yield in the present CRV, which is presently determined to be around 6\% per year, will significantly reduce its efficiency. The CRV and shielding designs need to meet the following requirements.

\begin{enumerate}
    \item Suppress cosmic ray background to a fraction of an event,
    \item Readout noise and experimental dead-time induced by beam activities must be $<1$ MHz and $<10$\% respectively,
    \item Radiation doses at the CRV readout must be lower than $<10^{10}$ neutrons (1 MeV eq./cm$^2$).
\end{enumerate}

\subsection{Meeting requirements: Design of Mu2e-II CRV}
\label{sec:CRVdesign}

The current Mu2e CRV detector will not be able to keep the conversion-like cosmic-ray induced background to less than one event for Mu2e-II.  Nor will it be able to keep the dead time to a reasonable level.  Increased radiation doses also pose a problem.  Hence a redesign of the CRV is needed for Mu2e-II.  We present R\&D efforts needed for such an upgraded CRV. 

Much of the proposed work lies in the realm of simulations of different designs.  These simulations are very time consuming: the probability of a cosmic-ray producing a conversion-like event is extremely small.  Fortunately, the Mu2e CRV group has devoted a considerable effort in developing and qualifying a sophisticated set of fast simulation tools.  These have given important additional insights into the nature of the cosmic-ray backgrounds and their sources; insights that were not available when the design of the original Mu2e CRV was finalized.  These tools should allow different designs to be evaluated expediently.

The work that needs to be done can be broken up into the following areas:
\begin{itemize}
    \item  Improving the shielding to minimize increases in the dead time and rates and radiation doses in hot areas.
    \item  Improving the existing CRV design to achieve the goal of less than one conversion-like background event while keeping the rates and dead time at a reasonable level.
    \item  Exploring ideas on how to veto background events due to cosmic-ray muons entering the apparatus through the muon beam gap. 
\end{itemize}

The background rates and radiation doses to the CRV come from two sources: the primary proton beam interactions in the production target and from stopped-muon produced secondaries. Shielding around the production target and around the detector solenoid in which the apparatus lies needs to be augmented. Much of that shielding is regular concrete. Improved shielding using Barite and boron-loaded concrete, as well as boron-loaded plastics, has been explored. This type of shielding proves to be very efficient in suppressing both read-out rates and experimental deadtime to negligible levels.

The present Mu2e CRV design has long rectangular counters.  We propose to investigate a CRV detector that is based on finer granularity counters in order to reduce the single-counter rates and to reduce the false coincidence rates in reconstructing track stubs from hits in different layers of the CRV. In addition, the Mu2e CRV inefficiency is driven by unavoidable gaps between counters.  Monte Carlo studies have shown that a large proportion of the muons not detected by the CRV comes from muons traversing the gaps, even though the scintillator layers are offset to minimize such effects. Improved designs, such as the use of triangular-shaped counters, will be studied in order to reduce such inefficiencies.  The aging of similar triangular counters has been measured by MINERvA \cite{minervaaging} and appears to be significantly less than that measured by Mu2e.  The source of the large aging measured by Mu2e is not understood and needs to be explored.

The efficiency of the CRV depends critically on the light yield of the counters.  The light yield can be improved by using SiPMs with a higher efficiency (PDE).  The technology has rapidly improved since the current Mu2e CRV devices were purchased: better SiPMs are available with higher PDEs and more radiation hard. 

Another way to increase the light yield is to pot the fibers in their channels with silicone resin, epoxy, or other materials. Preliminary studies show increases up to 50\% can be achieved. Further studies are needed to develop leak-free, fast filling techniques, and measurement of the aging rate for each filling material.

Using the present CRV design, the background component induced by cosmic ray muons that fall into the CRV coverage is estimated to be $0.22 \pm 0.15$ events at Mu2e-II. This background components can be suppressed to below 0.1 events with the enhanced CRV design. Cosmic ray neutrons will produce $0.62 \pm 0.04$ events. Background induced by cosmic ray neutrons can be easily suppressed to $0.02 \pm 0.002$ events with an addition of 6' of concrete shielding above the stopping target region. The component induced by cosmic ray muons entering through the uninstrumented CRV region is estimated to be $0.08 \pm 0.02$ events. This background component is challenging to suppress and assumed to be irreducible in this study. However, we plan to explore the opportunities to suppress this background component using a passive or active shielding options. 

In summary, we expect that the cosmic ray induced background at Mu2e-II can be suppressed to $0.20 \pm 0.08 \hbox{ (stat)}$ events over the 5-year (acquired live-time of $4.1\times10^7$ s) data taking period of the Mu2e-II experiment. Both CRV and shielding designs must be enhanced to achieve the requirements.

\section{Trigger and data acquisition}
\subsection{Requirements}
In order to set the requirements for the TDAQ system, we assume that Mu2e-II will adopt a similar experimental setup as Mu2e, but will improve granularity of detector elements up to a factor of two. The
direct consequences of these assumptions are:
\begin{itemize}
    \item 
    An increase in the event data size of a factor of approximately $6$; including a factor of $3$ due to the instantaneous rate and a factor of $2$ due to the number of channels,
     reaching a level of 1 MB/event; 
\item A reduced period when no beam is delivered to the apparatus, which in Mu2e is 1 s out of 1.4 s;
\item A factor of approximately $10$ larger dose on the electronics.
\end{itemize}

Assuming that the Mu2e-II storage capacity on tape will be twice that of Mu2e, reaching $\sim14$~PB/year (equivalent to a few kHz), the required trigger rejection needs to be a factor of $\sim5$ better than in Mu2e, which is at the level of a few hundreds.

\subsection{Architectures}

\begin{figure}[ht!]
    \begin{center}
        \includegraphics[width=0.48\textwidth]{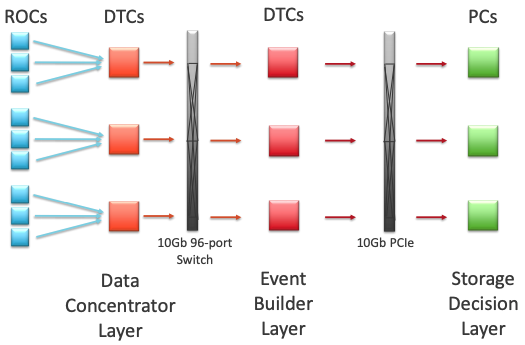}
        \caption{Mu2e data readout topology.}
   \label{fig:readout_topo}
   \end{center}
\end{figure}

Mu2e uses {\it artdaq}~\cite{ARTDAQ} and {\it art}~\cite{ART} software as event filtering and processing frameworks respectively. The detector Read Out Controllers (ROC), from the tracker and calorimeter stream out continuously the data, zero-suppressed, to the Data Transfer Controller units (DTC). The data of a given event is then grouped in a single server using a 10 GByte switch. Then, the online reconstruction of the event starts and makes a trigger decision. If an event gets triggered,  the data from the CRV is also pulled and everything is aggregated in a single data stream. Figure~\ref{fig:readout_topo} shows a scheme of the Mu2e data readout topology described above.

The Mu2e main physics triggers use the information of the reconstructed tracks to make the final decision. The Mu2e Online track reconstruction is factorized into three main steps~\cite{Mu2eTracking}: (i) hits preparation, where the digitized signals from the sub-detectors are converted into reconstructed hits, (ii) pattern-recognition to identify the group of hits that form helicoidal trajectories, and finally (iii) track fit through the hit wires, which performs a more accurate reconstruction of the track.

\subsubsection{2-level trigger, L1(FPGA)+HLT}
For Mu2e-II one of the ideas is to implement a L1 hardware trigger that exploits the first two stages of the online track reconstruction on a dedicated FPGA based board and then exploit the rest of the reconstruction on the commercial servers. The major challenges are represented by: (a) the amount of data that needs to be concentrated on a single board and (b) the migration of a non negligible part of the online reconstruction onto an FPGA. For the first challenge, the system will need to use more performant rad-hard optical transceivers (R\&D is already ongoing at CERN), which are needed to stream the data from the ROCs to the data-concentrator layer, and a more powerful switch (100 GB switches are already available). For the second challenge, it is important to realize that FPGA development can take place now – hardware is not needed! Starting now would help the understanding of required resources and in consideration of topology trade offs. For example, what size FPGA is best suited, or what are the advantages and disadvantages of commercially available hardware versus established custom boards in the community versus creating a new custom board. 

In the last decade, a new tool named High Level Synthesis (HLS)~\cite{Duarte_2018} has been developed to rival manual VHDL or Verilog algorithm development. The major HLS features are: (i) it allows non-specialists to easily understand and develop low and fixed latency FPGA algorithms, (ii) it simplifies offline emulation, (iii) it facilitates debug and verify in a software environment (often 10x faster iterations than firmware simulation tools). We also note that other HEP collaborations, like the CMS experiment at CERN, have been heavily investing in the approach to FPGA algorithm development. Interestingly, the Mu2e run plan offers the possibility to test (parasitically) a prototype of a L1 trigger board in the second phase (after the LBNF shutdown). Leveraging Mu2e as a live data source would give valuable feedback for advancing Mu2e II’s R\&D phase.

\subsubsection{Software trigger with GPUs}
For Mu2e-II another idea is to exploit the use of GPUs for implementing the reconstruction algorithms. Other experiments in HEP have been already using GPUs in the high level trigger (HLT), such as ALICE~\cite{Rohr_2012} and ATLAS~\cite{7581786}. There are also experiments that implemented L0 trigger on a dedicated GPU board~\cite{Ammendola_2017}. The major challenges are similar to those for the FPGA: (a) the amount of data that needs to be concentrated on a single board and (b) the migration of a non negligible part of the online reconstruction to the GPU. As above, the system will need to use more performant rad-hard optical transceivers (R\&D is already ongoing at CERN), which are needed to stream the data from the ROCs to the data-concentrator layer, and a more powerful switch (100 GB switches are already available). As we said for the FPGA, we note that GPU development can take
place now – hardware is not needed! Starting now would help decide which GPU is more suitable for the Mu2e-II needs.

\section{Backgrounds and Physics Sensitivity}
\label{sensitivity_estimate}

\subsection{Stopped Muon Rate}
\label{fluxes}

The production target design used in this study is the carbon conveyor with 28 spheres, each 0.75 cm in radius.
The stopped muon rate, defined as the number of muons stopped at the stopping target per proton on the production target, is: $9.1 \times 10^{-5}$ for this chosen design.
If we instead were to use the Mu2e target the stopping rate would be higher at $1.03\times10^{-4}$.
That target would not be feasible in Mu2e-II as it would require significantly more cooling during operation. Since R$\&$D is on-going it is likely that the stopped muon rate of $9.1 \times 10^{-5}$ is a conservative estimate.

\subsection{Conversion Signal}

In Mu2e-II low energy muons will be produced from the decays of pions produced when an 800 MeV proton beam interacts with the production target.
The resulting muons then traverse the transport solenoid, where collimators select the low momentum, negative muons.
These muons are then directed onto the stopping target located within the detector solenoid.

When low momentum, negatively charged muons are trapped in the field of the target nucleus, forming a muonic atom, the muon cascades down to the $1s$ bound state.
The three main processes considered are:

\begin{enumerate}
\item \textbf{decay in orbit (DIO):} $\mu^{-} \rightarrow \nu_{\mu}+\bar{\nu}_{e}+e^{-}$;
\item \textbf{muon capture:} $\mu^{-}+N(A,Z)\rightarrow \nu_{\mu}+N(A,Z-1)$;
\item \textbf{neutrinoless conversion:} $\mu^{-}+N(A,Z)\rightarrow e^{-}+N(A,Z)$;
\end{enumerate}

\noindent
where $N(A,Z)$ denotes the mass and atomic numbers of the target nuclei.
The event signature of $\mu^{-}N\rightarrow e^{-}N$~ in a muonic atom is, at first order, the emission of a mono-energetic electron with an energy:

\begin{equation}
  E_{\mu e} = m_{\mu} - E_{BE,1s} - E_{recoil},
\end{equation}

\noindent
where $m_{\mu}=105.66 $MeV$/c^{2}$) is the muon mass, $E_{BE,1s}$ is the binding energy of the $1s$ state, and $E_{recoil}$ denotes the nuclear recoil energy.
Radiative corrections have been calculated and are discussed in Ref. \cite{2017szafron}. $E_{\mu e}$ is nucleus dependent, for instance, $E_{\mu e}$ = 104.97 MeV for aluminum. The isotope dependence of the conversion electron energy is discussed in Sec.~\ref{sec:target_isotope_theory}.

\subsection{Backgrounds}

The following processes can provide potentially significant background contributions to the search for $\mu^{-}N \rightarrow e^{-}N$:

\begin{itemize}
\item \textbf{DIO} - free muon decay follows the Michel spectrum, peaking at an endpoint of $\sim$ 52 MeV/c, far from the signal region.
  The Mu2e-II tracker is expected to have a similar design to that used by Mu2e, an annular cylinder where the central region is  un-instrumented.
  This purposely blinds the detector to nearly all DIO electron backgrounds. However, when the muon decay takes place in a nuclear field, such as in the stopping target,
  electrons can be produced with an energy up to the kinematic endpoint at $\sim$ 105 MeV/c.
  Separating the signal electrons from these DIO recoil tail electrons necessitates excellent momentum resolution.

\item \textbf{Radiative Pion Capture (RPC)} - pions can contaminate the muon beam and generate a significant background which rapidly falls in time due to the short
  pion lifetime.
  Suppressing the RPC background requires the signal search measurement to be delayed with respect to the pion production time, referred to as a
  ``delayed live-gate." Fig.~\ref{fig:beam_livegate1} shows an illustration of the Mu2e-II beam.
  Proton pulses are indicated in gray.
  They are around 100 ns narrower than those in Mu2e.
  Due to the pion having a much shorter lifetime than the muonic atom, a time cut can be enforced
  to suppress the RPC background, where only data collected beyond this timing cut is used for the signal search.

  Late-arriving pions produced by protons not in the main beam pulse cannot be eliminated by the delayed live-gate, only extinction of out-of-time protons can do this.
  An extinction factor (ratio of out-of-time to in time protons) of $<10^{-11}$ is required to  suppress out-of-time backgrounds in Mu2e-II.

  \begin{figure}[ht]
    \centering
    \includegraphics[width=0.5\textwidth]{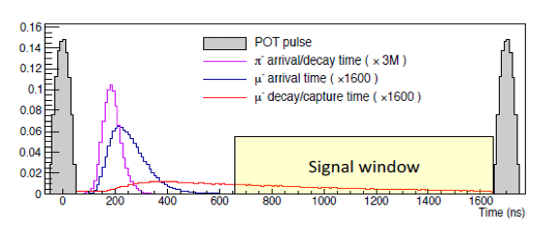}
    \caption{ \label{fig:beam_livegate1}Example of the delayed live gate.}
  \end{figure}

\item \textbf{Cosmic particles} - cosmic particles interacting and decaying in the detector volume are a source of electrons
  with a spectrum which covers the region around 100 MeV/c. Most cosmic particles reaching the detector are muons, therefore suppression of the cosmic background
  requires identifying muons and vetoing them.  This cosmic muon identification is performed using the CRV detector, which surrounds the experiment.

\end{itemize}

Optimization of the search sensitivity used in this paper is based on finding
the 2D momentum-time signal window maximizing the discovery potential of the experiment. 

\subsection{Modeling Backgrounds}
\subsubsection{ Muon decays in orbit}

The spectrum of the outgoing electrons from DIO has a recoil tail which rapidly decreases for $E>$ 100 MeV as the decay final state phase space shrinks as ($E_{CE}-E$)$^5$ when the DIO electron energy approaches the endpoint.

The momentum spectrum derived in Refs.~\cite{2016_Szafron_DIO_LL_PhysRev,2016_Szafron_DIO_LL_PhysLettB},
is used to parameterize DIO backgrounds in this study. This includes soft photon emission and vacuum polarization effects.

\subsubsection{Radiative pion capture}

Pions can contaminate the muon beam and potentially stop in the stopping target where they could undergo radiative pion capture,
$\pi^{-} + N(A,Z) \rightarrow \gamma^{(*)} + N(A, Z-1)$. This process, followed by an asymmetric $\gamma \ra e^+e^-$ conversion, produces electrons with energies
up to the charged pion mass, and is one of the main background sources to the $\mu^-\to e^-$ search.

This process is referred to as internal conversion. Conversion of on-shell photons in the detector material is referred to as the process of external conversion. Compton scattering of on-shell RPC photons in the detector also produces background electrons. This increases the RPC background in the electron channel and makes spectra of electrons
  and positrons produced in RPC different. The internal conversion fraction, $\rho$, the ratio of the off-shell and on-shell photon emission rates, is calculated in Refs. \cite{RPC_1955_Kroll_Wada_PhysRev.98.1355, RPC_1960_Joseph_IlNuovoCimento.16.6.997}. We assume that the internal conversion fraction does not depend on the photon energy and
  use the value of $\rho~=~0.0069~\pm~0.00031$ measured in \cite{samios}.

  The RPC background modeling in this study relies on the RPC measurements published in Ref. \cite{Bistrilich}.  There is no published data on Al, so the spectrum of RPC photons measured on a Mg target is used instead. For nuclei with the nuclear charge, $Z$, in the range $6 < Z < 20$,
 the measured RPC branching ratio varies within ${\sim} 10\%$, so
 the difference between Al and Mg should not introduce a significant additional systematic uncertainty.
 
 The pulsed timing structure of the proton beam leads to two distinct components of RPC background:

\begin{enumerate}
\item \textbf{In-time RPC:} radiative capture of pions produced by protons arriving in the beam pulse.
 The rate of in-time RPC will be highly suppressed by the pion lifetime and the choice of live-gate is optimized
  to minimize this background.
\item \textbf{Out-of-time RPC:} radiative capture of pions produced by out-of-time protons.
 A delayed live-gate cannot eliminate such pions, only extinction of out-of-time protons can do this.
\end{enumerate}

\subsubsection{Cosmic ray background}

The background component induced by cosmic ray particles is estimated to be $0.20 \pm 0.08$ events at Mu2e. See the Cosmic Ray Veto section~\ref{sec:CRVdesign} for details.

\subsubsection{Radiative muon capture}

Radiative muon capture (RMC), $\mu^{-} + N(A,Z) \rightarrow \gamma^{(*)} + \nu_\mu + N(A, Z-1)$, is an intrisic background at Mu2e and Mu2e-II.
It is similar to RPC, where the internal pair production spectrum used for RPC~\cite{RPC_1955_Kroll_Wada_PhysRev.98.1355}
is also applicable to describe internal pair production for nuclear RMC, as discussed in Ref. \cite{Plestid:2020irv}.
RMC measurements historically focused on the total rate, while the spectrum endpoint region is less well understood.
Most studies of nuclear RMC rely on approximations, such as the closure approximation, which are unreliable near this
endpoint \cite{PhysRevC.21.1951,CHRISTILLIN1980331}.
In order to understand any potential signals and describe the background electron and positron spectra at Mu2e-II it is important understand the
RMC (on- and off-shell) photon spectrum in the high energy region.
This is especially important for the $\mu^- N \to e^+ N'$ search, as the maximum kinematically allowed photon energy is $\sim$10 MeV above the conversion positron signal region.
For aluminum, the maximum allowed electron energy from an RMC interaction is $\sim$2.5 MeV below the signal region, and so the background contribution
in this region for the $\mu^- \to e^-$ search is negligible, as discussed in Ref. \cite{2022_Mu2e_SU2020}.
A measurement of the on- and off-shell RMC photon spectrum on alumimum (and potentially other medium heavy nuclei) is necessary to understand
the electron and positron background contributions from RMC. Data from Mu2e can help understand these aspects of RMC in aluminum.

\subsection{Beam Backgrounds}

Beam-related background contributions arise in our detectors when beam particles do not stop
in the stopping target. These particles originate from the initial protons arriving at the production
target in between the proton pulses and are suppressed by the proton beam extinction. Beam related backgrounds are expected to be very small:
\begin{itemize}
\item
 Beam electrons result from muons decaying in the beamline. If they reach the stopping target with a momentum close to 105 MeV/c
  they could fake a signal. As there is only a small probability of a large angle scattering in the stopping target and the beam extinction factor is small ($10^{-11}$) the expected
  contribution from beam electrons is $< 0.004$ events.
\item
  Decay in Flight of muons and pions in the DS can produce electrons with the momentum above 100 MeV/c which could get
  reconstructed without scattering in the stopping target.
  The estimated contribution from decays in flight is expected to be very small, at $< 0.007$ events.
\end{itemize}

\subsection{Stopping target studies}
\label{stopping_target_studies}

The muon stopping target provides the nuclear target for the $\mu^- \to e^-$ conversion process to take place; it also provides the normalization signal
  in the form of the muon capture process as well as contributes to the signal electron energy losses before reaching the detectors. The choice of target design can impact momentum resolution through energy straggling and multiple scattering of
the electron as it leaves. Consequently, the muon stopping target can significantly affect the achievable sensitivity of the overall Mu2e-II experiment. Its design is crucial
to achieving the physics goal.

In Mu2e the stopping target consists of 37 aluminum foils, 105 $\mu$m thick and 75 mm in radius, placed 22.22 mm apart and suspended in a frame. Each foil has a 21.5 mm radius hole in the center to help reduce the effects of beam flash. 

In this section preliminary studies are presented which look at optimization of the Mu2e-II muon stopping target design in terms of improving the single event sensitivity (SES) and achievable branching fraction upper limit (assuming no signal). Further R$\&$D will be required to finalize the Mu2e-II stopping target.

\subsubsection{Single Event Sensitivity Equation}

The single event sensitivity (SES) is the signal rate corresponding to an expectation of 1 event over the course of the experiment. The SES can be written as:

\begin{equation}
    SES = \frac{1}{POT \times \frac{stops}{POT} \times \frac{capture}{stop} \times \frac{N_{CE}^{reco}}{N_{CE}^{gen}}}
    \label{eq:SES}
\end{equation}

where:

\begin{itemize}
    \item $POT$ - number of protons at the production target;
    \item $\frac{stops}{POT}$ - number of stopped muons at the stopping target per number of simulated protons at the production target;
    \item $\frac{captures}{stop}$ - capture rate of the target nuclei, e.g. for aluminum this is 61 $\%$;
    \item $\frac{N_{CE}^{rec}}{N_{CE}^{gen}}$ - the reconstruction efficiency. 
\end{itemize}

\begin{figure}[ht]
     \centering
         \includegraphics[width=0.5\textwidth]{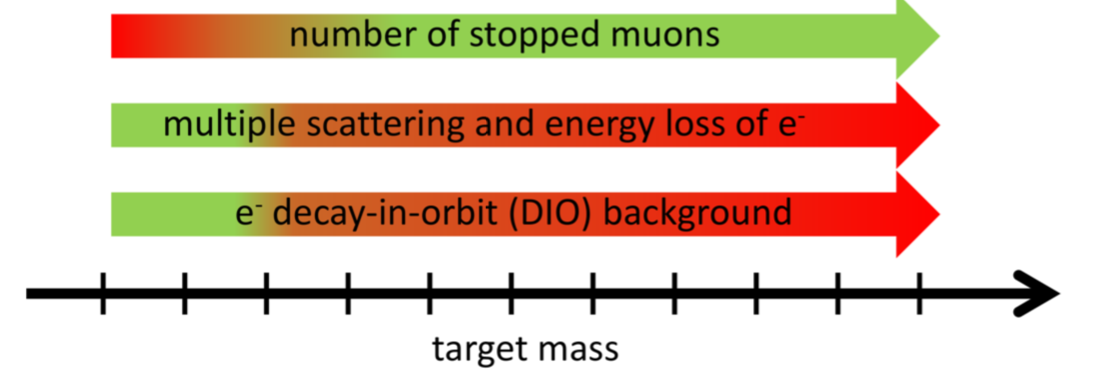}
         \caption{Relative changes in three processes as function of target mass. Here the green indicates a improvement in sensitivity and red a worsening sensitivity.}
         \label{fig:target_balance}
\end{figure}

Fig.~\ref{fig:target_balance} shows three effects and how their relative importance changes as we increase the target mass:

\begin{itemize}
    \item Number of stopped muons  - the more massive a target, the more muons we expect will be stopped;
    \item Multiple Scattering and energy loss of the outgoing electron - the more massive the target the more scattering it experiences as it attempts to leave;
    \item DIO background - the effects of the DIO background increase with mass
     due to the broadening of the signal energy spectrum after energy losses.
\end{itemize}

The studies aim to find a ``sweet spot'' in the target design where the number of stopped muons is large but where multiple
scattering and energy losses do not overwhelm the conversion electron search.

\subsubsection{Design Features}

There are a number of design features of the stopping target which may require optimization specifically for the Mu2e-II beam:

\begin{itemize}
\item Target Material - see Sec. \ref{sec:target_isotope_theory}. Depending on the results of Mu2e an alternative material, which complements aluminum may be used to
  help elucidate physics signal;
\item Target Mass - the choice of mass is expected to be a compromise. Larger target mass means higher stopped muon rate,
  but also increased scattering of the outgoing electron;
\item Target Geometry - for the same mass the physical design of the target can affect the amount of material the outgoing electron traverses,
  and both impact the SES and increase the background due to the broadening of the signal distribution;
\item Incoming Muon Momentum Structure - only around 37$\%$ of muons which enter the DS are stopped in the target. These are typically the lower momenta particles.
  If the incoming muon beam momenta distribution were altered, this could affect the momentum space of the muons stopped, and thus affect the yield.
\end{itemize}

\subsubsection{Alternative Materials}

 As discussed in Sec. \ref{sec:target_isotope_theory}, in the event that Mu2e has measured muon-to-electron conversion in Al, Mu2e-II should use an alternative target material. Measuring the relative conversion rates
in two complementary materials can help elucidate the underlying physics responsible for the conversion \cite{Kitano:2002mt,Heeck2022}. A preliminary study looked at titanium,
vanadium and lithium foil style targets and attempted to measure achievable SES in each. High-Z materials such as gold have the advantage of larger capture rates,
but the muonic atom lifetime is too short, and this makes the delayed live-gate highly inefficient,
thus beam-related backgrounds become an issue.

Theoretical work has suggested a low-Z material, such as lithium, could complement aluminium \cite{davidson2019,Heeck2022}.  From a practical point of view, a foil lithium-based target
is feasible. Lithium has been used in foil form by a previous experiment; Ref.~\cite{Brookhaven-CERN-Syracuse-Yale:1976aaq} details the practicalities of fabricating the foils.
From a physics perspective, the muonic lifetime in lithium is close to the free muon lifetime and the capture rate is small at just a few percent (see Tab.~\ref{tab:B_and_binding}).
Consequently, to reach the same upper limit as aluminum would require a much longer running time. So, choosing a lithium target would depend what conversion rate had been
observed in aluminum; if a rate of $\mathcal{O}(10^{-14} -10^{-16})$ had been measured at Mu2e or COMET, lithium could indeed be a good alternative material in Mu2e-II.

Another practicality is that, due to lithium's much lower density ($ 0.523$ g/cm$^{3}$), a very large number of foils would be required to get the same number of stopped muons
as in aluminum. Since there is limited space in the detector solenoid, there is a limit to how big the target can be in terms of horizontal length.
The dimensions of the foils were kept the same as those for the aluminium with the separation between foils being adjusted accordingly as more foils are added.
The minimum SES from those geometries simulated came from the $\sim$ 400 foil design which gave an SES of $\mathcal{O} (10^{-17})$.

Titanium has often been tipped as one of the best alternatives to aluminium. Titanium has a larger capture rate at $\sim$ 85\% but a shorter lifetime of $\sim 329$ ns (see Tab.~\ref{tab:B_and_binding}); however,
natural titanium has a mixed composition of isotopes. Ref.~\cite{Davidson2018Spin} discusses the difference between potential spin-dependent and spin-independent contributions.
Knowing the spin composition of the target might be useful, meaning a pure titanium target of a specific isotope would be necessary.
Obtaining pure samples of titanium is possible, but to get $\mathcal{O}(100g)$ of a given isotope would be difficult and costly.
Vanadium has an atomic number close to titanium and thus offers the same benefits but has just one main isotope.
A titanium and a vanadium target was studied using the same foil design as used for the aluminum target.
The optimum mass (with the lowest SES) was found for both elements to be $\sim$ 160 g.
The SESs with this mass were found to be slightly better than that achievable with aluminum due to the larger capture rate and were $\mathcal{O}(10^{-18})$.

Further study is ongoing and a subsequent publication is expected which will detail the studies described here.

\subsubsection{Alternative Geometries}

In this study the mass of the aluminium target was kept at a constant value of 171g. This is the approximate mass of the current Mu2e target.
Several alternative geometries were studied:

\begin{itemize}

    \item Screen Thick Mesh (SD) - circular screen made of mesh wires of 0.1143 mm thick.  Fig.~\ref{fig:screen} shows an example screen design.
    \item Screen Thick Mesh with Hole (SDH) - same as the SD design, but with a hole of 21.5 mm radius.
    \item Screen Fine Mesh (SM) - similar to  the SD design but with 0.02665 mm thick wires
    \item Screen Fine Mesh with Hole (SMH) -  the same as the SM design but with a 21.5 mm hole in the center.

    \begin{figure}[H]
     \centering
         \includegraphics[width=0.4\textwidth]{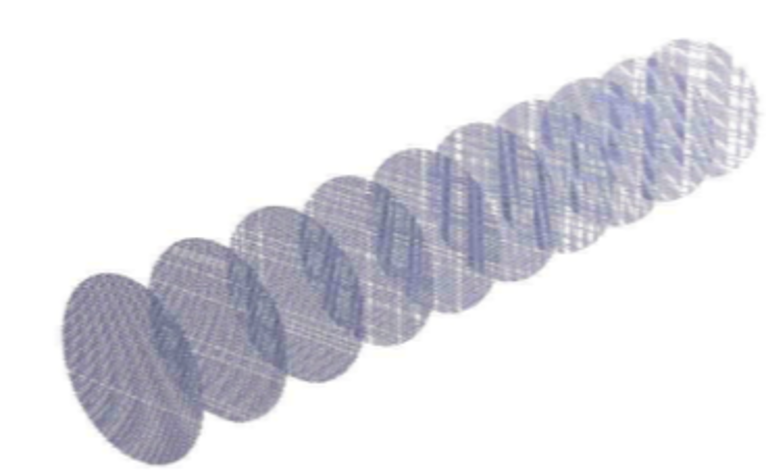}
         \caption{The screen designs are constructed from arrays of thin wires, with two commercially available wire sizes.
           The designs consist of grids of aluminum wires arranged in disks.}
      \label{fig:screen}
    \end{figure}

      \item Hexagonal Mesh (HM) - array of cylinders in hexagonal cross-section in XY.
      \begin{figure}[H]
     \centering
         \includegraphics[width=0.4\textwidth]{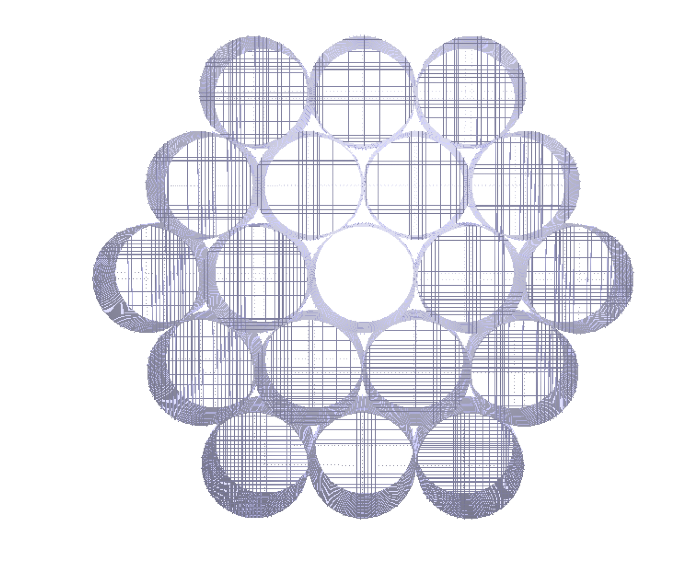}
         \caption{Design with hexagonal cross section, design comprises of screen based cylinders arranged in hexagonal shape.}
      \label{fig:hex}
    \end{figure}
    
      \item Cylinder Thick Mesh (CTM) - cylinder but made of mesh of 0.1143 mm thick strings.
    \item Cylinder Fine Mesh (CM) - cylinder but made of mesh of 0.02665 mm thick strings.

     \begin{figure}[ht]
\centering
\subfloat{
      \includegraphics[width=0.4\textwidth]{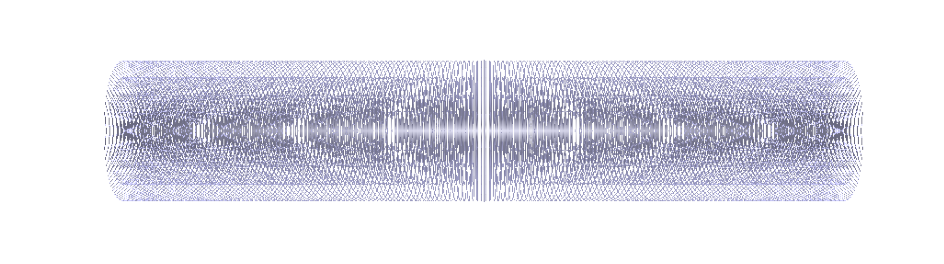}
}
\hfill
\subfloat{
      \includegraphics[width=0.3\textwidth]{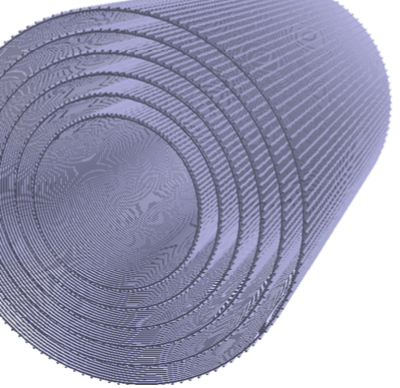}
}

\caption{Longitudinal (top) and transverse (bottom) view of cylindrical design made of strings of 0.02665 mm arranged in a mesh.} \label{fig:cylmesh}
\end{figure}

\end{itemize}

\begin{table}[H]
\centering
\caption{Relative changes (as fraction of what can be achieved with the Mu2e target design) in SES and median upper limit on $R_{\mu e}$ at 90\% CL ($R_{\mu e}$(90\% CL)) (BFUL) for alternative geometries
  with respect to the Mu2e era geometry.}
\begin{tabular}{c c  c} 
\hline\hline
Design Name &  SES & BFUL (90$\%$ CL) \\ [0.5ex] 
\hline\hline
Mu2e Default Design&1&1\\
CTM&1.03&-\\
HM &1.06&-\\
SD &0.93& 0.96\\
SDH &0.95&0.96\\
SM&0.99&1.0\\
SMH & 1.03&1.02\\
\hline\hline
\end{tabular}
\label{table:target_design_results}
\end{table}

Table~\ref{table:target_design_results} lists the ratio of SES (relative to the default) for each alternative design.
The screen designs are shown to improve the SES by $\sim 5\%$ with the thinner strings having the better performance.
The shift in projected branching fraction upper limit is calculated using a background only hypothesis and the work of Feldman-Cousins \cite{FELDMANCOUSINS1998}
in a counting experiment. DIO and RPC are the only background processes considered.

Since the improvements seen are minimal, in the remainder of this analysis the 37 foil, Mu2e target design is used.

\subsection{Experimental Assumptions}

The simulation and reconstruction software used in this study was taken from the framework developed for the Mu2e experiment, where the
geometry was updated for the Mu2e-II design.
The Mu2e simulation framework is based on Geant4 (see Appendix \ref{sec:GEANT4}). It takes into consideration cross sections and time dependencies of the physics processes as well as the
responses of the sub-detectors, and effects of hit readout and digitization. The Mu2e software environment is described in detail in
Ref. \cite{2022_Mu2e_SU2020}.
The reconstruction algorithms are also described in this reference. We do not update, or re-optimize, these algorithms for the Mu2e-II environment, meaning our reconstruction efficiency with pileup will likely be improved with further study.

The  geometry includes:

\begin{itemize}
    \item A carbon conveyor production target with 28 spheres, each 0.75 cm in radius;
    \item No antiproton windows in the TS since the beam energy is too low to produce antiprotons;
    \item An aluminium stopping target with design identical to that used in Mu2e;
    \item A straw tube tracker similar to the Mu2e tracker with thinner straws with 8~$\mu$m thick walls (compared to 15~$\mu$m in Mu2e)
      and reduced mass through the removal of the gold layer in straws;
    \item The crystal calorimeter has the same number of crystals as in Mu2e and they have the same geometry ($34 \times 34 \times 200$ mm$^{3}$).
      The major difference is that they are made of barium fluoride instead of cesium iodide.
      The waveform response of the SiPMs with the barium fluoride is an educated guess. Further study is required to understand the actual response function.

\end{itemize}

The extinction factor (i.e., the ratio of out of time to in time protons) is assumed to be $<1 \times 10^{-11}$.

\subsection{Selection Criteria}
\label{selection cuts}

The following selection cuts are applied in this study:

\begin{itemize}
\item N(hits) $\ge 20$: Requires a sufficient number of hits along the track for the momentum to be reconstructed well;
\item $|{\rm D}_0| < 100$ mm: Requires the helical track to be consistent with the 75 mm in radius stopping target, where the 100 mm threshold
  accounts for tracks with energy loses before reaching the tracker; 
\item ${\rm R_{max}} < 680$ mm: Ensures tracks are well contained within the 700 mm radius tracker;
\item $0.5 < \tan({\rm dip}) < 1$: Requires the topology to be consistent with helical tracks originating from the stopping target;
\item $\sigma_{\rm T_0} < 0.9$ ns: Rejects tracks with high $\rm T_{0}$ uncertainty, which are often tracks with the calorimeter cluster not included
  in the track reconstruction;
\item $E_{\rm cluster} > 10$ MeV and $E_{\rm cluster} / P_{\rm track} < 1.05$: Requires a calorimeter cluster associated with the track with a $E/P$ ratio consistent
  with an electron;
\item Track quality ANN selection: Rejects poorly reconstructed tracks;
\item $T_0 < 1650$ ns: Avoid early flash from the next beam pulse.
\end{itemize}

\subsection{Efficiencies}
\label{efficiency}

For a pure conversion electron sampled (without pileup) the reconstruction efficiency per simulated event is 36.7$\%$ and the selection efficiency using the cuts defined
in Sec.~\ref{selection cuts} is 72.7$\%$, meaning the total efficiency is 26.7$\%$.
Current studies indicate the trigger efficiency at Mu2e for the electron signal will be higher than 98\%. We therefore assume a trigger efficiency of 95\%
at Mu2e-II. The dominant effect of pileup on the simulated track reconstruction is the worsening of the track momentum resolution due to both background
hits added along the reconstructed track trajectory and the loss of true track hits due to tighter hit selection requirements. The reconstruction efficiency
decreases by 2.5\% after the introduction of pileup at Mu2e, where we assume a reconstruction efficiency drop of 5\% at Mu2e-II.

\subsection{Tracker Resolution}

The tracker momentum resolution is crucial to discriminating the conversion signal from DIO background.
The distribution of the difference between the reconstructed momentum and the MC true momentum of the electron at the front of the tracker
is studied to understand the tracker reconstruction resolution. Furthermore, the effect of energy losses before entering
the tracker is also important as this can broaden the signal momentum distribution, so the full width at half maximum (FWHM) of the
reconstructed momentum spectrum for the conversion electron signal is another important metric for how well the signal can be separated
from the rapidly falling DIO background.

\begin{figure} [t]
    \centering
    \includegraphics[width=3in]{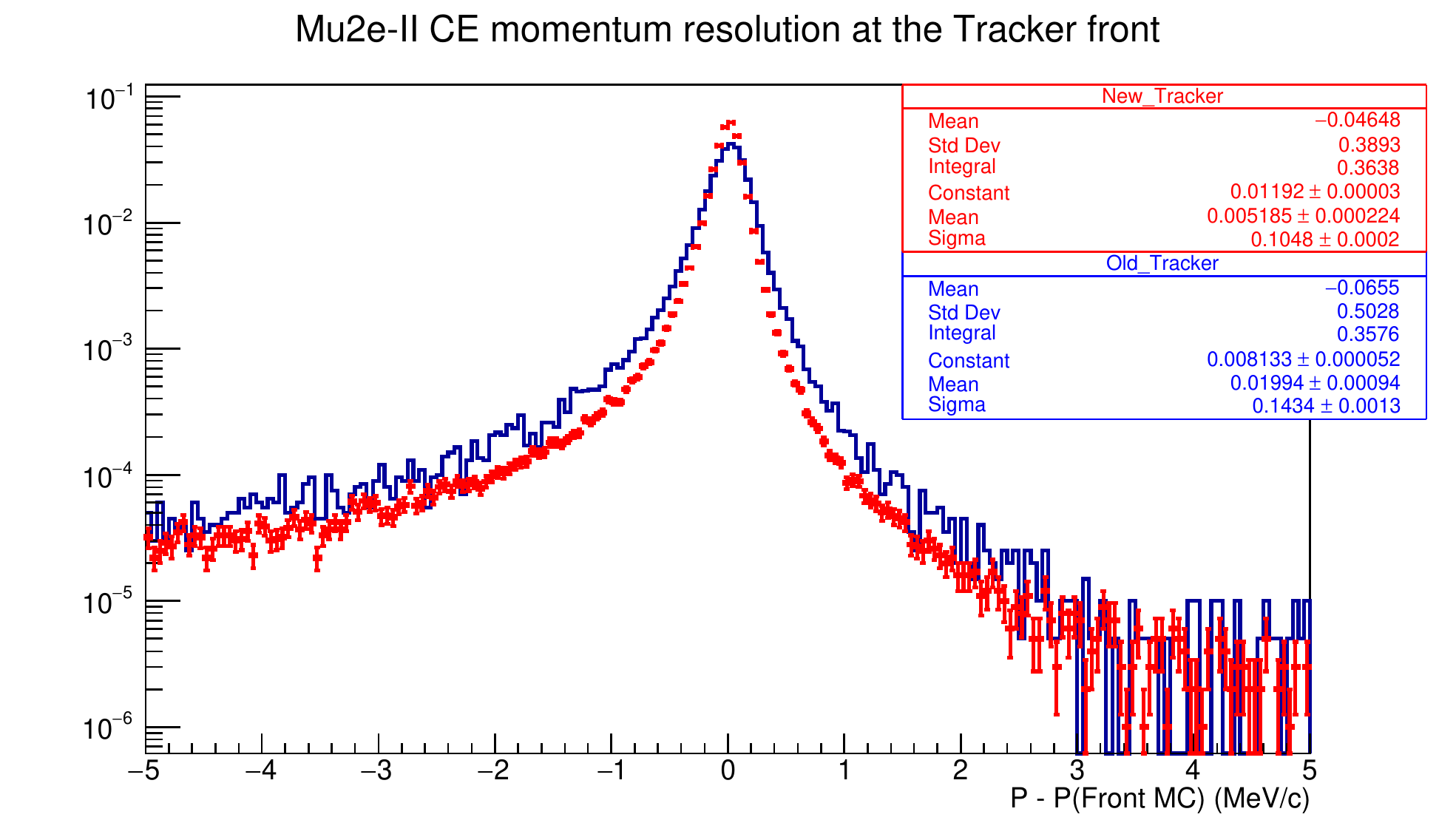}
    \caption{
      Tracker resolution for Mu2e-II (in red) compared to the Mu2e (in blue) era tracker simulation. The
      reconstructed track momentum at the front of the tracker is compared to the MC true momentum
      at the front of the tracker, without any selection cuts applied. The distributions are normalized
      to per simulated conversion electron event.
    }
    \label{fig:tracker_res_noTrkID}
\end{figure}

\begin{figure} [t]
    \centering
    \includegraphics[width=3in]{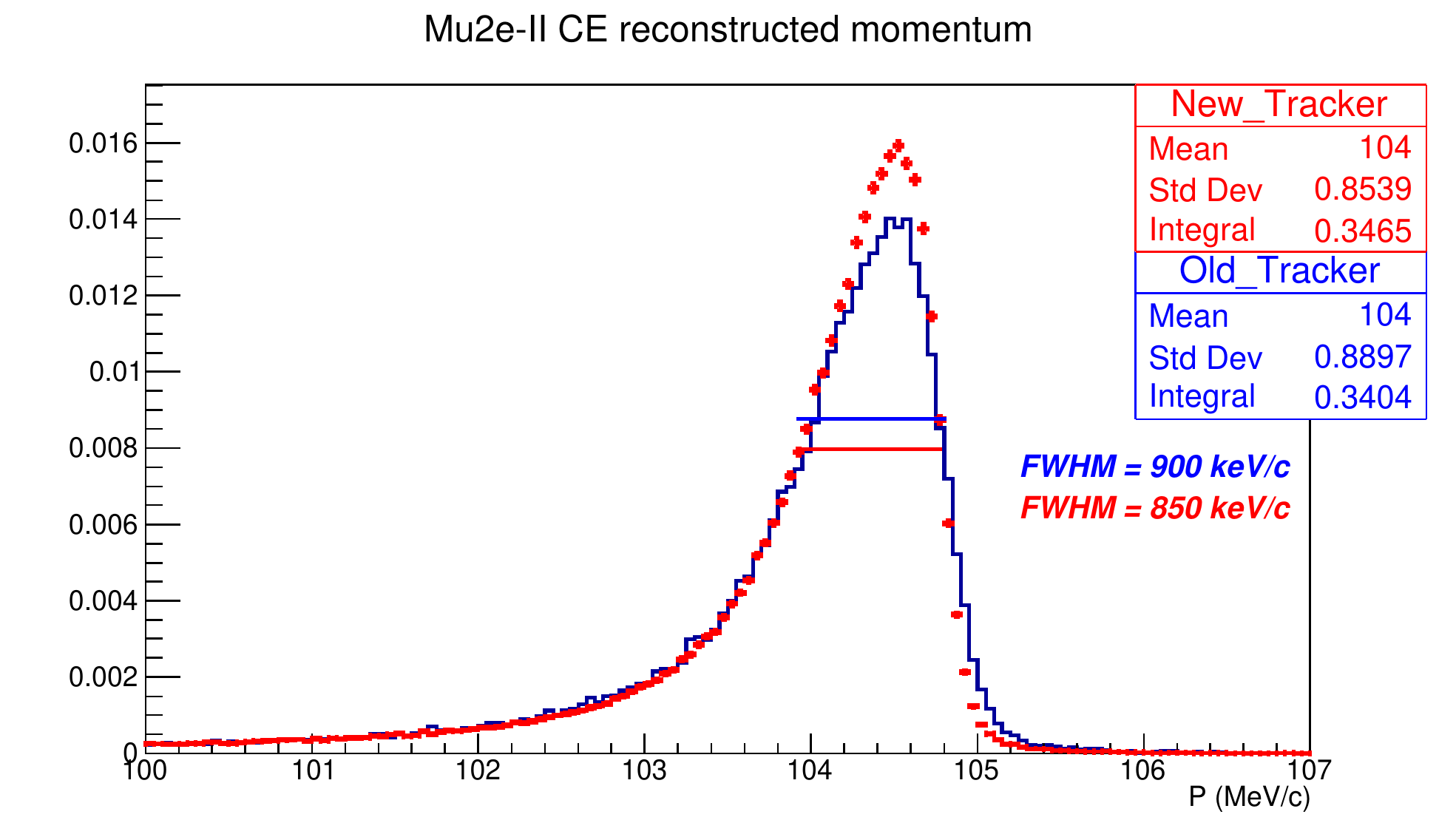}
    \caption{
      Reconstructed track momentum for the conversion electron signal for Mu2e-II (in red) compared to the
      Mu2e (in blue) era tracker simulation, without any selection cuts applied. The distributions are normalized
      to per simulated conversion electron event.
    }
    \label{fig:tracker_p_noTrkID}
\end{figure}

Figs.~\ref{fig:tracker_res_noTrkID} and \ref{fig:tracker_p_noTrkID} show how the resolution and conversion electron momentum distributions look before
and after updating the tracker geometry to use thinner straw walls.
The thinner straw tracker has a better core resolution than the Mu2e era tracker design. The FWHM of the reconstructed momentum spectrum
is not significantly improved due to the large contribution from the energy losses in the stopping target and the inner proton absorber, which protects
the tracker from the significant flux of protons emitted in muon captures in the stopping target.

\begin{figure} [t]
    \centering
    \includegraphics[width=3in]{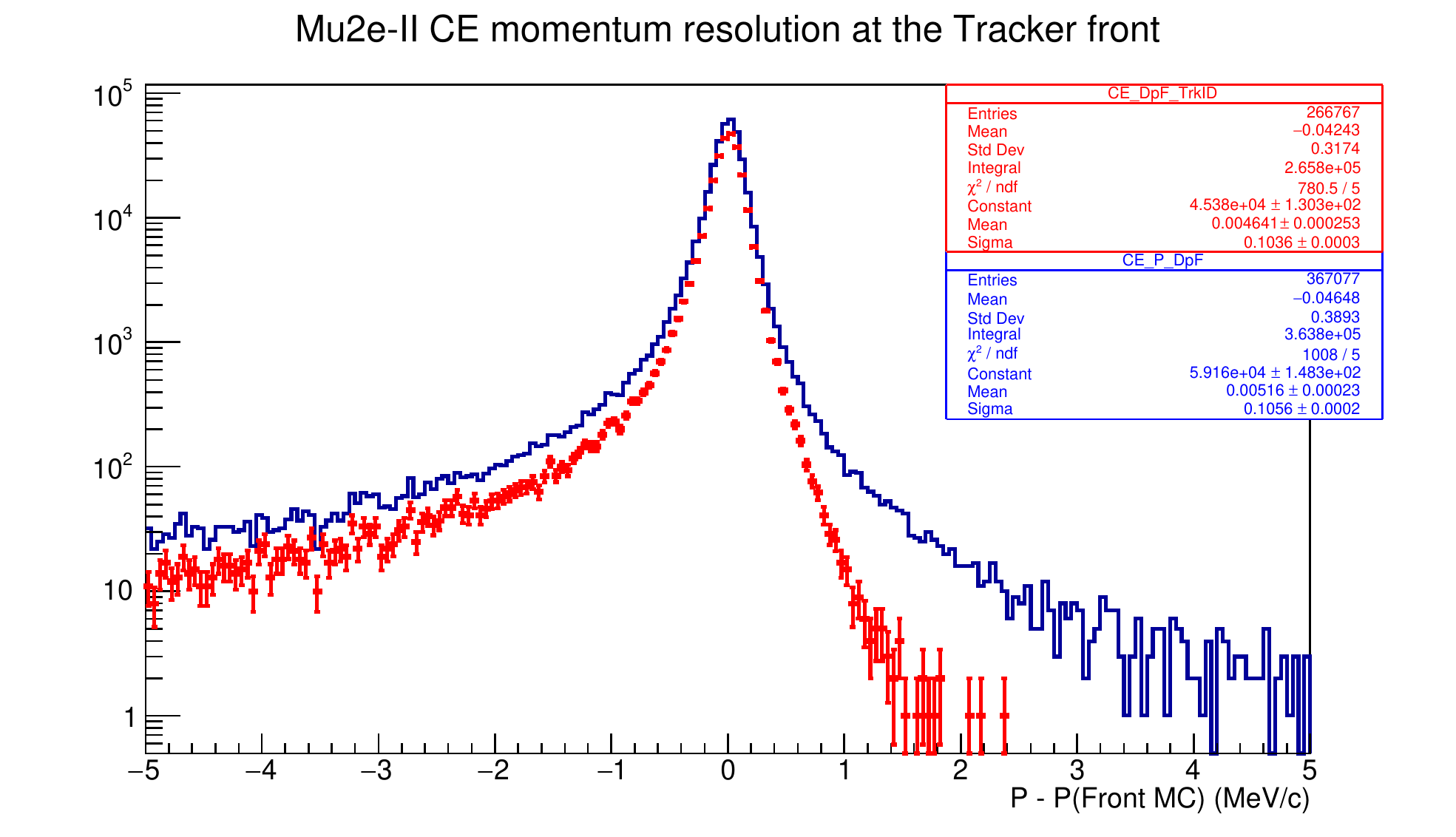}
    \caption{
      Tracker resolution for the conversion electron signal at Mu2e-II. The
      reconstructed track momentum at the front of the tracker is compared to the MC true momentum
      at the front of the tracker, before (in blue) and after (in red) selection cuts are applied.
      The distributions are normalized to per simulated conversion electron event.
    }
    \label{fig:ce_dpf_TrkID}
\end{figure}

Fig.~\ref{fig:ce_dpf_TrkID} shows the comparison between the resolution of the tracker for the Mu2e-II era tracker geometry (with 8 $\mu$m, lower mass straws) before
and after the selection cuts outlined in Sec.~\ref{selection cuts} were applied. The right-side tail of the resolution distribution is the
most dangerous, as it leads to lower momentum DIO electrons entering the signal window due to a higher reconstructed track momentum. The selection
cuts, mostly the ANN based track quality selection, significantly suppress these poorly reconstructed events.

\begin{figure} [t]
    \centering
    \includegraphics[width=3in]{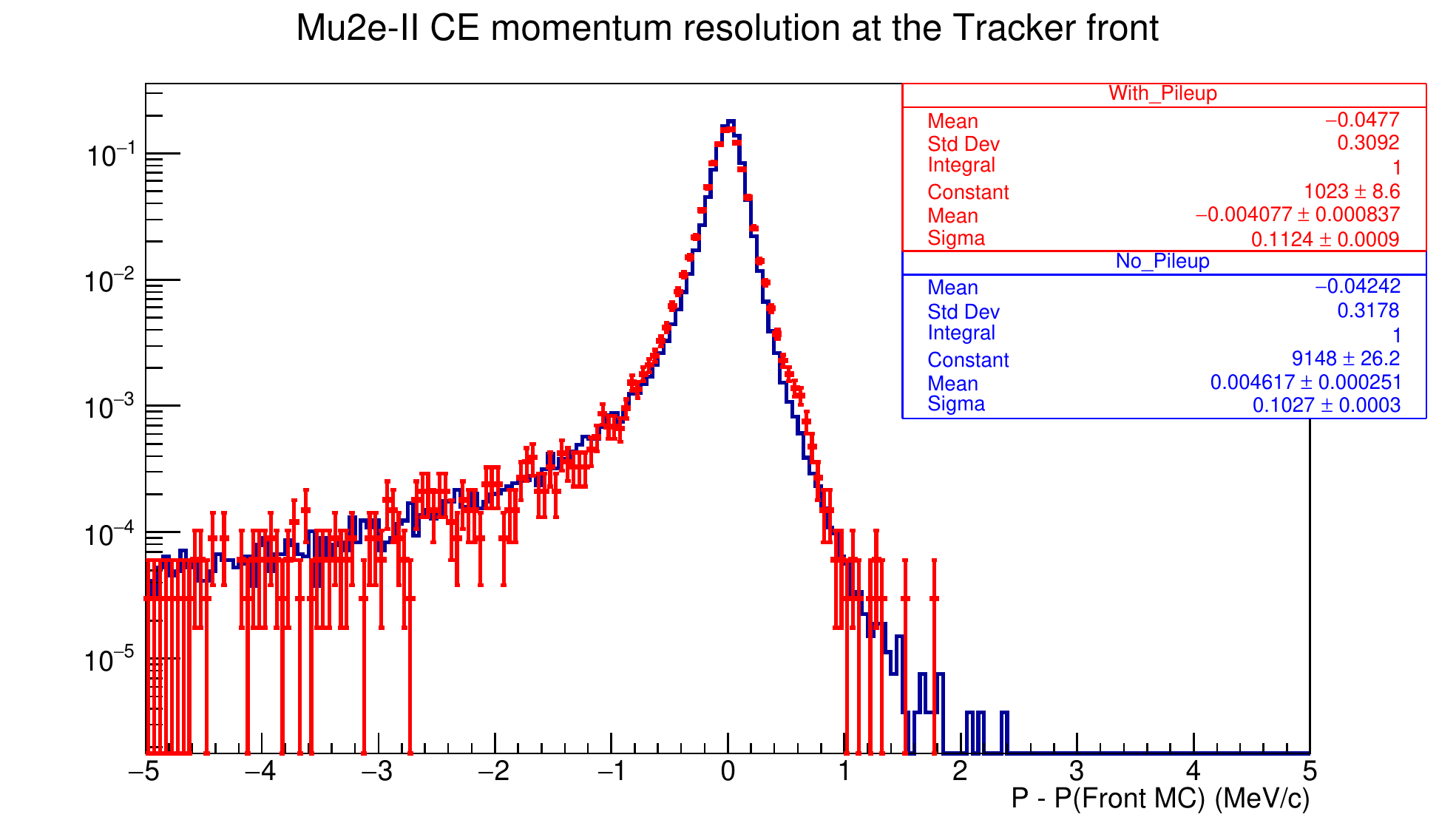}
    \caption{
      Tracker resolution for the conversion electron signal at Mu2e-II before (in blue) and after (in red) the
      introduction of pileup. The reconstructed track momentum at the front of the tracker is compared to the MC true momentum
      at the front of the tracker, after selection cuts are applied. The distributions are normalized to unity.
    }
    \label{fig:ce_mix_dpf_TrkID}
\end{figure}

\begin{figure} [t]
    \centering
    \includegraphics[width=3in]{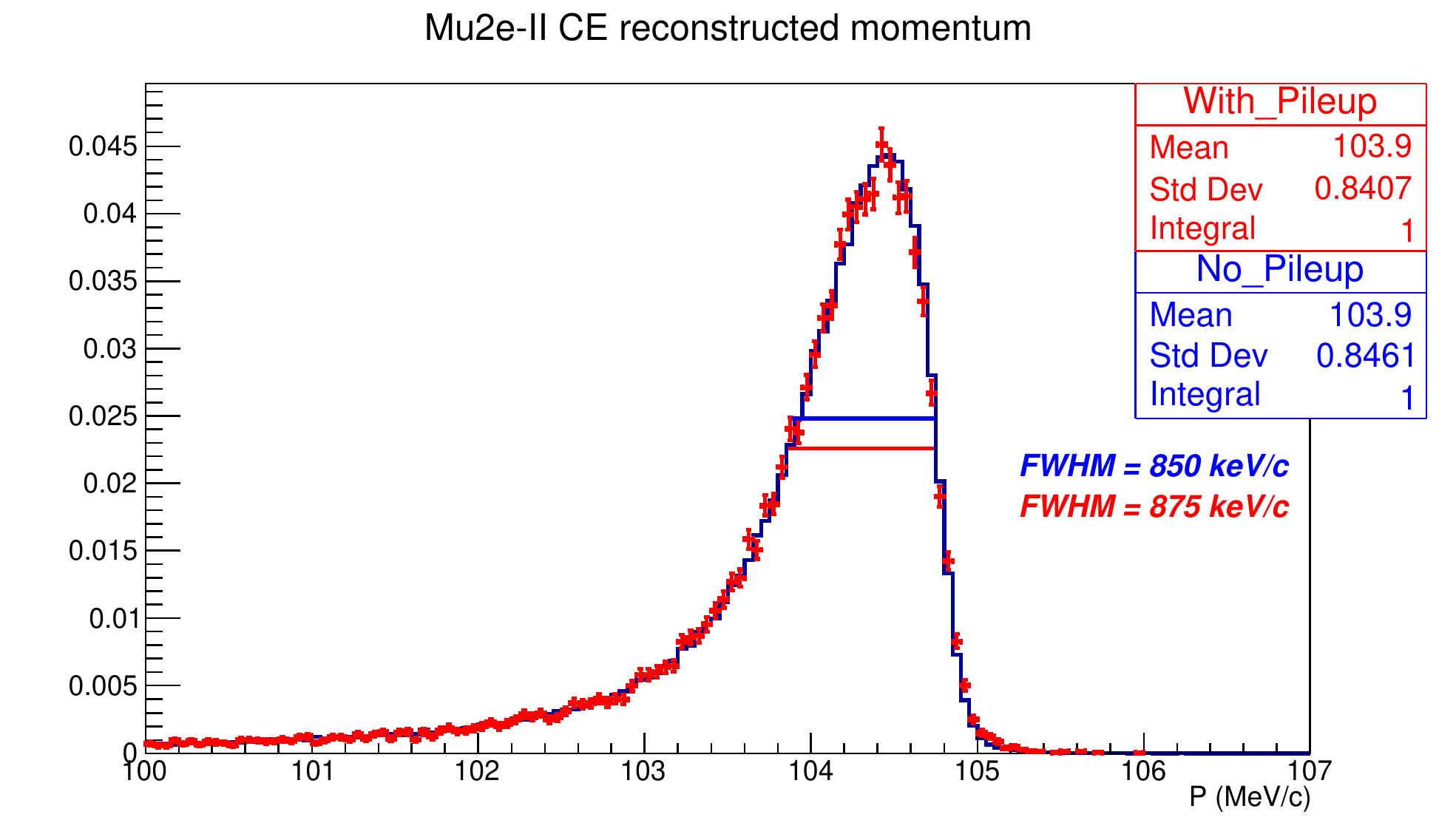}
    \caption{
      Reconstructed track momentum for the conversion electron signal at Mu2e-II,
      before (in blue) and after (in red) the introduction of pileup, after selection cuts are applied.
      The distributions are normalized to unity.
    }
    \label{fig:ce_mix_p_TrkID}
\end{figure}

Figs~\ref{fig:ce_mix_dpf_TrkID} and \ref{fig:ce_mix_p_TrkID} show how the resolution and electron momentum distributions changed with pileup.
The impact on the resolution and FWHM is minor, with a small degradation of both after pileup is introduced to the simulation.
This shows the robustness of the track reconstruction algorithms to changes in the detector occupancies.

\subsection{DIO background estimate}

\begin{figure} [t]
    \centering
    \includegraphics[width=3in]{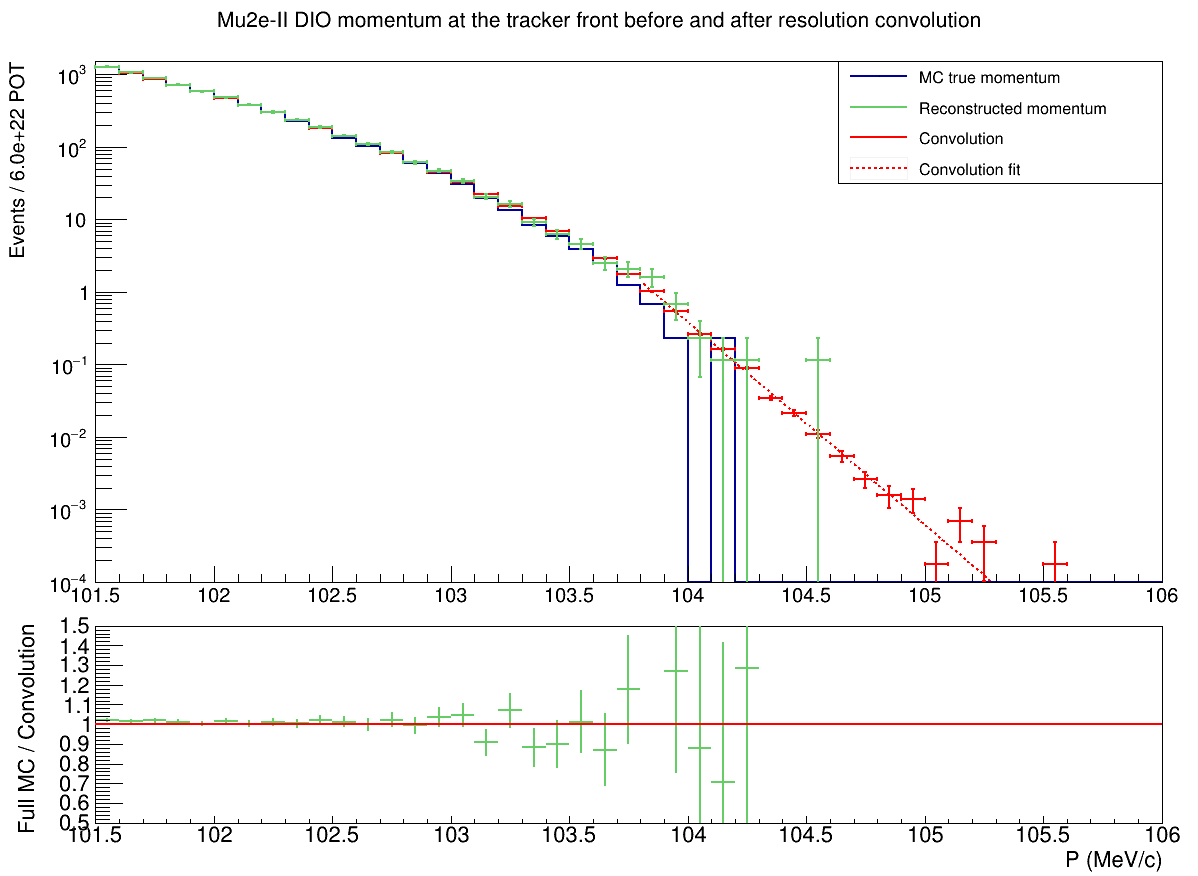}
    \caption{
      The top plot compares the MC true DIO electron momentum to the reconstructed momentum using the full MC simulation
      as well as to a convolution of the true momentum at the tracker with the tracker resolution. The right side tail of the convolution
      is fit with an exponential. The bottom plot is a ratio of the full simulation over the expectation from the convolution.
    }
    \label{fig:dio_bkg_fit}
\end{figure}

The DIO background drives the choice of the low edge of the signal window, as the background
rapidly increases with decreasing momentum. The DIO background is sensitive to low probability tails
in the tracker resolution, which can lead to lower momentum DIO tracks entering the signal window. Due
to the computational requirements to generate a sufficiently large DIO dataset with pileup and in the
interest of time, we estimate this background using a toy MC technique. A DIO electron dataset was generated
without pileup, and the true electron momentum at the tracker was convolved with the tracker resolution to
estimate the reconstructed DIO spectrum. The high momentum tail of this reconstructed spectrum was fit with
an exponential, which was used to estimate the background for the signal window optimization. The final
fit is shown in Figure \ref{fig:dio_bkg_fit}, where the convolved spectrum agrees well with the lower statistics
full MC simulation.

\subsection{Sensitivity Estimate and Justification}

To estimate the Mu2e-II sensitivity to $\mu^- \rightarrow e^-$, the two-dimensional time and momentum signal window is first
optimized assuming the two-dimensional expected background and signal distributions. The optimization requires a metric
to optimize, where the \textit{mean} discovery potential of the experiment is used, as was done in Ref.~\cite{2022_Mu2e_SU2020}.
The mean discovery potential is the value of $R_{\mu e}$ corresponding to an \textit{average} deviation of $5\sigma$ from
the background-only hypothesis.
The discovery threshold is defined as at least a $5\sigma$ deviation (p = $2.867\times 10^{-7}$) from the background-only
hypothesis, or $P(n \geq n_{\rm obs} | \mu_{\rm bkg}) < 2.867\times 10^{-7}$.
This is not as clear of a sensitivity value as the \textit{median} discovery potential,
which is the minimum value of $R_{\mu e}$ such that at least 50\% of identical experiments would claim a discovery,
but has the advantage of being less sensitive to the discrete nature of the measurement and avoids known pathologies of optimizations using median values.

The optimized signal window is $104.05 < p < 104.90$ MeV/c and $690 < T_{0} < 1650$ ns.
For the 5-year Mu2e-II data taking run, the total expected background in the signal window is 0.47 events. The SES
in this window is $R_{\mu e} = 3.3\times 10^{-18}$, over five orders of magnitude below the current limit from SINDRUM II,
$R_{\mu e} < 7\times 10^{-13}$ at 90\% CL. For an expectation of 0.47 events, the median expected 90\% CL upper limit on the
mean expected number of signal events using the Feldman-Cousins confidence belt construction \cite{FELDMANCOUSINS1998} is 1.97 events,
or $R_{\mu e} < 6.4\times 10^{-18}$. This is a five order of magnitude improvement over the SINDRUM II limit and an order of magnitude improvement
on the sensitivity of the Mu2e experiment.
For the Mu2e-II background expectation of 0.47 events, a discovery requires an observation of at least 8 events,
and the median discovery potential is $R_{\mu e} = 2.3\times 10^{-17}$.

\subsection{Summary table}

Table~\ref{tab:background_summary} gives a break-down of the expected yields for each background process for Mu2e (Runs I and II combined) and the
Mu2e-II study presented here, with the assumptions described above. In this table we assume 5 years of data taking, equal to $6 \times 10^{22}$ protons on
target and $5.5 \times 10^{18}$ stopped muons.
The upper and lower edges of the momentum and time selection windows are optimized using the mean discovery potential,
as described previously.
The optimal window is determined to be $104.05 < p < 104.90$ MeV/c and $690 < T_0 < 1650$ ns.

\begin{table*}[t!]
 \renewcommand{\baselinestretch}{1.0}\normalsize
  \caption{Summary of the expected background rates for the Mu2e experiment compared to the expected background rates for the Mu2e-II experiment.
    The Mu2e signal window is $103.85 < p < 104.90$ MeV/c and $700 < t < 1695$ ns and the Mu2e-II signal window optimized
    for the 5-year running plan is $104.05 < p < 104.90$ MeV/c and $690 < t < 1650$ ns. It is important to note that the Mu2e-II numbers are for
    the geometry described which includes the carbon ball production target and aluminum foil stopping target designs.}
  \centering
  \begin{tabular}[t]{l|c|c}
    Results                 &   Mu2e              &  Mu2e-II (5-year)    \\
    \hline
    Backgrounds             &                     &                      \\
    \hline
    DIO                     &    $ 0.144 $        &  $ 0.263$            \\
    Cosmics                 &    $ 0.209 $        &  $ 0.171$            \\
    RPC (in-time)           &    $ 0.009 $        &  $ 0.033$            \\
    RPC (out-of-time)       &    $ 0.016 $        &  $<0.0057$           \\
    RMC                     &    $<0.004 $        &  $<0.02  $           \\
    Antiprotons             &    $ 0.040 $        &  $ 0.000 $           \\
    Decays in flight        &    $<0.004 $        &  $<0.011 $           \\
    Beam electrons          &    $0.0002 $        &  $<0.006 $           \\
    \hline
    Total                   &     0.41            &    0.47             \\
    \hline
    \hline
    N(muon stops)           &   $6.7 \times 10^{ 18}$   &   $5.5 \times 10^{ 19}$  \\
    SES                     &   $3.01\times 10^{-17}$   &   $3.25\times 10^{-18}$  \\
    $R_{\mu e}$(discovery)   &   $1.89\times 10^{-16}$   &   $2.34\times 10^{-17}$  \\
    $R_{\mu e}$(90\% CL)     &   $6.01\times 10^{-17}$   &   $6.39\times 10^{-18}$  \\
    \hline
  \end{tabular}
  \label{tab:background_summary}
\end{table*}

\subsection{Discussion and Future Improvements}

We have presented a preliminary, but detailed, study to determine the estimated sensitivity to neutrinoless muon to electron conversion
in the Mu2e-II experiment.
We have assumed a carbon based, conveyor style production target, and an aluminum, foil stopping target. The tracker mass has been
reduced by using thinner straws of just 8 $\mu$m thick walls. A barium fluoride calorimeter replaces the Mu2e cesium iodide calorimeter.
The study took advantage of the sophisticated simulation and reconstruction software devised for the Mu2e experiment, further optimization
of the reconstruction is required to deal with tracker occupancies seen at Mu2e-II.
The estimated SES  is found to be $3.25\times 10^{-18}$ for a 5 year running period ($ 6 \times 10^{22}$ POT and $5.5 \times 10^{18}$
stopped muons).
This is a factor of $10^{5}$ below the current upper limit, and at least an order of magnitude lower than the Mu2e projection.
There is still a large amount of R\&D and software optimization to be done, and it is likely that this is a conservative estimate.

Further optimization of the production and stopping targets could improve the stopped muon yields.

A wedge absorber could be deployed in the transport solenoid to effectively cool the incoming muons, pushing them to lower momentum, and increasing
the fraction which are stopped in the stopping target. Around 37$\%$ of muons entering the DS are stopped in the target.
These are the lower momentum muons ($< 50$ MeV/c). One way to improve the achievable SES is to increase the number of stopped muons.
A low-Z disk absorber, made of lithium hydride (LiH) or beryllium, within the TS or at the entrance to the DS could distort
the spectrum such that higher momentum muons are ``pushed" into the $< 50$ MeV/c region and thus are stopped.
Preliminary studies showed that although higher momentum muons could be
pushed to lower momenta, the effect was counteracted by the lower momentum muons being lost to the new absorbers.
At too low thickness the effect on the incoming momenta is too small to make any significant change in the number of stops and for
anything $> 0.1 $ mm too many lower momentum muons are lost, decreasing the overall number of stopped muons.

Further optimization and studies are required to understand how best to alter the incoming momentum distribution to achieve the desired improvement
in number of stops.
Wedge absorbers have been used in other muon experiments: MICE \cite{neuffer2017use} and the muon g - 2 experiment \cite{PhysRevAccelBeams.22.053501}.
This design takes advantage of the correlation in the muon position in the XY plane and its momentum after traversing the TS.
The wedge shape can be used to slow the higher momentum muons, which pass through the thickest part, while the lowest momentum muons pass through the thinnest
region and thus experience little energy loss.

The momentum resolution could be improved through removal of the inner proton absorber within the detector solenoid.
The occupancies are highly impacted by beam electrons, shielding in the transport solenoid could reduce the flux of these electrons, again improving
our resolution.
There's also substantial tracker R\&D where alternative gases, for example, could be used within the straws, again improving resolution through better
timing separation.

\begin{acknowledgments}
We are grateful for the vital contributions of the Fermilab staff and the technical staff of the participating institutions. This work was supported by the US Department of Energy; the Istituto Nazionale di Fisica Nucleare, Italy; the Science and Technology Facilities Council, UK; the Ministry of Education and Science, Russian Federation; the National Science Foundation, USA; the National Science Foundation, China; the Helmholtz Association, Germany; and the EU Horizon 2020 Research and Innovation Program under the Marie Sklodowska-Curie Grant Agreement Nos. 101006726, 101003460, 734303, 822185, and 858199. Members of the Mu2e Collaboration were involved in the preparation of this document, using the resources of the collaborating institutions, including the Fermi National Accelerator Laboratory (Fermilab), a U.S. Department of Energy, Office of Science, HEP User Facility. Fermilab is managed by Fermi Research Alliance, LLC (FRA), acting under Contract No. DE-AC02-07CH11359.
\end{acknowledgments}

\appendix*

\section{Software Tools}
Several different software simulation tools were used in the studies done for this paper. We discuss these tools here.

\subsection{FLUKA}
\label{sec:FLUKA}

The FLUKA~\cite{Ferrari2005, BOHLEN2014211, FLUKAweb} radiation transport and interaction code is a fully integrated particle physics Monte Carlo simulation package containing implementations of sound and modern physical models. FLUKA simulates the interaction and transport of 60 different particles, and also enables an estimate of the residual radiation and activation. Editable {\tt FORTRAN} routines allow including user-defined settings and modifications. A powerful graphical interface~\cite{Vlachoudis2009} facilitates the editing of FLUKA input, execution of the code. and visualization of the output.
The Mu2e magnetic field maps have also been interfaced to FLUKA and allow a realistic modeling of the trajectories of charged particles.

\subsection{GEANT4}
\label{sec:GEANT4}
Geant4 is a toolkit for simulating the passage of particles through matter \cite{Geant4}. It includes particle tracking and hits, provides for complex experimental geometries, and offers a variety of physics models that cover many orders of magnitude of energy. Geant4 serves as the simulation tool for the Mu2e/Mu2e-II particle event chain starting with particle production resulting from the proton beam interacting with the production target, continuing with particle transport through the solenoids, and lastly modeling interactions of particles in the stopping target and the subsequent interactions in the detectors. In the Geant4 studies done for this paper, we used version 10.7 patch-02. The hadronic interactions are modeled according to the "ShieldingM" model provided by Geant4.

\subsection{MARS15}
\label{sec:MARS15}
The MARS15 code~\cite{ref:mokhov_striganov_2007, 
ref:mokhov_james_2016} is a fully integrated particle transport and
interaction code consisting of a set of Monte Carlo programs
(written in Fortran and C++) that allow modeling and simulation
of interactions of all types of leptons, hadrons, photons, and
heavy ions with matter. Interactions in the energy range spanning
from a fraction of an electronvolt up to hundreds of TeV can be simulated. A model geometry can be described in several ways
including a built-in extended geometry description language, 
ROOT geometry, and the import of external gdml files. MARS15
incorporated CEM and LAQGSM~\cite{Mashnik:2005fx} generators for the description of
particle-nuclei and particle-particle interactions above ~100 MeV.
Also, MARS15 includes the DeTra~\cite{ref:aarnio_1998} code to model nuclear decays and transmutations as well as EGS5~\cite{ref:hirayama_2005} code to model low-energy electromagnetic showers.

\subsection{TrackToy}
\label{sec:TrackToy}

The TrackToy hybrid Monte Carlo is used to quickly estimate the relative
importance of different tracker and tracker-region parameters on the Mu2e-II signal sensitivity.
In TrackToy, muon particle 4-vectors from the Geant4 simulation (see Appendix~\ref{sec:GEANT4}) are propagated as quasi-helical trajectories through a simplified model of the Mu2e Detector Solenoid (DS) magnetic field to the stopping target, using algorithms for helix propagation in non-uniform fields from the KinKal package.
The stopping target is modeled as a hollow cylinder with an adjustable but uniform mass density.
The stopping position and time of each beam muon is modeled from its predicted helical path through the target material, using the muon stopping ranges tabulated by the Particle Data Group (PDG) \cite{MuStopRange}.
The TrackToy stopping rate, times, and positions were found to be in good agreement with predictions from the Geant4 simulation.

Daughter electrons from stopped muons which Decay in Orbit (DIO) or convert are generated from the muon stop positions and times assuming an isotropic angular distribution, and a decay time distribution sampled from the material-dependent muon lifetime exponential.
Conversion electrons are given the target material endpoint energy, DIO electron energies are generated by sampling the distribution tabulated in \cite{2017szafron}.
The average daughter electron energy loss in the target material is estimated using the EStar \cite{EStar} tables from NIST.
Electrons emitted backwards (away from the tracker) may reflect in the magnetic gradient and thus re-enter the target, resulting in additional energy loss.  The actual daughter electron energy loss is modeled by sampling a Moyal distribution with average and RMS set to the EStar table values.

A thin polyethylene cylindrical shell between the target and the tracker is used to stop or degrade protons emitted following nuclear muon capture in Mu2e before they reach the tracker.
This proton absorber (IPA) protects the tracker sense wires from the charge load and large pulses generated by highly-ionizing protons.
It as assumed this shield, or something like it, will also be needed for Mu2e-II.
Several models of the IPA geometry and material are supported in TrackToy.
Daughter electron energy loss in the IPA is modeled the same way as target energy loss.

The tracker active volume is modeled as a hollow cylinder.
The density of cells, the cell mass, and the hit efficiency and resolution properties are all parameters.
Daughter electrons trajectories are propagated geometrically through the tracker active volume to predict the number of cells traversed.
The straw azimuthal distribution is selected randomly, based on the distribution of straws angles in the Mu2e tracker.  The straw material traversed is predicted by sampling a flat impact parameter distribution.
The energy loss and multiple scattering angle mean and RMS from the straw material are predicted using the KinKal material model.
Tracker hit times are simulated assuming a constant drift velocity and signal propagation velocity, smeared using a Gaussian resolution model.

Daughter electrons exiting the Tracker intersect a 2-disk Calorimeter, with the disk geometry taken from Mu2e.
Daughters trajectories that traverse more than a few mm of a disk generate a signal at shower max (or the disk exit position for short paths), with position and time Gaussian smeared according to spatial and timing resolution parameters.
These calorimeter signals are used as time constraint hits in the KinKal track fit described below.

Simulated hits are fit using a multi-stage configuration of the KinKal fit.
The initial stage uses hits as digital signals, constraining the track according to the straw size and the signal time.
The final stage uses the drift information to constrain the track, assigning left-right ambiguities according to the sign of the tracks estimated angular momentum about the wire, for hits at least 0.5 mm from the wire.
Simulated annealing is used to find the global optimum ambiguity assignment.

\begin{figure*} [t]
    \centering
    \includegraphics[width=6.5in]{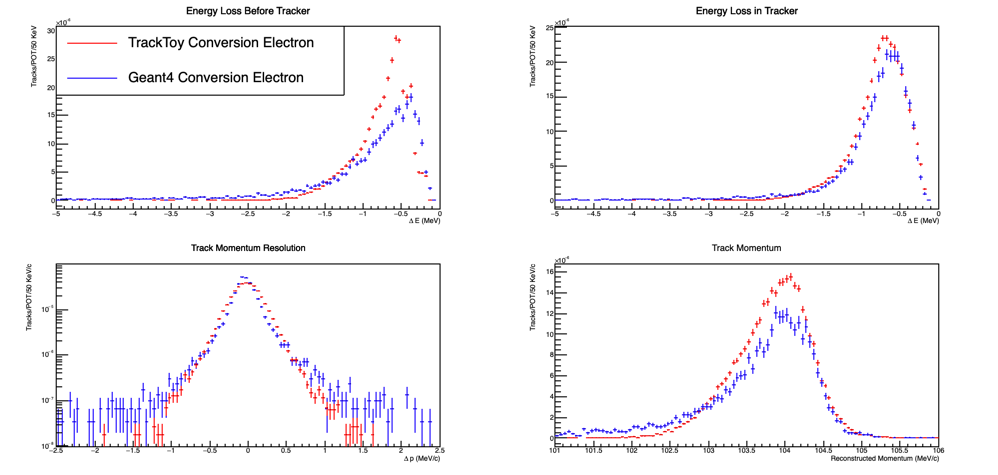}
    \caption{
      Comparison of conversion electron tracks simulated by the TrackToy hybrid Monte Carlo and the Mu2e Geant4 Monte Carlo.  The top row of plots compares the energy loss predictions in the passive material before the tracker (left) and the tracker itself (right).  The bottom row of plots compare the reconstructed track momentum resolution (left) and the reconstructed momentum distribution (right).  The vertical axes are scaled to the number of Protons On Target (POT) simulated.
    }
    \label{fig:TrackToy}
\end{figure*}

Figure \ref{fig:TrackToy} plots compare the TrackToy conversion electron track fits with tracks reconstructed in Mu2e Geant4 simulations, with TrackToy configured to emulate the Mu2e experiment.
The TrackToy simulation with KinKal fit runs at approximately 100 Hz on a macbook pro.

\bibliographystyle{utcaps_mod}
\bibliography{mu2eii}

\end{document}